\documentclass[aps,prd,superscriptaddress,nofootinbib,tighten,preprint]{revtex4}
\pdfoutput=1
\usepackage[english]{babel}
\usepackage{amsmath}
\usepackage{amssymb}
\usepackage{color}
\usepackage{subfigure}
\usepackage{hyperref}
\usepackage{amsfonts}
\usepackage{amssymb}
\usepackage{multirow}
\usepackage{graphicx}
\usepackage{slashed}
\usepackage{float}
\usepackage{pstricks}
\usepackage[toc,page]{appendix}

\newcommand{\be}{\begin{equation}}
\newcommand{\ee}{\end{equation}}
\newcommand{\bea}{\begin{eqnarray}}
\newcommand{\eea}{\end{eqnarray}}

\catcode`@=12
\def\la{\mathrel{\mathchoice {\vcenter{\offinterlineskip\halign{\hfil
$\displaystyle##$\hfil\cr<\cr\sim\cr}}}
{\vcenter{\offinterlineskip\halign{\hfil$\textstyle##$\hfil\cr<\cr\sim\cr}}}
{\vcenter{\offinterlineskip\halign{\hfil$\scriptstyle##$\hfil\cr<\cr\sim\cr}}}
{\vcenter{\offinterlineskip\halign{\hfil$\scriptscriptstyle##$\hfil\cr<\cr\sim
\cr}}}}}
\def\ga{\mathrel{\mathchoice {\vcenter{\offinterlineskip\halign{\hfil
$\displaystyle##$\hfil\cr>\cr\sim\cr}}}
{\vcenter{\offinterlineskip\halign{\hfil$\textstyle##$\hfil\cr>\cr\sim\cr}}}
{\vcenter{\offinterlineskip\halign{\hfil$\scriptstyle##$\hfil\cr>\cr\sim\cr}}}
{\vcenter{\offinterlineskip\halign{\hfil$\scriptscriptstyle##$\hfil\cr>\cr\sim
\cr}}}}}

\def\mET{E_T \hspace{-1.2em}/\;\:}

\begin{document}  
\title{Explaining Dark Matter and Neutrino Mass
in the light of TYPE-II Seesaw Model} 

\author{Anirban Biswas}
\email{anirbanbiswas@hri.res.in}
\affiliation{Harish-Chandra Research Institute, Chhatnag Road,
Jhunsi, Allahabad 211 019, India}
\affiliation{Homi Bhabha National Institute,
Training School Complex, Anushaktinagar, Mumbai - 400094, India}

\author{Avirup Shaw}
\email{avirup.cu@gmail.com}
\affiliation{Department of Theoretical Physics, Indian
Association for the Cultivation of Science,\\
2A $\&$ 2B Raja S.C. Mullick Road, Jadavpur,
Kolkata 700 032, India}


\begin{abstract}  
\noindent 
With the motivation of simultaneously explaining dark matter and
neutrino masses, mixing angles, we have invoked the Type-II seesaw model
extended by an extra $SU(2)$ doublet $\Phi$. Moreover, we have imposed
a $\mathbb{Z}_2$ parity on $\Phi$ which remains unbroken as the
vacuum expectation value of $\Phi$ is zero.\,\,Consequently, the lightest
neutral component of $\Phi$ becomes naturally stable
and can be a viable dark matter candidate.\,\,On
the other hand, light Majorana masses for neutrinos have been
generated following usual Type-II seesaw mechanism.
Further in this framework, for the first
time we have derived the full set of vacuum stability and unitarity
conditions, which must be satisfied to obtain a stable vacuum as well
as to preserve the unitarity of the model respectively. 
Thereafter, we have performed extensive phenomenological studies
of both dark matter and neutrino sectors considering all possible
theoretical and current experimental constraints. Finally,
we have also discussed a qualitative collider signatures of
dark matter and associated odd particles at the 13 TeV Large
Hadron Collider.    
\vskip 5pt \noindent  
\end{abstract}  

\maketitle

\section{Introduction}
\label{intro}
The observation of various satellite borne experiments
namely WMAP \cite{Hinshaw:2012aka} and more
recently Planck \cite{Ade:2015xua}, establish
firmly the existence of dark matter in the Universe over the ordinary
luminous matter. The results of these experiments are indicating that
more than 80\% matter content of our Universe has been made of an unknown
non-luminous matter or dark matter. In terms of cosmological language,
the amount of dark matter present at the current epoch is expressed
as $\Omega\,h^2 = 0.1199 \pm 0.0027$ \cite{Ade:2015xua} where
$\Omega$ is known as the relic density of dark matter and
$h$ is the present value of Hubble parameter $H_0$ normalised by 100. In spite
of this precise measurement, the particle nature of dark matter still
remains an enigma. The least we can say about a dark matter candidate
is that it is electrically neutral and must have a lifetime greater
than the present age of the Universe. Moreover, N-body simulation
requires dark matter candidate to be non-relativistic ({\it cold})
at the time of its decoupling from the thermal plasma to explain small
scale structures of the Universe \cite{Frenk:2012ph}. Unfortunately, none of the
Standard Model (SM) particles can fulfil all these properties and
hence there exist various beyond Standard Model (BSM) theories in the
literature \cite{Jungman:1995df, Bertone:2004pz, Hooper:2007qk,
ArkaniHamed:2008qn, Kusenko:2009up, Feng:2010gw}
containing either at least one or more dark matter
candidates. Among the different kinds of dark matter candidates, Weakly
Interacting Massive Particle (WIMP) \cite{Gondolo:1990dk,
Srednicki:1988ce} is the most favourite class
and so far, neutralino \cite{Jungman:1995df} in the
supersymmetric extension of the SM is
the well studied WIMP candidate. There are also a plethora of well motivated
non-supersymmetric BSM theories which have dealt with WIMP type dark matter
candidate \cite{Silveira:1985rk, Burgess:2000yq, McDonald:1993ex,
Barbieri:2006dq, LopezHonorez:2006gr, Kim:2008pp, Hambye:2008bq}.
Since the interaction strength of a WIMP is around week
scale hence various experimental groups \cite{Akerib:2016vxi,
Amole:2017dex, Armengaud:2016cvl, Agnese:2014aze}
have been trying to detect it
directly over the last two decades by measuring the recoil energies of
detector nuclei scattered by WIMPs. However, no such event has been
found and as a result, dark matter nucleon elastic scattering cross
section is getting severely constrained. Currently most stringent bounds on
dark matter spin independent scattering cross section have been reported
by the XENON 1T collaboration \cite{Aprile:2017iyp}\footnote{Recently,
PandaX-II  collaboration \cite{1708.06917} has published their results on
the exclusion limits of WIMP-nucleon spin independent scattering
cross section ($\sigma_{\rm SI}$). Although, their results are
most stringent for a WIMP of mass larger than 100 GeV, are very similar
with the upper limits of XENON 1T.}. Future direct detection experiment
like DARWIN \cite{Aalbers:2016jon} is expecting to detect or
ruled out the WIMP hypothesis by exploring the entire experimentally
accessible parameter space of a WIMP (just above the neutrino floor). 


On the other hand, neutrinos remain massless in the SM as there
is no right handed counterpart of each ${\nu_{\alpha}}_{L}$
where $\alpha$ is the generation index. However, the existence
of a tiny nonzero mass difference between $\nu_{\mu}$ and $\nu_{\tau}$
has been first confirmed by the atmospheric neutrino data
of Super-kamiokande collaboration \cite{Fukuda:1998mi} from
neutrino oscillation. Thereafter, many experimental groups
\cite{Ahmad:2002jz, Araki:2004mb, Abe:2011sj, Ahn:2012nd} have
precisely measured the mass squared differences and mixing angles
among different generations of neutrinos. In spite of these wonderful
experimental achievements, we still have not properly understood the
exact method of neutrino mass generation. There exist various mechanisms
for generating tiny neutrino masses at tree level (via seesaw mechanisms)
\cite{Minkowski:1977sc, Mohapatra:1979ia, Magg:1980ut, Ma:1998dx, Foot:1988aq,
Ma:2002pf, Mohapatra:1986bd} and beyond \cite{Zee:1985id, Ma:2006km,
Gustafsson:2012vj} by adding extra bosonic or fermionic degrees
of freedom in the particle spectrum of SM. Moreover,
the exact flavour structure in the neutrino sector, which is responsible
for generating such a mixing pattern, still remains unknown to us.
Furthermore, there are other important issues which are yet
to be resolved. For example the particle nature of neutrinos
(i.e. Dirac or Majorana fermion), mass hierarchy
(i.e.\,{\it Normal} or {\it Inverted}),
determination of octant for the atmospheric mixing
angle $\theta_{23}$, CP violation in the leptonic sector
(i.e. measurement of Dirac CP phase $\delta$) etc. More recently,
T2K collaboration \cite{Abe:2017vif} has reported their analysis
of neutrino and antineutrino oscillations where they have excluded
the hypothesis of CP conservation in the leptonic sector (i.e. $\delta
=0$ or $\pi$) at 90\% C.L. Their preliminary result indicate a
range for $\delta$ lies in between third and fourth quadrant. Other neutrino
experiments like DUNE \cite{Acciarri:2015uup},
NO$\nu$A \cite{Adamson:2017gxd} etc. will
address some of these issues in near future.

In the present article we try to cure both of these lacunae of the SM by introducing a Higgs triplet and an extra Higgs doublet to the particle spectrum of SM. Furthermore, we impose a discrete $\mathbb{Z}_2$ symmetry in addition to the SM gauge symmetry. Under this $\mathbb{Z}_2$ symmetry  the triplet field and the SM particles are even while the extra doublet field is odd\footnote{Here the $\mathbb{Z}_2$ odd doublet is analogous to the one in Inert Doublet Model (IDM) ~\cite{Barbieri:2006dq, LopezHonorez:2006gr, Ma:2006km}.}. This kind of BSM scenario has been studied earlier in \cite{Chen:2014lla}. To the best of our knowledge in such set up first time we derive the vacuum stability and unitarity constraints and use these constraints in our phenomenological study. This set up can serve our two fold motivations. First of all, as we have demanded that the extra doublet is odd under $\mathbb{Z}_2$ symmetry, consequently the lightest particle of neutral component of this doublet can play the role of viable dark matter candidate in this scenario. Secondly, with the small vacuum expectation value (VEV) of Higgs triplet field, required to satisfy the electroweak precision test, we can explain small neutrino masses by the Type-II seesaw mechanism \cite{Magg:1980ut, Cheng:1980qt, Lazarides:1980nt, Mohapatra:1980yp, Dev:2013ff, Dev:2013hka} without introducing heavy right handed neutrinos. In the present work, we have explored both the normal and inverted hierarchies of neutrino mass spectra. At this point, we would like to mention that all the  possible current experimental constraints have been taken into account while we investigate the dark matter related issues as well as the generation of neutrino masses and their mixings.

Apart from providing a viable solution to dark matter problem and neutrino mass generation, this scenario contains several non-standard scalars which can be classified into two categories. In one class we have $\mathbb{Z}_2$ even scalars originate from the mixing between triplet fields and SM scalar doublet fields while the three different components of the extra scalar doublet can be represented as $\mathbb{Z}_2$ odd scalars. Therefore, one has the opportunity to explore these non-standard scalars at the current and future collider experiments. In literature one can find several articles where the search of $\mathbb{Z}_2$ even scalars have been explored in context of the Large Hadron Collider (LHC)
\cite{Chun:2003ej, Han:2007bk, Perez:2008ha, Han:2015hba, Han:2015sca, Mitra:2016wpr} as well as at the International Linear Collider (ILC) \cite{Shen:2015pih, Cao:2016hvg, Blunier:2016peh}. However, in this work instead of $\mathbb{Z}_2$ even scalars, we have performed collider search of dark matter and the associated $\mathbb{Z}_2$ odd scalars at the 13 TeV LHC. Among the different final states, we find an optimistic result for $2\ell+\mET$ signal at the 13 TeV LHC with an integrated luminosity of 3000$fb^{-1}$.
 
One should note that, relying on the value of triplet VEV, decay modes
of different non-standard scalars show distinct behaviour. From the
consideration of electroweak precision test the triplet VEV can not be
larger than a few GeV \cite{Aoki:2012jj, Patrignani:2016xqp}. However,
it can vary from $10^{-9}$ GeV to  ${\cal O} (1)$ GeV. Within this
range the non-standard Higgs bosons decay in several distinct channels.
To be more specific, for $ {\rm triplet~VEV} < 10^{-4} ~\rm GeV$,
the doubly charged Higgs dominantly decays into two same-sign leptonic
final state. The latest same-sign dilepton searches at the LHC  
have already put strong lower limit on doubly charged Higgs mass
($>$ 770 - 800 GeV)~\cite{ATLAS:2017iqw}. On the other hand for
${\rm triplet~VEV}  > 10^{-4} ~\rm GeV$, only gauge boson final
state or cascade decays of singly charged Higgs
(if they are kinematically allowed) are possible~\cite{Perez:2008ha,
Melfo:2011nx,Aoki:2011pz,Han:2015hba, Han:2015sca}. The collider search
becomes more involved in this region of triplet VEV due to more
complicated decay patterns of the doubly charged Higgs. As a result,
the lower bound on the mass of the doubly charged Higgs is very
relaxed. Therefore, in this region one can find scenarios where
the mass of doubly charged Higgs may goes down to about 100 GeV 
\cite{Melfo:2011nx, Chabab:2016vqn}. In this article, for all
practical purposes we have considered the triplet VEV greater
than $10^{-4}$  GeV. For example, for the generation of neutrino
mass we set triplet VEV at $10^{-3}$ GeV. Whereas, for the purpose
of dark matter analysis we show our results for two different
values of triplet VEV e.g., $10^{-3}$ GeV and 3 GeV respectively. This is in
stark contrast to the Ref.\;\cite{Chen:2014lla} where the triplet
VEV has been considered less than $10^{-4}$  GeV. Further, for
collider study we have fixed the value of triplet VEV at 3 GeV
and hence the doubly charged Higgs decays into $W^\pm W^\pm$ with
100\% branching ratio.

We organise this article as follows. First we introduce the model with possible interactions and set our conventions in Sec. \ref{2}. Within this section we have also evaluated the vacuum stability and unitarity conditions in detail.
In Sec. \ref{neu}, we discuss the neutrino mass generation via Type-II seesaw mechanism and explain neutrino oscillation data for normal and inverted hierarchies at $3\sigma$ range. The viability of dark matter candidate proposed in this work has been extensively studied in Sec. \ref{dm}, considering all possible bounds from direct and indirect experiments. In Sec. \ref{cldr}, we show the prospects of collider signature of the dark matter candidate of the present model at 13 TeV LHC.
Finally in Sec. \ref{con} we summarize our results.

\section{Type-II Seesaw with Inert Doublet}\label{2}
In this section, we discuss the model briefly. In order to produce a
viable dark matter candidate, we introduce a $\mathbb{Z}_2$ symmetry
in the SM gauge symmetry $SU(2)_{\mathbb{L}}\times U(1)_{\mathbb{Y}}$.
Moreover, to generate the neutrino masses and also having a stable
dark matter candidate, we incorporate a scalar triplet $\Delta$
with hypercharge two and a scalar doublet $\Phi$ with hypercharge
one in the SM fields. Further, we demand that the SM particles and
the triplet $\Delta$ are even under $\mathbb{Z}_2$ parity
while the new doublet $\Phi$ is odd under $\mathbb{Z}_2$ parity.
The field $\Phi$ cannot develop a VEV at the time of electroweak
symmetry breaking as this will break the $\mathbb{Z}_2$ symmetry
spontaneously, which will jeopardize the dark matter stability.
With this newly added $\mathbb{Z}_2$ symmetry, we discuss different
interaction terms involving SM fields, $\Delta$ and $\Phi$. The total
Lagrangian which incorporates all possible interactions can
be written as:
\begin{eqnarray}
{\cal L} = {\cal L}_{\rm Yukawa}+{\cal L}_{\rm Kinetic}-V(H,\Delta,\Phi),
\label{lag}
\end{eqnarray}
where the relevant kinetic and Yukawa interaction terms are respectively
\begin{eqnarray}
\label{y2kinetic}
{\cal L}_{\rm kinetic} &=&\left(D_\mu H\right)^\dag \left(D^\mu H\right)
+{\rm Tr}\left[\left(D_\mu \Delta\right)^\dag
\left(D^\mu\Delta\right)\right]+\left(D_\mu \Phi\right)^\dag
\left(D^\mu \Phi\right), \\
\label{y2Yukawa}
{\cal L}_{\rm Yukawa} &=& {\cal L}_{\rm Yukawa}^{\rm SM}-
\dfrac{Y^\nu_{ij}}{2} L_i^{\sf T}\mathcal{C}i\sigma_2
\Delta L_j+{\rm h.c.}\, .
\end{eqnarray}
The first two terms of ${\cal L}_{\rm kinetic}$ generate the masses
of gauge bosons $W^\pm$ and $Z$ by electroweak symmetry breaking
mechanism (EWSB), however the third term does not contribute to
gauge boson masses as $\Phi$ does not possess any VEV. Here $L_i$
represents $SU(2)_{\mathbb{L}}$ doublet of left handed leptons
where $i$ being the generational index, $Y^\nu$ represents
Yukawa coupling and $\mathcal{C}$ is the
charge conjugation operator. Further, ${\cal L}_{\rm Yukawa}^{\rm SM}$
denotes the Yukawa interactions for all SM fermions. Later, we
will discuss the second term of Yukawa interactions in detail
in the neutrino section (Section \ref{neu}). There is no term
which involves the coupling between $\Phi$ and the SM fermions as
$\Phi$ is odd under $\mathbb{Z}_2$ parity while the SM fermions
are even under $\mathbb{Z}_2$ symmetry. Representations for
the doublets $H$ and $\Phi$ are chosen as
$H^{\sf T}\equiv(h^+~~ (v_d+\eta^0+i z_1)/\sqrt{2})$ and
$\Phi^{\sf T}\equiv (\phi^+~~(\phi^0+i a^0)/\sqrt{2})$ respectively.
The triplet field $\Delta^{\sf T}(\equiv(\Delta^1~~\Delta^2~~\Delta^3))$
transforms as $({\mathbf{3}},\,2)$ under the
$SU(2)_{\mathbb{L}}\times U(1)_{\mathbb{Y}}$ gauge group,
so one can write $\Delta=\frac{\sigma^i}{\sqrt{2}}\Delta^i\,\,(i=1,\,2,\,3)$,
which gives a $2\times 2$ representation given in the following:
\begin{eqnarray}
\Delta=\left( \begin{array}{cc} 
    \delta^+/\sqrt{2} & \delta^{++}  \\ 
    \delta^0 & -\delta^+/\sqrt{2} \\ 
 \end{array}  \right)\,. \
\label{delta}
\end{eqnarray}
In the above $\Delta^1=(\delta^{++}+\delta^0)/\sqrt 2,~\Delta^2=i(\delta^{++}-\delta^0)/\sqrt 2,~\Delta^3=\delta^+$. The neutral component of the triplet field can be expressed as $\delta^0= (v_t + \xi^0 + iz_2)/\sqrt{2}$ where
$v_d$ and $v_t$ are vacuum expectation values of the doublet
$H$ and triplet $\Delta$ respectively. The covariant derivative of
the scalar field $\Delta$ is given by,
\begin{equation} 
D_\mu \Delta = \partial_\mu \Delta + i\frac{g_2}{2}
[\sigma^a W_\mu^a,\Delta]+ig_1B_\mu \Delta \qquad (a=1,\,2,\,3)\,.
\end{equation}
Here $\sigma^i$'s are the Pauli matrices while $g_2$ and $g_1$ are 
coupling constants for the gauge groups $SU(2)_{\mathbb{L}}$
and $U(1)_{\mathbb{Y}}$ respectively.  

Let us discuss the scalar potential given in the following \cite{Chen:2014lla}:
\begin{eqnarray} 
V(H,\Delta,\Phi) &=& -m^2_H(H^\dag H)+\frac{\lambda }{4}(H^\dag H)^2+M^2_\Delta {\rm Tr}(\Delta ^\dag \Delta)+ \left(\mu H^{\sf T}i\sigma_2\Delta^\dag H+{\rm h.c.}\right)\,\nonumber\\
&&+\lambda _1(H^\dag H){\rm Tr}(\Delta ^\dag \Delta)+\lambda _2\left[{\rm Tr}(\Delta ^\dag \Delta)\right]^2+\lambda _3{\rm Tr}(\Delta ^\dag \Delta)^2+\lambda _4 (H^\dag\Delta\Delta^\dag H )\,\nonumber\\
&&+m^2_\Phi(\Phi^\dag \Phi)+\lambda_\Phi(\Phi^\dag \Phi)^2+ \lambda_5 (H^\dag H)(\Phi^\dag \Phi)+\lambda_6 (H^\dag \Phi\Phi^\dag H)\,\nonumber\\
&&+\lambda_7(\Phi^\dag \Phi){\rm Tr}(\Delta ^\dag \Delta)+\lambda_8 (\Phi^\dag \Delta\Delta^\dag \Phi)+\lambda_9\left[(\Phi^\dag H)^2+{\rm h.c.}\right]\,\nonumber\\
&&+\left(\tilde{\mu} \Phi^{\sf T}i\sigma_2\Delta^\dag \Phi+{\rm h.c.}\right).
\label{eq:Vpd}
\end{eqnarray}
 

Here, $\lambda$, $\lambda_\Phi$ and $\lambda _i$ ($i=1,\ldots 9$)
are dimensionless coupling constants, while $m_H$, $m_\Phi$, $M_\Delta$,
$\mu$ and $\tilde{\mu}$ are mass parameters of the above potential.
Whereas $\lambda_9$, $\mu$ and $\tilde{\mu}$ are the only terms which
can generate CP phases, as the other terms of the potential are
self-conjugate. However, two of them can be removed by redefining
the fields $H$, $\Phi$ and $\Delta$. Furthermore, we assume
that $m^2_H > 0$ for the spontaneous breaking of above
mentioned gauge group. 

After EWSB we obtain a doubly charged scalar $H^{\pm\pm}$ including a
singly charged scalar, $H^\pm$, a pair of neutral CP even Higgs ($h^0,H^0$),
a CP odd scalar ($A^0$) and as usual three massless Goldstone bosons
($G^\pm,G^0$). Further, we also have three particles
($\phi^\pm$, $\phi^0$, and $a^0$) which are members of
the inert $SU(2)$ doublet. The mass eigenvalues for the
$\mathbb{Z}_2$ even physical scalar are given by \cite{Arhrib:2011uy}:
\begin{eqnarray}
M^2_{H^{\pm\pm}} &=& \frac{\sqrt{2}\mu v^2_d-\lambda _4v^2_d
v_t-2\lambda _3v^3_t}{2v_t},\\
M^2_{H^\pm} &=& \frac{(v_d^2+2v^2_t)(2\sqrt{2}\mu-\lambda _4 v_t)}
{4v_t},\label{sing-mass}\\
M^2_{A^0} &=& \frac{\mu(v_d^2+4v^2_t)}{\sqrt{2}v_t},\\
M^2_{h^0} &=& \frac{1}{2}\left(A+C-\sqrt{(A-C)^2+4B^2}\right),\\
M^2_{H^0} &=& \frac{1}{2}\left(A+C+\sqrt{(A-C)^2+4B^2}\right),
\label{eq:scamass1}
\end{eqnarray}
{\rm with}~~
\begin{eqnarray}
A &=& \frac{\lambda }{2} v_d^2,\\
B &=& v_d[-\sqrt{2}\mu+(\lambda _1+\lambda _4)v_t],\\
C &=& \frac{\sqrt{2}\mu v^2_d+4(\lambda _2+\lambda _3)v^3_t}{2v_t},
\label{eq:abc}
\end{eqnarray}
while the mass eigenvalues of $\mathbb{Z}_2$ odd scalars are:
\begin{eqnarray}
M^2_{\phi^0} &=& m^2_\Phi+\frac12(\lambda_5+\lambda_6)v^2_d
+\frac12(\lambda_7+\lambda_8)v^2_t+\lambda_9v^2_d-\sqrt{2}\tilde{\mu}v_t,
\label{eq:dmmass}\\
M^2_{a^0} &=& m^2_\Phi+\frac12(\lambda_5+\lambda_6)v^2_d
+\frac12(\lambda_7+\lambda_8)v^2_t-\lambda_9v^2_d+\sqrt{2}\tilde{\mu}v_t,
\label{eq:a0mass}\\
M^2_{\phi^\pm}    &=& m^2_\Phi+\frac12\lambda_5v^2_d+\frac12\lambda_7v^2_t.
\label{eq:phipmass}
\end{eqnarray}

The mixing between the SM doublet and the triplet scalar fields
in the charged, CP even as well as CP odd scalar sectors are
respectively given by: 
\begin{eqnarray}
\left(\begin{array}{c}
G^\pm \\H^\pm \end{array}\right) &=& \left(\begin{array}{cc}
\cos\beta' & \sin\beta'\\
-\sin\beta' & \cos\beta'
\end{array}\right)
\left(\begin{array}{c}
h^\pm \\ \delta^\pm \end{array}\right),\\
\left(\begin{array}{c} 
h^0 \\H^0 \end{array}\right) &=& \left(\begin{array}{cc}
\cos \alpha & \sin \alpha\\
-\sin \alpha & \cos \alpha
\end{array}\right)
\left(\begin{array}{c}
\eta^0 \\ \xi^0 \end{array}\right),\\
\left(\begin{array}{c}
G^0 \\ A^0 \end{array}\right) &=& \left(\begin{array}{cc}
\cos\beta & \sin\beta\\
-\sin\beta & \cos\beta
\end{array}\right)
\left(\begin{array}{c}
z_1 \\ z_2 \end{array}\right),
\end{eqnarray} 
and the respective mixing angles are given by:
\begin{eqnarray}
\tan\beta' &=& \frac{\sqrt{2}\,v_t}{v_d},\label{mix1}\\
\tan\beta &=& \frac{2\,v_t}{v_d}=\sqrt{2} \tan\beta',\label{mix2}\\
\tan{2 \alpha} &=& \frac{2B}{A-C}\,, \label{mix3}
\end{eqnarray} 
where the expressions of $A$, $B$ and $C$ are already given
in Eq.\,\ref{eq:scamass1}.

\subsection{Different constraints}\label{constaints}
Before going to study the phenomenological aspects of neutrino
and dark matter sectors, it is necessary to check various
constraints from theoretical considerations like vacuum
stability, unitarity of the scattering matrices and perturbativity.
Further, the model parameters also need to satisfy the
phenomenological constraints arising from electroweak precision test and
Higgs signal strength. Therefore, to serve the purposes
we need to choose a set of free parameters of this model.
In practice, a convenient set of free parameters 
are given in the following, however some of them are not
independent:
\begin{equation}
\{ \tan{\alpha}, M_{H^{\pm\pm}}, M_{H^{\pm}}, M_{H^0}(=M_{A^0}),
M_{\phi^0}, M_{\phi^\pm}, \lambda_\Phi, \lambda_5, \lambda_6,
\lambda_7, \lambda_8, \lambda_9 \}\,.
\label{eq:para}
\end{equation}

\subsubsection {\underline {Vacuum stability bounds:}}
This section has been dedicated to derive the necessary and
sufficient conditions for the stability of the vacuum. These
conditions come from requiring that the potential given in
Eq.\;\ref{eq:Vpd} be bounded from below when the scalar fields
become large in any direction of the field space. The constraints
ensuring boundedness from below (BFB) of the present potential
have not been studied in the literature so far.
It would thus be very relevant to derive these constraints
in the present model. For large field values, the potential given
in Eq.\;\ref{eq:Vpd} is generically dominated by the quartic part
of the potential. Hence, in this limit we can ignore any terms
with dimensionful couplings, mass terms or soft terms. So the general
potential given in Eq.\;\ref{eq:Vpd} can be written as in the following
way which contains only the quartic terms,

\begin{eqnarray} 
V^{(4)}(H,\Delta,\Phi) &=&\frac{\lambda }{4}(H^\dag H)^2+\lambda _1(H^\dag H){\rm Tr}(\Delta ^\dag \Delta)+\lambda _2\left[{\rm Tr}(\Delta ^\dag \Delta)\right]^2+\lambda _3{\rm Tr}(\Delta ^\dag \Delta)^2\nonumber\\
&&+\lambda _4 (H^\dag\Delta\Delta^\dag H )\,+\lambda_\Phi(\Phi^\dag \Phi)^2+ \lambda_5 (H^\dag H)(\Phi^\dag \Phi)+\lambda_6 (H^\dag \Phi\Phi^\dag H)\,\nonumber\\
&&+\lambda_7(\Phi^\dag \Phi){\rm Tr}(\Delta ^\dag \Delta)+\lambda_8 (\Phi^\dag \Delta\Delta^\dag \Phi)+\lambda_9\left[(\Phi^\dag H)^2+{\rm h.c.}\right].
\label{eq:V4pd}
\end{eqnarray}

To determine the BFB conditions we have used {\it copositivity} criteria
as given in Ref.\;\cite{Kannike:2012pe}. For this purpose we need
to express the scalar potential $V^{(4)}$ in a biquadratic form
$A_{ij}\psi^2_i\psi^2_j$, where $\psi_i \equiv H, \Delta, \Phi$.
If the matrix $A_{ij}$ is copositive then we can demand that the
potential is bounded from below. Let us write down the matrix in our case:

\begin{equation}
A=\left( \begin{array}{ccc} 
    \frac14{\lambda}& \frac12{(\lambda_1+\xi\lambda_4)} & \frac12{[\lambda_5+\rho^2(\lambda_6-2|\lambda_9|)]}\\ 
    \frac12{(\lambda_1+\xi\lambda_4)} & (\lambda_2+\zeta\lambda_3) & \frac12{(\lambda_7+\xi'\lambda_8)}\\
    \frac12{[\lambda_5+\rho^2(\lambda_6-2|\lambda_9|)]}& \frac12{(\lambda_7+\xi'\lambda_8)}& \lambda_\Phi 
\end{array}  \right)\,.
\end{equation}

The parameters $\xi$, $\xi'$, $\zeta$ and $\rho$ appearing in the
matrix elements are required to determine all the necessary and
sufficient BFB conditions. The detail illustrations of the parameters
can be found in \cite{Arhrib:2011uy} where two fields (one doublet and a
triplet) have been considered. However, in our case we have three
different fields (two doublets and a triplet). Using the prescription
given in Ref.\,\cite{Arhrib:2011uy}, we have defined the parameters
in the following way, 
\begin{eqnarray}
&&\zeta \equiv  {\rm Tr} (\Delta^{\dagger}{\Delta})^2/[{\rm Tr}(\Delta^{\dagger}{\Delta})]^2,\\
&&\rho\equiv|H^\dagger\Phi|/|H||\Phi|,\\
&&\xi \equiv (H^\dagger\Delta\Delta^{\dagger}H)/(H^\dagger{H}~{\rm Tr}(\Delta^{\dagger}{\Delta}) ), \\
&&\xi'\equiv(\Phi^\dagger\Delta\Delta^{\dagger}\Phi)/(\Phi^\dagger{\Phi}~{\rm Tr}(\Delta^{\dagger}{\Delta}) ). 
\end{eqnarray}
The $upper$ and $lower$ limits of these parameters are given
as [$\frac{1}{2},1$], [$0,1$], [$0,1$] and [$0,1$] respectively
\cite{Arhrib:2011uy}. To determine the all possible BFB conditions
of the scalar potential, we consider both the limits of these parameters
and respect the copositivity criteria. Finally, we can write down the
following BFB conditions by demanding the symmetric matrix $A_{ij}$ is
copositive \cite{Kannike:2012pe}. 
\begin{eqnarray}
&&\lambda\geq 0, \\
&&(\lambda_2+\zeta\lambda_3)\geq 0, \\
&&\lambda_\Phi\geq 0,\\ 
&&{(\lambda_1+\xi\lambda_4)}+\sqrt{\lambda (\lambda_2+\zeta\lambda_3)}\geq 0,\\
&&{\lambda_5+\rho^2(\lambda_6-2|\lambda_9|)}+\sqrt{\lambda \lambda_\Phi}\geq 0,\\
&&{(\lambda_7+\xi'\lambda_8)}+2\sqrt{(\lambda_2+\zeta\lambda_3)\lambda_\Phi}\geq 0,
\label{eq:copo}
\end{eqnarray}

{\small
\begin{eqnarray}
&&\hspace*{-1cm}\sqrt{\lambda(\lambda_2+\zeta\lambda_3)\lambda_\Phi}+(\lambda_1+\xi\lambda_4)\sqrt{\lambda_\Phi}+[\lambda_5+\rho^2(\lambda_6-2|\lambda_9|)]\sqrt{\lambda_2+\zeta\lambda_3}+\frac{(\lambda_7+\xi'\lambda_8)}{2}\sqrt{\lambda}\\
&&\hspace*{-1cm}+{\tiny \sqrt{\left\{(\lambda_1+\xi\lambda_4)+\sqrt{\lambda(\lambda_2+\zeta\lambda_3)}\right\}\left\{(\lambda_7+\xi'\lambda_8)+2\sqrt{\lambda_\Phi(\lambda_2+\zeta\lambda_3)}\right\}\left\{[\lambda_5+\rho^2(\lambda_6-2|\lambda_9|)]+\sqrt{\lambda\lambda_\Phi}\right\}} \geq 0}.\nonumber  
\end{eqnarray}
}

Substituting the lower and upper limits of the parameters, one can get the
full set of vacuum stability conditions given in Appendix \ref{vac_cond}
(see Eq.\;\ref{eq:lstab1} to Eq.\;\ref{eq:lstab3}).






\subsubsection {\underline {Unitarity bounds:}}
\label{sec:unitarity}
In this section we discuss the unitarity constraints on the parameters of scalar potential by using the tree-level unitarity of various scattering processes. One can find the scalar-scalar scattering, gauge boson-gauge boson scattering and scalar-gauge boson scattering in the context of SM in \cite{Appelquist:1971yj,Cornwall:1974km,Lee:1977eg}. In the case of various extended Higgs sector scenario, the generalizations of such constraints can be found in literature \cite{Kanemura:1993hm, Akeroyd:2000wc, Aoki:2007ah, Gogoladze:2008ak}. It has been a well known fact that in the high energy limit using equivalence theorem \cite{Lee:1977eg, Cornwall:1973tb, Chanowitz:1985hj} one can replace longitudinal gauge bosons by those of the corresponding Nambu-Goldstone bosons in $2 \rightarrow 2$ scattering. Hence, following this prescription in the current model, our main focus is to consider only the Higgs-Goldstone interactions of the scalar potential given in Eq.\;\ref{eq:Vpd}. Furthermore, under this situation the $2$-body scalar scattering processes are dominated by the quartic interactions only.

To determine the unitarity constraints, it has been a usual trend to calculate the $\mathcal{S}$-matrix amplitude in the basis of unrotated states, corresponding to the fields before electroweak symmetry breaking. Because, in this situation the quartic scalar vertices have a much simpler form with respect to the complicated functions of $\lambda_i$, $\lambda_\Phi$, $\alpha$ and $\beta$ involved in the physical basis\footnote{For the inert Higgs doublet, the physical basis are equivalent to the gauge basis as in this case the vacuum expectation value is zero.} ($H^{\pm\pm}$, $H^\pm$, $G^\pm$, $h^0$, $H^0$, $A^0$, $G^0$, $\phi^0$, $\phi^\pm$ and $a^0$). So in the unrotated basis ($\delta^{\pm\pm}$, $\delta^\pm$, $h^\pm$, $\phi^\pm$, $\eta^0$, $\xi^0$, $z_1$, $z_2$, $\phi^0$ and $a^0$), we study full set of $2$-body scalar scattering processes which lead to
a $ 68\times68$ $\mathcal{S}$-matrix. This matrix can be decomposed into 7 block submatrices with definite charge. For example, $\mathcal{M}_1(18 \times 18)$, $\mathcal{M}_2(10\times 10)$ and $\mathcal{M}_3(3\times 3)$ corresponding to 
neutral charged states, $\mathcal{M}_4(21\times 21)$ corresponding to the
singly charged states, $\mathcal{M}_5(12 \times 12)$ corresponding to the
doubly charged states, $\mathcal{M}_6(3 \times 3)$ corresponding to the
triply charged states and finally $\mathcal{M}_7(1 \times 1)$ corresponding
to the unique quartic charged state. These submatrices are hermitian,
so the eigenvalues will always be real-valued. 

To this end, we would like to mention that in the following cases we will determine the eigenvalues of the above mentioned submatrices. However, there is a caveat. The structure of some of the submatrices are very challenging, so it is not possible
to find out the analytic form of all the eigenvalues of those matrices. However, using numerical technique given in \cite{Adhikary:2013bma} we can derive
the remaining eigenvalues. Eventually, we will have all the full set of
eigenvalues by which we will put the unitarity constraints on the model
parameters. 

The first submatrix ${\mathcal{M}}_1$ corresponds to the scatterings whose
initial and final states are one of the following:

\begin{eqnarray}
&&\bigg\{h^+\delta^-, \delta^+ h^-, \phi^+ \delta^-, \delta^+\phi^-, h^+\phi^-,\phi^+h^-,\eta^0z_2,\xi^0z_1, z_1z_2, \eta^0\xi^0, \phi^0 \eta^0, \phi^0 z_1, \eta^0a^0, a^0 z_1,\phi^0 \xi^0, \phi^0 z_2, \nonumber \\ &&
a^0 \xi^0, a^0 z_2 \bigg\},\nonumber 
\end{eqnarray}


\begin{figure}[!h]
\begin{center}
\includegraphics[height=12cm,width=18cm,angle=0]{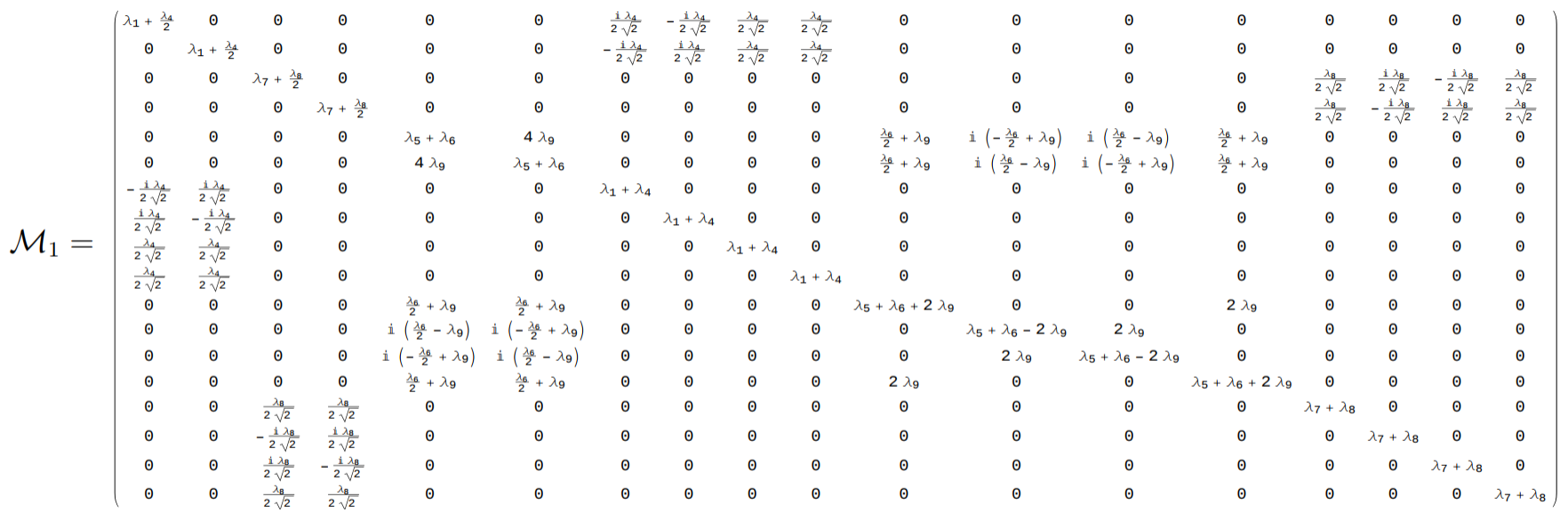}
\end{center}
\end{figure}

Eigenvalues of ${\mathcal{M}}_1$ are: 
{
\begin{eqnarray}
&\bigg\{\lambda_1,\lambda_1,\lambda_1+ \lambda_4,\lambda_1+\lambda_4,\lambda_1+\frac{3\lambda_4}{2},\lambda_1+\frac{3 \lambda_4}{2},\lambda_5+\lambda_6,\lambda_5+\lambda_6,\lambda_7,\lambda_7,\lambda_7+\lambda_8,\nonumber \\ & \lambda_7+\lambda_8,\lambda_7+\frac{3 \lambda_8}{2},\lambda_7+\frac{3\lambda_8}{2},\lambda_5+2 \lambda_6-6 \lambda_9,\lambda_5-2 \lambda_9,
\lambda_5+2\lambda_9,\lambda_5+2\lambda_6+6 \lambda_9\bigg\}. \nonumber
\end{eqnarray}
}

The second submatrix ${\mathcal{M}}_2$ corresponds to the scatterings whose
initial and final states are one of the following:
\begin{eqnarray}
\bigg\{h^+h^-, \delta^+ \delta^-, \frac{z_1z_1}{\sqrt{2}}, \frac{z_2z_2}{\sqrt{2}}, \frac{\eta^0 \eta^0}{\sqrt{2}}, \frac{\xi^0 \xi^0}{\sqrt{2}}, \phi^+ \phi^-, \frac{\phi^0 \phi^0}{\sqrt{2}},   
\frac{a^0 a^0}{\sqrt{2}}, \delta^{++} \delta^{--} \nonumber
\bigg\},
\end{eqnarray}

\begin{figure}[H]
\begin{center}
\includegraphics[height=8cm,width=17.8cm,angle=0]{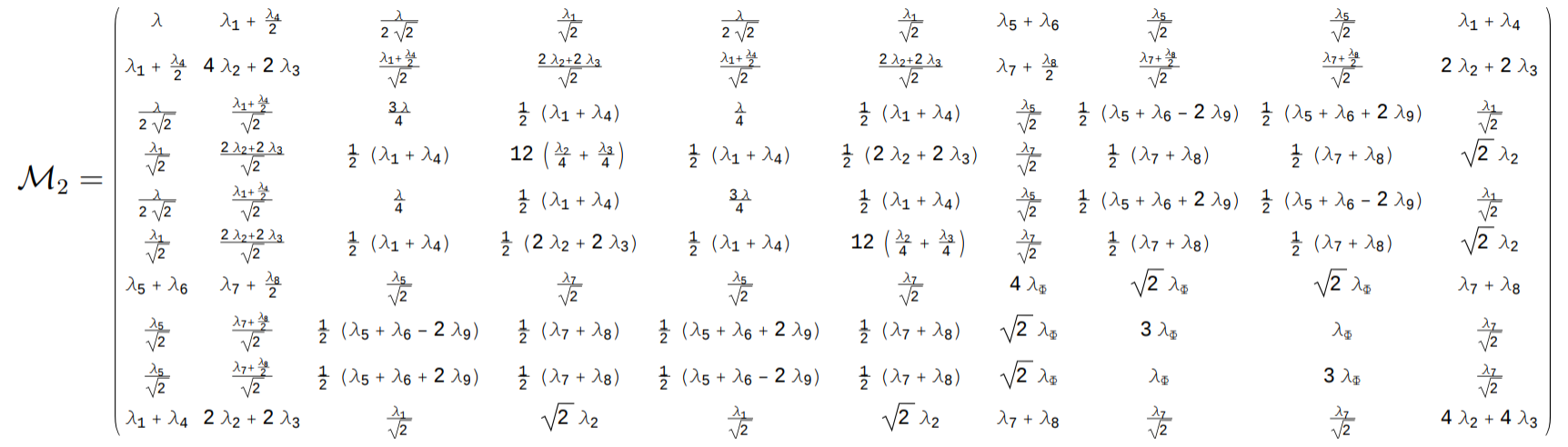}
\end{center}
\end{figure}

Eigenvalues of ${\mathcal{M}}_2$ are:
{
\begin{eqnarray}
&&\bigg\{ 2 \lambda_2,2 (\lambda_2+\lambda_3),\frac{1}{4} \left(\lambda -\sqrt{64 \lambda^2_9+(\lambda -4 \lambda_\Phi )^2}+4 \lambda_\Phi \right),\frac{1}{4} \left(\lambda +\sqrt{64 \lambda^2_9+(\lambda -4 \lambda_\Phi )^2}+4 \lambda_\Phi \right)\bigg\}. \nonumber 
\end{eqnarray}}

Rest of the six eigenvalues have been obtained by numerically solving the
cubic Eqs.\;\ref{eq:qubic1} and \ref{eq:qubic2} given in Appendix \ref{vac_neu}. 

The third submatrix ${\mathcal{M}}_3$ corresponds to the scatterings whose
initial and final states are one of the following:
\begin{eqnarray}
\bigg\{\eta^0 z_1, \xi^0 z_2, \phi^0 a^0 \bigg\}, \nonumber
\end{eqnarray} 

\begin{figure}[!h]
\begin{center}
\includegraphics[height=2cm,width=7cm,angle=0]{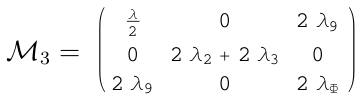}
\end{center}
\end{figure}

Eigenvalues of ${\mathcal{M}}_3$ are:
{
\begin{eqnarray}
\bigg\{ 2 (\lambda_2+\lambda_3),\frac{1}{4} \left(\lambda -\sqrt{64 \lambda^2_9+(\lambda -4 \lambda_\Phi )^2}+4 \lambda_\Phi \right),\frac{1}{4} \left(\lambda +\sqrt{64 \lambda^2_9+(\lambda -4 \lambda_\Phi )^2}+4 \lambda_\Phi \right)\bigg\} \nonumber. 
\end{eqnarray}
}

The fourth submatrix ${\mathcal{M}}_4$ corresponds to the scatterings,
where one charge channels occur for $2 \rightarrow 2$ scattering
between the 20 charged states:

\begin{eqnarray}
&&\bigg\{\eta^0 h^+, \xi^0 h^+, z_1h^+, z_2 h^+, h \delta^+, \xi^0\delta^+, z_1\delta^+, z_2\delta^+, \eta^0\phi^0, \xi^0\phi^+, z_1\phi^+
z_2\phi^+, h^+\phi^0, h^+a^0, \delta^+\phi^0, \delta^+ a^0, \nonumber \\ &&\phi^+ \phi^0, \phi^+ a^0, \delta^{++} \delta^-, \delta^{++} h^-, \delta^{++}\phi^-
\bigg\}, \nonumber
\end{eqnarray}

\begin{figure}[!h]
\begin{center}
\includegraphics[height=11.5cm,width=18cm,angle=0]{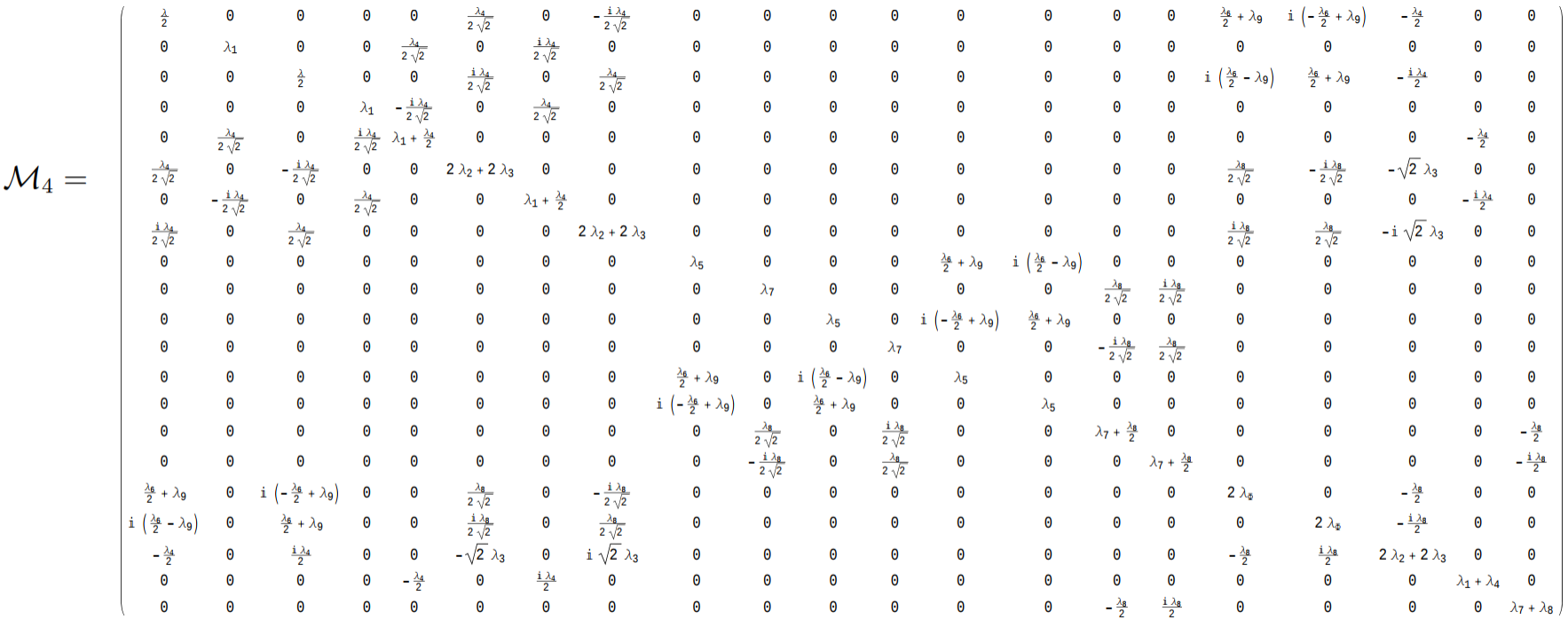}
\end{center}
\end{figure}

Eigenvalues of ${\mathcal{M}}_4$ are:
{
\begin{eqnarray}
&&
\bigg\{\text{$\lambda_1$},\text{$\lambda_1$},2 \text{$\lambda_2$},2 (\text{$\lambda_2$}+\text{$\lambda_3$}),\text{$\lambda_1$}-\frac{\text{$\lambda_4$}}{2},\text{$\lambda_1$}+\text{$\lambda_4$},\text{$\lambda_1$}+\frac{3 \text{$\lambda_4$}}{2},\text{$\lambda_5$}-\text{$\lambda_6$},\text{$\lambda_5$}+\text{$\lambda_6$},\text{$\lambda_7$},\text{$\lambda_7$},\text{$\lambda_7$}-\frac{\text{$\lambda_8$}}{2}, \nonumber \\ && \text{$\lambda_7$}+\text{$\lambda_8$},\text{$\lambda_7$}+\frac{3 \text{$\lambda_8$}}{2}, \text{$\lambda_5$}-2 \text{$\lambda_9$},\text{$\lambda_5$}+2 \text{$\lambda_9$}, \frac{1}{4} \left(\lambda +\sqrt{64 \text{$\lambda^2_9$}+(\lambda -4 \lambda_\Phi )^2}+4 \lambda_\Phi \right), \nonumber \\ && \frac{1}{4} \left(\lambda -\sqrt{64 \text{$\lambda^2_9$}+(\lambda -4 \lambda_\Phi )^2}+4 \lambda_\Phi \right)\bigg\}. \nonumber
\end{eqnarray}
Remaining three eigenvalues have been obtained from the
cubic Eq.\;\ref{eq:qubic2} (see Appendix \ref{vac_neu})
using numerical technique.}

The fifth submatrix ${\mathcal{M}}_5$ corresponds to the scatterings,
where double charge channels occur for $2 \rightarrow 2$ scattering
between the 12 charged states:
\begin{eqnarray}
&&\bigg\{\frac{h^+ h^+}{\sqrt{2}}, \frac{\delta^+ \delta^+}{\sqrt{2}}, \delta^+ h^+, \frac{\phi^+\phi^+}{\sqrt{2}}, \phi^+\delta^+, \phi^+ h^+, \delta^{++} \xi^0, \delta^{++} z_2, \delta^{++} z_1, \delta^{++} \eta^0,
\delta^{++} \phi^0, \delta^{++} a^0 \nonumber
\bigg\},
\end{eqnarray}

\begin{figure}[!h]
\begin{center}
\includegraphics[height=5cm,width=14cm,angle=0]{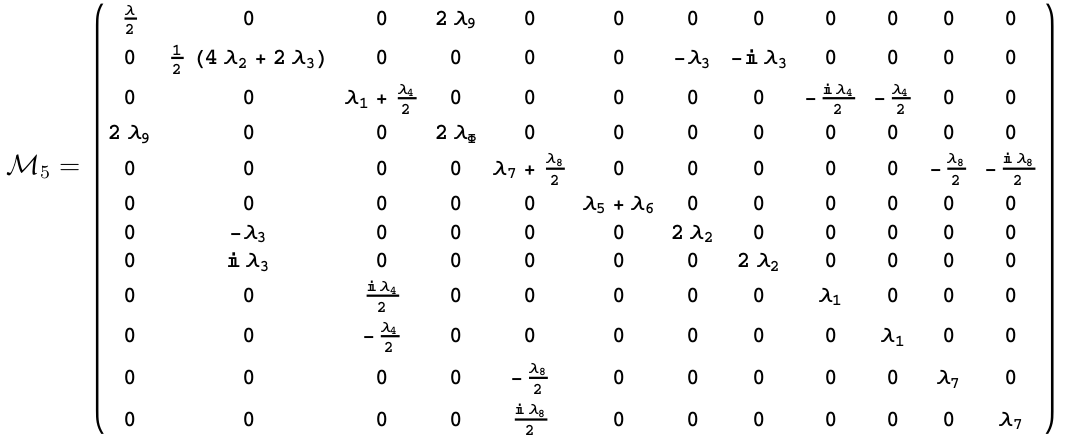}
\end{center}
\end{figure}

Eigenvalues of ${\mathcal{M}}_5$ are:
{
\begin{eqnarray}
&&\bigg\{\text{$\lambda_1$},2 \text{$\lambda_2$},2 \text{$\lambda_2$}-\text{$\lambda_3$},2 (\text{$\lambda_2$}+\text{$\lambda_3$}),\text{$\lambda_1$}-\frac{\text{$\lambda_4$}}{2},\text{$\lambda_1$}+\text{$\lambda_4$},\text{$\lambda_5$}+\text{$\lambda_6$},\text{$\lambda_7$},\text{$\lambda_7$}-\frac{\text{$\lambda_8$}}{2},\text{$\lambda_7$}+\text{$\lambda_8$}, \nonumber \\ &&
\frac{1}{4} \left(\lambda +\sqrt{64 \text{$\lambda^2_9$}+(\lambda -4 \lambda_\Phi )^2}+4 \lambda_\Phi \right), 
\frac{1}{4} \left(\lambda -\sqrt{64 \text{$\lambda^2_9$}+(\lambda -4 \lambda_\Phi )^2}+4 \lambda_\Phi \right)\bigg\}. \nonumber
\end{eqnarray}}
The sixth submatrix ${\mathcal{M}}_6$ corresponds to the scatterings,
where triple charge channels occur for $2 \rightarrow 2$ scattering between
the 3 charged states:
\begin{eqnarray}
\bigg\{ \delta^{++} h^+, \delta^{++} \delta^+, \delta^{++} \phi^+
\bigg\}, \nonumber 
\end{eqnarray}

\begin{figure}[!h]
\begin{center}
{\includegraphics[scale=0.4,angle=0]{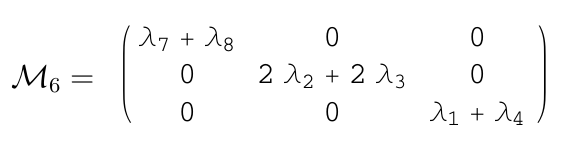}}
\end{center}
\end{figure}
Eigenvalues of ${\mathcal{M}}_6$ are: 
{\{2(\text{$\lambda_2$}+\text{$\lambda_3$}),\,
\text{$\lambda_1$}+\text{$\lambda_4$},\,
\text{$\lambda_7$}+\text{$\lambda_8$}\}}.
Finally, there is unique quadruple charged state
$\frac{\delta^{++} \delta^{++}}{\sqrt{2}}$ which leads to eigenvalue
\begin{equation}
{\mathcal{M}}_7= 2(\lambda_2+\lambda_3). \nonumber
\end{equation}

These eigenvalues, can be labelled as $a_i$, then the $\mathcal{S}$-matrix
unitarity constraint for elastic scattering demands $|{\rm Re}(a_i)|\leq\frac12$ \cite{Lee:1977eg}. Using this condition we generate the following relations. However, these conditions are not the full set of unitarity conditions as we have already mentioned that some of the eigenvalues of few submatrices are evaluated numerically. Hence, using the following conditions,
\begin{eqnarray}
&&\hspace{-0.75cm}
|\lambda_1|\leq 8\pi,\,|2\lambda_2|\leq 8\pi,\,
|\lambda_1+\lambda_4|\leq 8\pi,\,|2(\lambda_2+\lambda_3)|\leq 8\pi,\,
\left|\lambda_1+\frac{3 \lambda_4}{2}\right|\leq 8\pi,\,
\left|\lambda_1-\frac{\lambda_4}{2}\right|\leq 8\pi,
\nonumber \\
&&\hspace{-0.75cm}
|2\lambda_2-\lambda_3|\leq 8\pi,\,
|\lambda_5+2\lambda_6-6\lambda_9|\leq 8\pi,\,
|\lambda_5+2\lambda_6+6\lambda_9|\leq 8\pi,\,
|\lambda_5+2\lambda_9|\leq 8\pi,\,|\lambda_5-2\lambda_9|\leq 8\pi,
\nonumber \\
&&\hspace{-0.75cm}|
\lambda_5-\lambda_6|\leq 8\pi,\,|\lambda_5+\lambda_6|\leq 8\pi,\,
|\lambda_7|\leq 8\pi,\,|\lambda_7+\lambda_8|\leq 8\pi,\,
\left|\lambda_7+\frac{3\lambda_8}{2}\right|\leq 8\pi,\,
\left|\lambda_7-\frac{\lambda_8}{2}\right|\leq 8\pi,
\nonumber \\
&&\hspace{-0.75cm}
\bigg|\frac{1}{4} \left(\lambda +\sqrt{64 \text{$\lambda^2_9$}
+(\lambda -4 \lambda_\Phi )^2}+4 \lambda_\Phi \right)\bigg|\leq 8\pi,\,
\bigg|\frac{1}{4} \left(\lambda -\sqrt{64 \text{$\lambda^2_9$}
+(\lambda -4 \lambda_\Phi )^2}+4 \lambda_\Phi \right)\bigg|\leq 8\pi,
\label{eq:luni}
\end{eqnarray}
and numerically evaluated six eigenvalues (whose absolute value should be $\leq 8\pi$)
we have imposed full set of unitarity constraints on the model parameters. 

\subsubsection {\underline {Perturbativity:}}
If we demand that the model in the present work behaves as a
perturbative quantum field theory at any energy scale,
then we have to ensure the following conditions. For the
scalar quartic coupling $\lambda, \lambda_\Phi,\lambda_i (i = 1-9)$,
the perturbativity criterion is,
\begin{equation}
|\lambda|, |\lambda_\Phi|,|\lambda_i| < 4\pi.
\end{equation}
The corresponding constraints for the gauge and Yukawa
interactions are,
\begin{equation}
g_i, y_i < \sqrt{4\pi},
\end{equation}
where, $g_i$'s and $y_i$'s are the gauge and Yukawa
coupling constants respectively.

\subsubsection {\underline {Constraints from electroweak precision test:}}
Electroweak precision test (EWPT) can be considered as a very useful
tool in constraining any BSM scenario. As the current scenario contains several non-standard scalars, hence they contribute to
the electroweak precision observables, the $S, T, U$ parameters
\cite{Lavoura:1993nq, Barbieri:2006dq, Chun:2012jw, Aoki:2012jj}.
The stringent bound comes from the $T$-parameter
which imposes strict limit on the mass splitting
between the non-standard scalars. Therefore, we tune the relative mass
splitting between the non-standard scalars in such a way for which the present scenario satisfy the constraints from EWPT \cite{Baak:2014ora}.
Further, the electroweak precision data constraint
the $\rho$-parameter to be very close to its SM value of
unity and from the latest data \cite{Patrignani:2016xqp} one gets
an upper bound on $v_t\la 4$ GeV which we maintain
in our analysis.

\subsubsection {\underline {Constraints from Higgs signal strength
\bigg({$\mu_{\gamma\gamma}=\frac{\sigma(pp\rightarrow h)_{\rm BSM}\times
{\rm BR}(h\rightarrow\gamma\gamma)_{\rm BSM}}{\sigma(pp\rightarrow h)_{\rm SM}
\times {\rm BR}(h\rightarrow\gamma\gamma)_{\rm SM}}$}\bigg):}} Moreover, apart
from the above mentioned theoretical constraints, it is necessary
to incorporate the constraints from LHC data in the model. As in
the present model all the decay widths and cross sections are
modified with respect to that of the SM predictions so in
our analysis we have constrained the parameter space of
this model by the present LHC Higgs data \cite{ATLAS:2016nke}. 
\section{Neutrino masses and mixings}
\label{neu}

In this section, we have tried to explain the origin of neutrino
masses and their intergenerational mixing angles. In the present
model, as we have one $SU(2)_{\mathbb{L}}$ scalar triplet $\Delta$
(Eq.\;\ref{delta}), hence one can generate Majorana mass term
for the SM neutrinos using Type-II seesaw mechanism
\cite{Magg:1980ut, Ma:1998dx}.
The Yukawa interaction term which is responsible for the Majorana
masses of SM neutrinos is given by
\begin{eqnarray}
{\cal L}_{\rm Yukawa} \supset -
\dfrac{Y^\nu_{ij}}{2} L_i^{\sf T}\mathcal{C}\,i\sigma_2
\,\Delta\,L_j+{\rm h.c.}\, 
\label{int:yukawa}
\end{eqnarray}
where $Y^\nu_{ij}$ is the Yukawa coupling and $i,\,j=1,2,3$
are generational indices of the SM leptons. When the scalar
triplet acquires a VEV $v_t$, Majorana masses for the SM
neutrinos are generated at tree level, which is
\begin{eqnarray}
{M_{\nu}}_{ij} = \frac{Y^\nu_{ij}}{\sqrt{2}}\,v_t\,.
\end{eqnarray}
Since this a Majorana type mass term for the SM neutrinos,
${M_{\nu}}_{ij}$ must be a symmetric matrix.
Therefore, for three generations of the SM neutrinos
the Majorana mass matrix $M_{\nu}$ has the following form
\begin{eqnarray}
M_{\nu} = \frac{v_t}{\sqrt{2}}\left(\begin{array}{ccc}
y_1 ~~&~~ y_2
~~&~~y_3 \\
~~&~~\\
y_2 ~~&~~ y_4
~~&~~ y_5\\
~~&~~\\
y_3 ~~&
~~ y_5 ~~&~~ y_6 \\
\end{array}\right) \,,
\label{mnu}
\end{eqnarray}  
where, for notational simplicity we have redefined the Yukawa couplings
as $Y^{\nu}_{11}=y_1$, $Y^{\nu}_{12}=Y^{\nu}_{21}=y_2$,
$Y^{\nu}_{13}=Y^{\nu}_{31}=y_3$, $Y^{\nu}_{22}=y_4$,
$Y^{\nu}_{23}=Y^{\nu}_{32}=y_5$ and $Y^{\nu}_{33}=y_6$.
Now, our goal is to diagonalise the above mass matrix
and find the mass eigenvalues and mixing angles. To diagonalise a complex
symmetric matrix $M_{\nu}$ (all six independent elements of $M_{\nu}$ can be
in general complex) we need a unitary matrix
$U$ so that $U^{\dagger} M_{\nu} U^{\star}$
is a diagonal matrix ($M_{dia}$). This is however not the eigenvalue equation,
which has usually been solved for the case of matrix diagonalisation. Therefore,
instead of diagonalising a complex symmetric matrix $M_{\nu}$, one can easily
construct a hermitian matrix $h=M^{\dagger}_{\nu} M_{\nu}$ using $M_{\nu}$,
such that $U^{\dagger} h U=M^2_{dia}$ is a diagonal
matrix with real non-negative entities
at the diagonal positions. The unitary matrix $U$ is the usual PMNS matrix which
has the following form
\begin{eqnarray}
U_{\rm PMNS} = U_{\rm CKM} \left(\begin{array}{ccc}
1 ~&~ 0
~&~0 \\
0 ~&~ \exp{i\frac{\alpha}{2}}
~&~ 0\\
0 ~&
~ 0 ~&~ \exp{i\frac{\beta}{2}} \\
\end{array}\right)\,,
\label{upmns}
\end{eqnarray}
where $U_{\rm CKM}$ is the usual CKM matrix
containing three mixing angles $\theta_{12}$, $\theta_{23}$,
$\theta_{23}$ and one phase $\delta$, called the Dirac CP phase
\footnote{Because, any nonzero value of $\sin\delta$ can generate
CP violating effects in vacuum neutrino oscillations
if $\theta_{13}\neq 0$, i.e.
$P({\nu_{\alpha}\rightarrow \nu_{\beta}})
\neq P({\bar{\nu}_{\alpha}\rightarrow \bar{\nu}_{\beta}})$,
($\alpha,\, \beta=e,\,\mu,\,\tau$.) in vacuum oscillation
when $\sin{\delta}\neq 0$ and $\theta_{13}\neq 0$ \cite{Akhmedov:1999uz}.}
while $\alpha$, $\beta$ are known as the Majorana phases.
If SM neutrinos are Dirac fermions then $\alpha=\beta=0$.

We have diagonalised the hermitian matrix $h$ by the
unitary matrix $U_{\rm PMNS}$ and find the mass square
differences and mixing angles between different generations
of SM neutrinos. Dirac phase $\delta$ can be found by
using a quantity known as Jarlskog Invariant ($J_{\rm CP}$)
\cite{Jarlskog:1985ht}, which is related to the elements of
$h$ matrix as,
\begin{eqnarray}
J_{\rm CP} = \dfrac{{\rm Im}(h_{12}\,h_{23}\,h_{31})}
{\Delta m^2_{21}\,\Delta m^2_{32}\,\Delta m^2_{31}\,}\,,
\label{jcph}
\end{eqnarray}
where numerator represents the imaginary part of
the product $h_{12}\,h_{23}\,h_{31}$ while in the denominator
$\Delta m^2_{ij}=m^2_j - m^2_i$. One the other hand 
$J_{\rm CP}$ can also be written in terms of mixing
angles and Dirac CP phases, i.e. 
\begin{eqnarray}
J_{\rm CP} = {\rm Im}(U_{\mu 3} U^{\star}_{e3} U_{e2} U^{\star}_{\mu 2})
=\dfrac{1}{8}\sin 2\theta_{12}\sin 2\theta_{23}
\sin 2\theta_{13}\cos\theta_{13}\sin\delta\,.
\label{jcpa}
\end{eqnarray}
Equating Eq.\,\ref{jcph} and Eq.\,\ref{jcpa}, one can
easily find the value of Dirac CP phase $\delta$.

As mentioned earlier, Yukawa couplings in the
neutrino mass matrix $M_{\nu}$ (Eq.\,\ref{mnu}) can be
in general complex numbers. Therefore, in Eq.\,\ref{mnu}
we have 12 independent parameters. We have varied all
Yukawa couplings (both real and imaginary parts) in the
following range
\begin{eqnarray}
10^{-13}\,{\rm GeV}&\leq & {\rm Re}(y_i) \times {v_t},
\,\leq\,10^{-9}\,{\rm GeV}\,,(i=1\,{\rm to}\,6)\,,\\
10^{-13}\,{\rm GeV}&\leq & {\rm Im}(y_i)
\times {v_t}\,\leq\, 10^{-9}\,{\rm GeV}\,,(i=1\,{\rm to}\,6)\,,
\end{eqnarray}
where we have chosen $v_t=10^{-3}$ GeV, which is consistent with
all the present bounds \cite{Baak:2014ora}. To find
the allowed values of Yukawa couplings by diagonalising the
neutrino mass matrix ($M_{\nu}$), we have considered following
experimental/observational results.
\begin{itemize}
\item Allowed values of three mixing angles in $3\sigma$ range
\cite{Capozzi:2016rtj} from neutrino oscillation data,\\
i.e. $30^{\circ}\leq\theta_{12}\leq36.51^{\circ}$, $37.99^{\circ}(38.23^\circ)
\leq\theta_{23}\leq 51.71^{\circ}(52.95^\circ)$ 
\\ and $7.82^{\circ}(7.84^\circ)\leq \theta_{13}\leq 9.02^{\circ}
(9.06^{\circ})$ for NH(IH).
\item Allowed values of mass squared differences in $3\sigma$ range
\cite{Capozzi:2016rtj} from neutrino oscillation data,\\
i.e. $6.93\,\,{\rm eV}^2\leq \dfrac{\Delta m^2_{21}}{10^{-5}}
\leq 7.97\,\,{\rm eV}^2$
and $2.37 (2.33)\,\,{\rm eV}^2\leq \left|\dfrac{\Delta m^2_{31}}{10^{-3}}\right|
\leq 2.63 (2.60)\,\,{\rm eV}^2$ \\ for NH(IH).
\item Cosmological upper limit on sum over all three neutrino masses
in $2\sigma$ range, \\ i.e. $\sum_i m_i <0.23$ eV \cite{Ade:2015xua}.
\item Current values of mixing angles from neutrino
oscillation data also put upper limit on the absolute value
of $J_{\rm CP}$ which is $|J_{\rm CP}|\leq 0.039$ \cite{Petcov:2013poa}.
\item Allowed region of Dirac CP phase $\delta$ obtain from
the T2K experiment at 90\% C.L. \cite{Abe:2017vif},\\
i.e. $-2.789(-2.296) \leq \delta\,({\rm rad})
\leq -0.764(-0.524)$ with best fit value $\delta = -1.791
(-1.382)$ rad for NH(IH). 
\item Upper bound on effective Majorana mass
$m_{\beta\beta}<(0.15-0.33)\,{\rm eV}$ at 90\% C.L.
from GERDA phase II experiment \cite{Agostini:2017iyd}.
The bound on $m_{\beta\beta}$
is obtained from the non-observation of neutrinoless
double beta decay from $^{76}$Ge ($^{76}{\rm Ge}\rightarrow ^{76}{\rm Se}
+ 2\,e^-$) source at GERDA phase II \cite{Agostini:2017iyd}
experiment and thus consequently reported a lower limit
on the half life $T^{0\nu}_{1/2}(^{76}{\rm Ge}) > 5.3\times10^{25}$
yr at 90\% C.L. 
\end{itemize}

\begin{figure}[h!]
\centering
\includegraphics[height=5cm,width=5.5cm,angle=0]{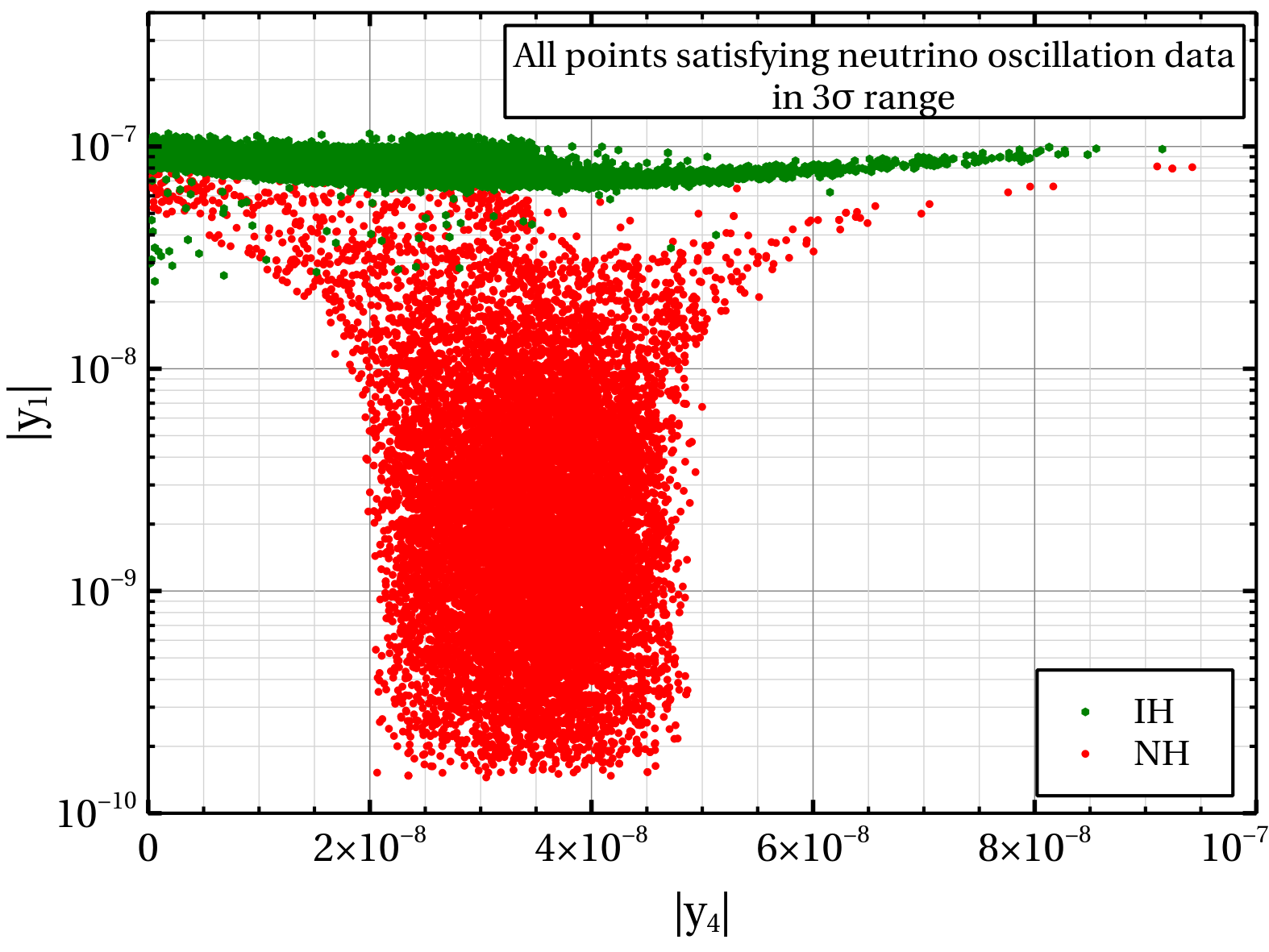}
\includegraphics[height=5cm,width=5.5cm,angle=0]{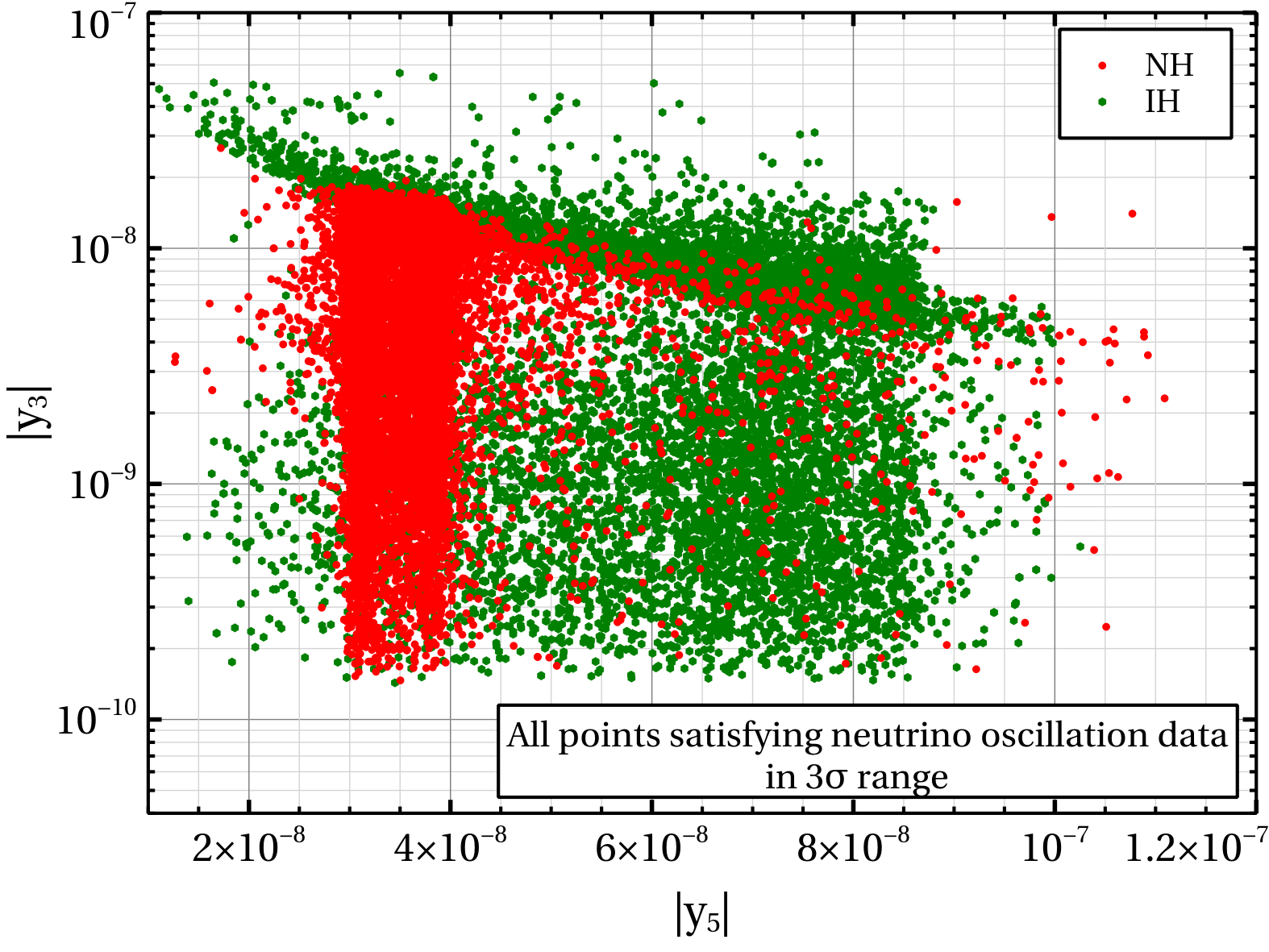}
\includegraphics[height=5cm,width=5.5cm,angle=0]{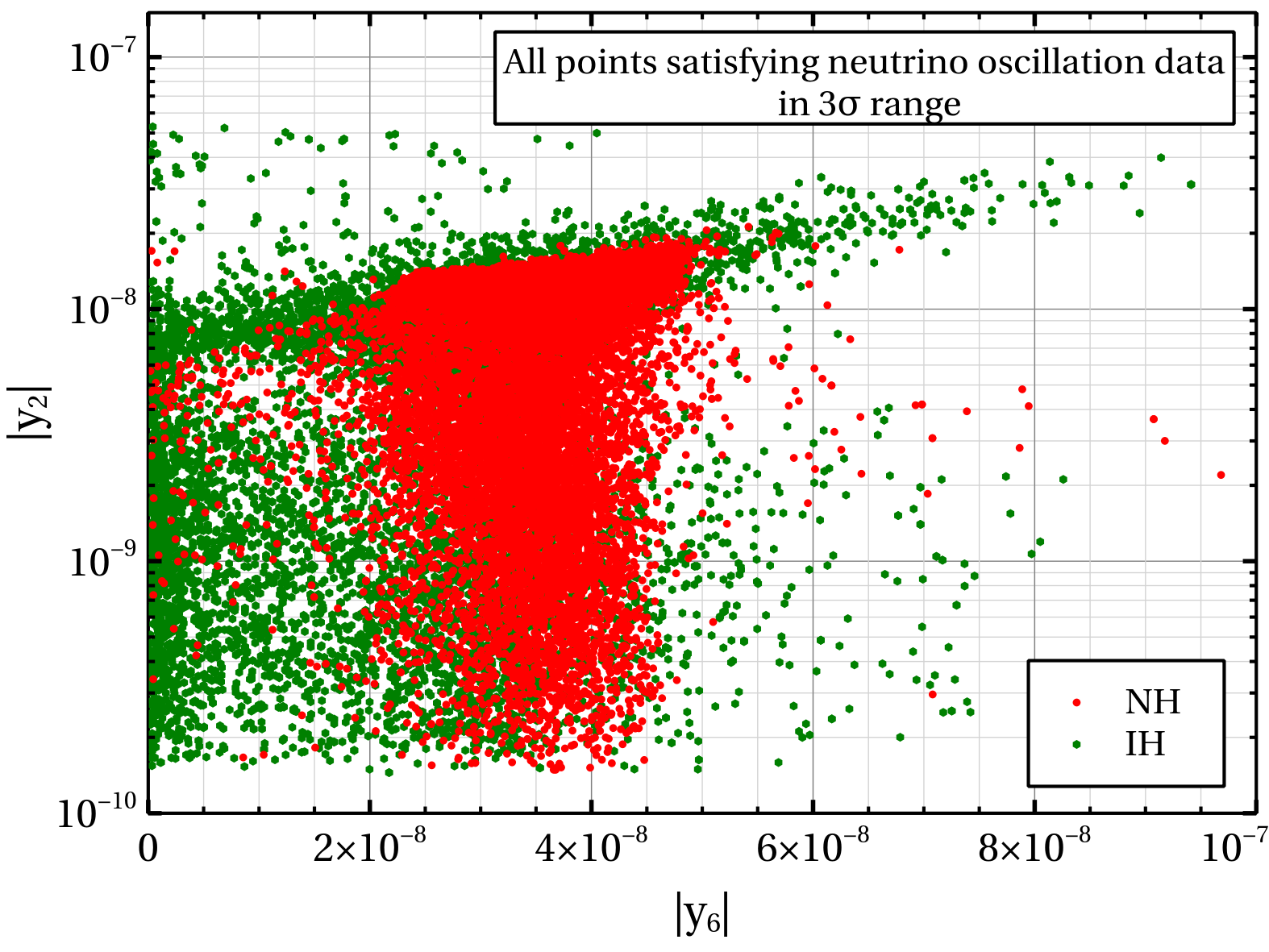}
\caption{Allowed ranges of absolute values of Yukawa couplings satisfying
neutrino oscillation data in 3$\sigma$ limit for both normal (red coloured
region) and inverted (green coloured region) hierarchies. All three plots
have been generated for $v_{t}=10^{-3}$ GeV.}
\label{plot:yukawa}
\end{figure}

In order to obtain the allowed values of Yukawa couplings, defined in
Eq.\,\ref{int:yukawa}, we have diagonalised the neutrino mass matrix
Eq.\,\ref{mnu} following the diagonalisation procedure
given in Ref. \cite{Adhikary:2013bma}
and find the physical masses and intergenerational mixing angles
of SM neutrinos. The corresponding ranges for the absolute values
of Yukawa couplings which reproduce the neutrino oscillation data
in $3\sigma$ range and other experimental results as well
(mentioned above) are shown in all the three panels of
Fig.\,\ref{plot:yukawa}. The red coloured patches in
Fig.\,\ref{plot:yukawa} representing the allowed regions
for the normal hierarchical scenario while the green
coloured regions are for the inverted mass ordering of neutrinos.
All three plots of Fig.\,\ref{plot:yukawa} and also other
plots in the present section (Section \ref{neu}) have
been generated for the triplet scalar VEV $v_{t}=10^{-3}$
GeV.

\begin{figure}[h!]
\centering
\includegraphics[height=5cm,width=7.5cm,angle=0]{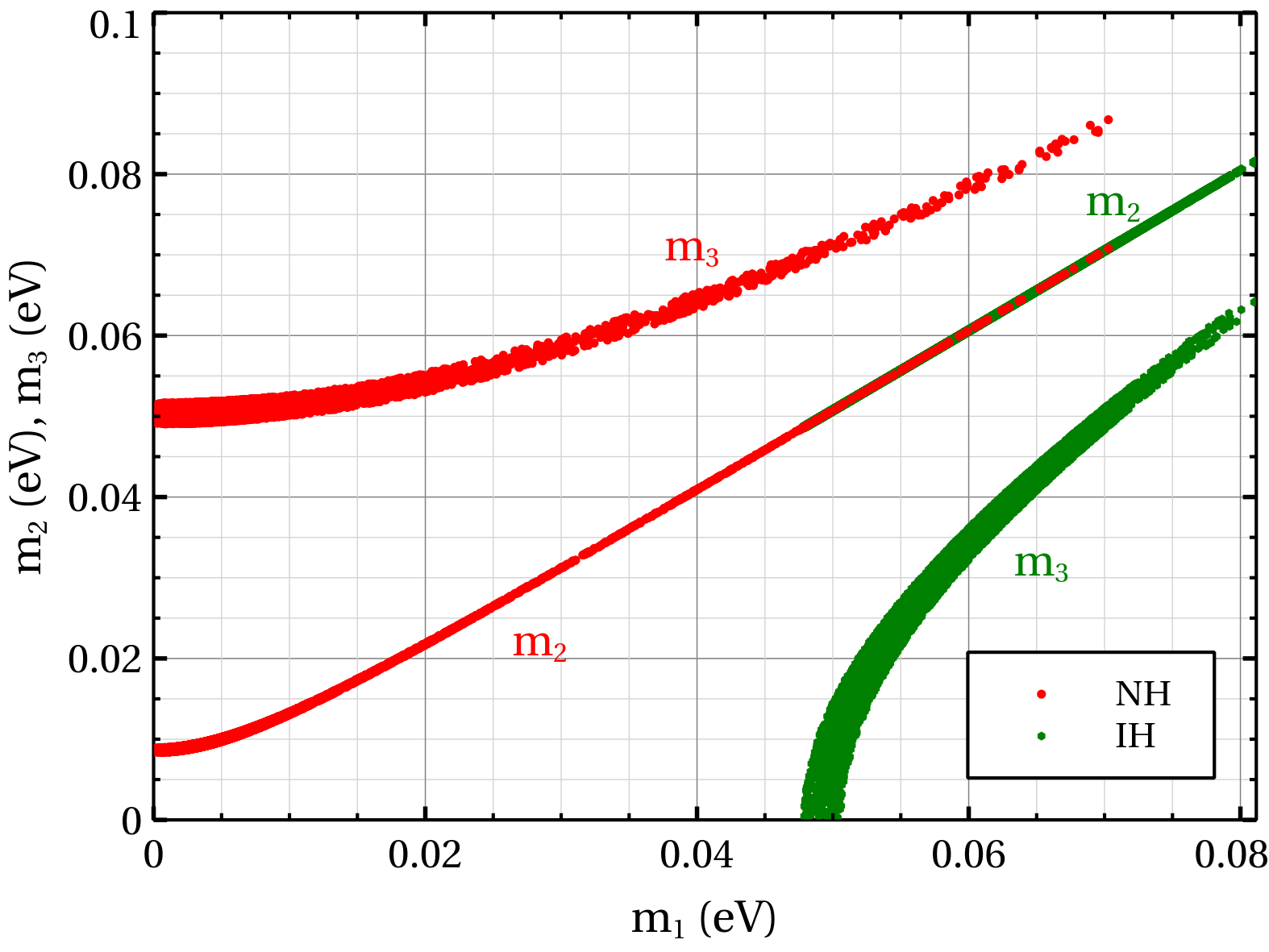}
\includegraphics[height=5cm,width=7.5cm,angle=0]{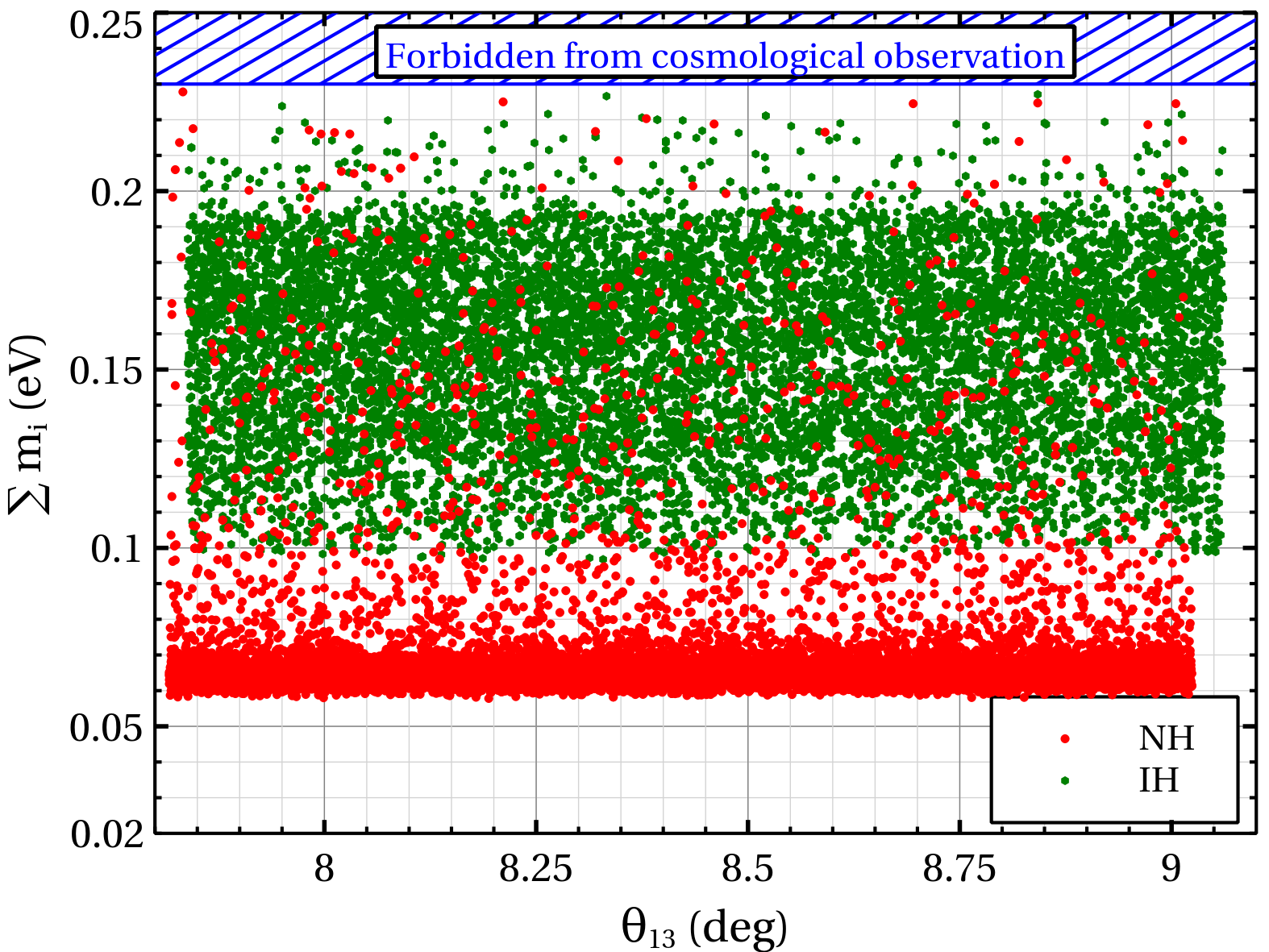}
\caption{Left panel: Absolute values of neutrino masses allowed by
neutrino oscillation data. Right panel: Allowed values
of $\sum m_i$ with respected to the measured values of $\theta_{13}$.}
\label{plot:numass-th13}
\end{figure}

In the left panel of Fig.\,\ref{plot:numass-th13}, we show
the absolute values of neutrino masses allowed from the
neutrino oscillation data for both normal (red coloured points)
and inverted (green coloured points) hierarchies. As the solar
neutrino data suggests extremely small splitting between
$m_2$ and $m_1$ ($\Delta m^2_{21} \sim 10^{-5}$ eV$^2$), 
the $m_1-m_2$ parameter space is very narrow 
and almost aligned along the line $m_1=m_2$ for both
hierarchical scenarios. Moreover, as expected
for the case of inverted mass ordering ($m_2\ga m_1>m_3$), the allowed
values of $m_1$ is larger compared to that of
normal mass ordering ($m_3>m_2 \ga m_1$). Furthermore,
from the left panel of Fig.\,\ref{plot:numass-th13},
it is also seen that for NH, the allowed values of $m_3$
lie above the line $m_1=m_3$ in $m_1-m_3$ parameter
space indicating $m_3>m_1$ while the exactly opposite
nature has been observed for the inverted hierarchical case.
In the right panel of Fig.\,\ref{plot:numass-th13}, we
plot the sum over all three neutrino masses ($\sum m_i$)
with the allowed values of reactor mixing angle $\theta_{13}$.
From this figure one can clearly see that for the normal
hierarchical scenario, $\sum m_i$ in the present case
is mainly concentrated around $\sim$ 0.06 eV to 0.1 eV.
On the other hand, the sum of all three neutrino masses
for the inverted ordering mostly lie between 0.1 eV to
0.2 eV. The blue dashed region corresponds to
$\sum m_i>0.23$ eV which is excluded from the cosmological
observation at 95\% C.L. Also, in the present model irrespective
of neutrino mass ordering, we find that $\sum m_i$ is uniformly
distributed over the entire experimentally allowed values
of $\theta_{13}$. 

\begin{figure}[h!]
\centering
\includegraphics[height=5cm,width=7.5cm,angle=0]{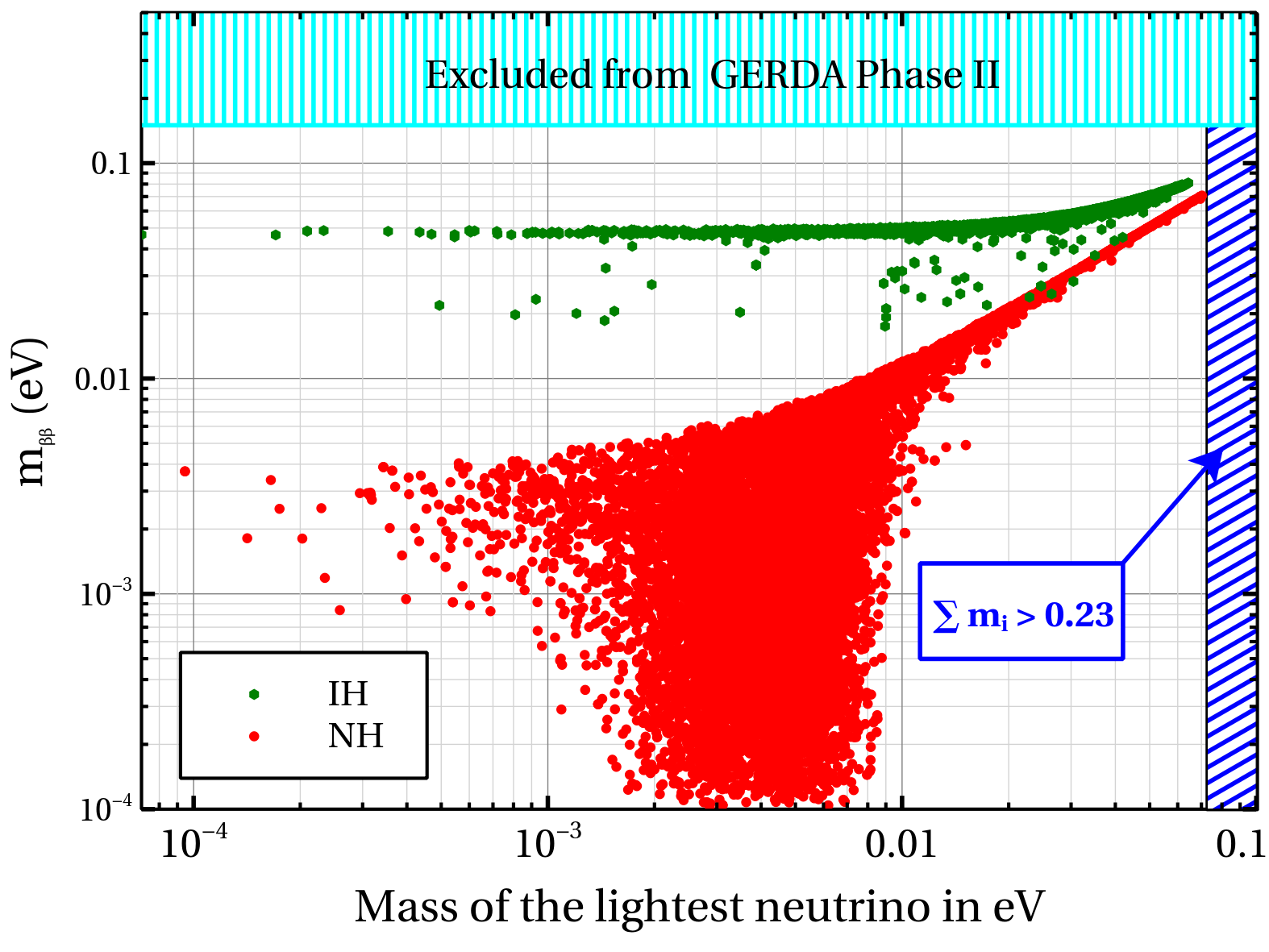}
\includegraphics[height=5cm,width=7.5cm,angle=0]{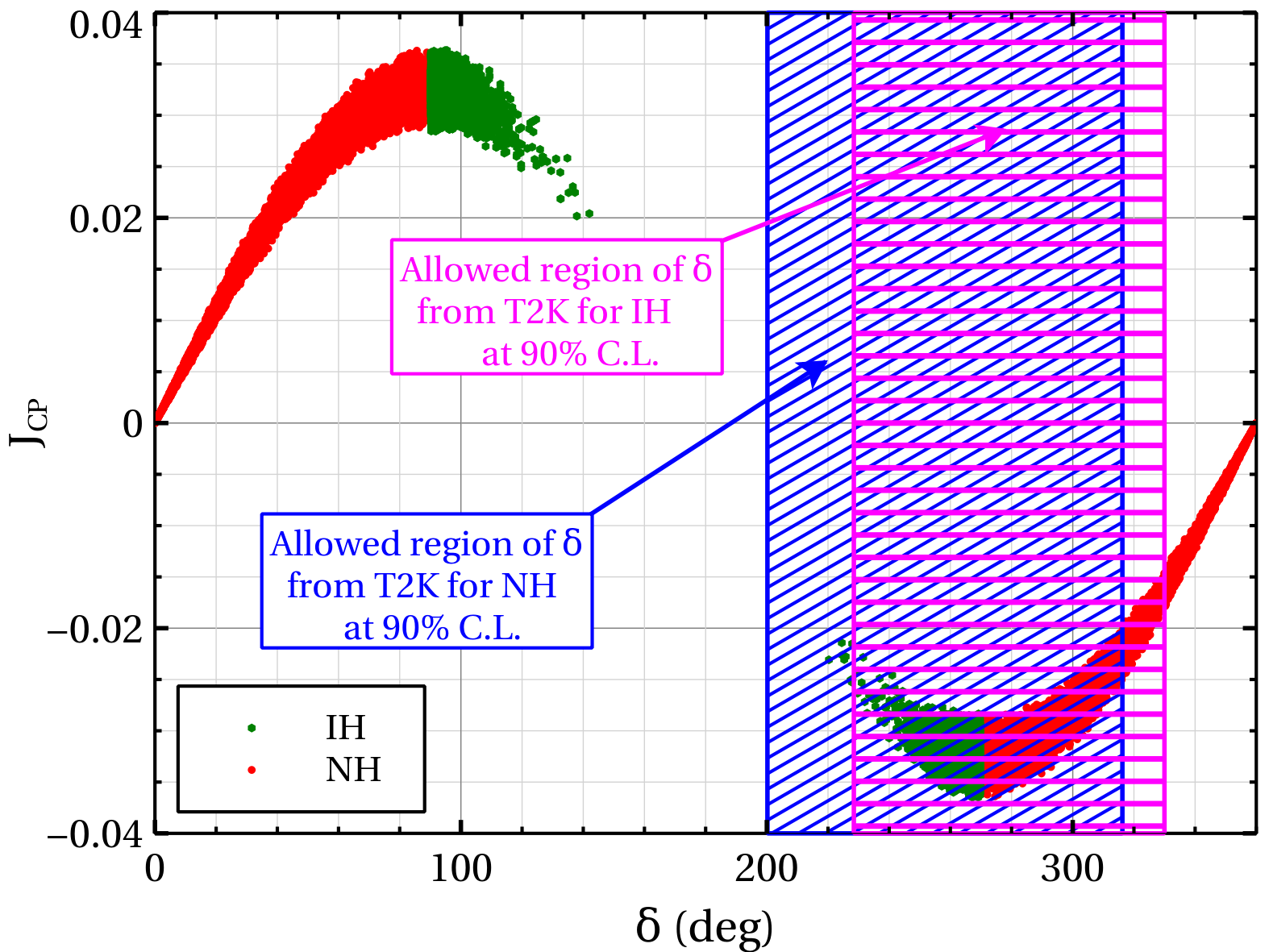}
\caption{Left panel: Variation of effective neutrino mass parameter with
mass of the lightest neutrino for both normal and inverted hierarchies.
Right panel: Predicted values of $J_{CP}$ and $\delta$ from the
present model. Current limits on $\delta$ from the T2K experiment
are indicated by the dashed regions.}
\label{plot:mbb-jcp}
\end{figure}

Variation of effective Majorana mass $m_{\beta\beta}$
with respect to the lightest neutrino mass has been shown
in the left panel of Fig.\,\ref{plot:mbb-jcp}. The effective
Majorana mass parameter $m_{\beta\beta}=\left|
\sum_k m_k\,(U_{\rm PMNS})^2_{1k}\right| = (M_{\nu})_{11}$ is an
important quantity as it enters into the expression of
lifetime of neutrinoless double $\beta$ decay i.e.
$2n\rightarrow 2p + 2e^{-}$. This process violates
lepton number by 2 units and is possible only
if the neutrinos are Majorana fermions. Like the previous
plots, here also we have indicated the values of $m_{\beta\beta}$
by red(green) coloured points for NH(IH) in $m_{1(3)}-m_{\beta\beta}$
plane. The most stringent bound on $m_{\beta\beta}$ comes from
GERDA phase II experiments which has reported an upper
bound on the effective Majorana mass $m_{\beta\beta}<0.15-0.33$ eV
at 90\% C.L. \cite{Agostini:2017iyd}. This upper bound on $m_{\beta\beta}$
has been shown by the turquoise dashed region in $m_{1(3)}-m_{\beta\beta}$
plane while the blue dashed region indicating the upper bound on the
mass of the lightest neutrino $m_{1}\la0.0713$ eV (for NH)\footnote{The
corresponding upper limit on $m_3$ for IH is $\sim 0.0654$ eV.}
obtained by combining the cosmological upper limit on $\sum m_i$
and neutrino oscillation data. Moreover, from this plot it also appears
that the allowed values of $m_{\beta\beta}$ for IH are larger
compared to that of NH. This can be understood from the right panel
of Fig.\,\ref{plot:numass-th13}, where one can easily see that
for most of the allowed region of $\theta_{13}$, the values of $\sum m_i$
are larger for the case of inverted mass ordering. Consequently,
the effective Majorana mass parameter $m_{\beta \beta}
= \left|\sum_k m_k\,(U_{\rm PMNS})^2_{1k}\right|$ for IH appears
to be large as the elements of $U_{\rm PMNS}$ matrix are
nearly identical for both the mass hierarchies. Furthermore,
since the effective Majorana mass parameter is related to
the $(1,1)$ elements of neutrino mass matrix $M_{\nu}$,
absolute values of the Yukawa coupling $y_1$ ($(1,1)$
element of $M_{\nu}$, Eq.\,\ref{mnu}) are large for IH and mainly
concentrated around $\sim 10^{-7}$
(see left most plot of Fig.\,\ref{plot:yukawa}).

In the right panel of Fig.\,\ref{plot:mbb-jcp}, we
show the predicted values of $J_{\rm CP}$ and Dirac
CP phase $\delta$ from the present model, which have
been computed using the relevant model parameters
satisfying neutrino oscillation data and other experimental
bounds mentioned above. From this plot it is clearly seen
the Dirac CP phase $\delta$ has two allowed regions regardless
of neutrino mass hierarchies. However, for the normal ordering
the predicted ranges of $\delta$ are larger compared to that for
the inverted mass ordering. For NH, $\delta$ spans the entire
first and fourth quadrant while it lies between $90^\circ-140^\circ$
and $220^\circ-270^\circ$ for IH. Recently T2K experiment has
reported a 90\% C.L. allowed region for $\delta$ which is
$200.20^{\circ}<\delta<316.23^{\circ}$($228.45^{\circ}
<\delta<329.98^{\circ}$) for normal(inverted) mass ordering
\cite{Abe:2017vif}. In the right panel
of Fig.\,\ref{plot:mbb-jcp}, these results have been
indicated by the blue and pink dashed regions respectively.
This plot indicates that the
T2K results prefer the values of Dirac CP phase $\delta$
lying in the third and fourth quadrant instead of other
two remaining quadrants. 
Moreover, for these values of $\delta$
we have also computed $J_{CP}$ and from the right panel
of Fig.\,\ref{plot:mbb-jcp} one can easily notice that
irrespective of neutrino mass ordering the absolute
values of $J_{CP}$ always lie below 0.039
\cite{Petcov:2013poa}.

On top of these, the neutrino mixing matrix
($U_{\rm PMNS}$, Eq.\,\ref{upmns}) introduces flavour
violating decays in the leptonic sector
such as $\mu \rightarrow e\,\gamma$ etc., which remain absent
in the SM and occur at one loop level in the present model
due to contributions from virtual $W^\pm$, $H^\pm$ and $H^{\pm\pm}$
loops. The expression for the Branching ratio of
$\mu \rightarrow e\,\gamma$ in the present scenario is given by
\cite{1205.4671},
\begin{eqnarray}
{\rm BR}(\mu \rightarrow e\,\gamma) &\simeq&
\dfrac{\Gamma(\mu \rightarrow e\,\gamma)}
{\Gamma(\mu \rightarrow e\,\nu_{\mu}\,\bar{\nu}_{e})}\,,\nonumber\\
&=& \dfrac{\alpha_{\rm em}}{48\,\pi}
\dfrac{\left|\left(M^\dagger_{\nu}\,M_{\nu}\right)_{e\,\mu}\right|^2}
{G_F^2\,v^4_t}\,\left(\dfrac{1}{M^2_{H^\pm}}+\dfrac{8}
{M^2_{H^{\pm\pm}}}\right)^2\,,
\label{eq:mutoegamma}
\end{eqnarray}
where $G_F$ is the Fermi constant and
$\alpha_{\rm em} = \dfrac{{\bf e}^2}{4\,\pi}$, ${\bf e}$ being
the magnitude of electric charge of electron. The non-observation
of this flavour violating decay imposes a strong upper limit
on the branching ratio of this decay mode. Currently the most
stringent upper bound on ${\rm BR}(\mu \rightarrow e\,\gamma)$
has been reported by the MEG collaboration \cite{TheMEG:2016wtm} which is
${\rm BR}(\mu \rightarrow e\,\gamma)<4.2\times 10^{-13}$
at 90\% C.L. We have checked that for our allowed parameter
space, which reproduces the neutrino oscillation data and
also satisfies all the other relevant constraints considered
in this section, the quantity ${\rm BR}(\mu \rightarrow e\,\gamma)$
comes out to be many orders of magnitude less than the
present experimental bound. 
\section{Dark Matter}\label{dm}
We have already mentioned in the Section \ref{2} that besides
the usual SM gauge symmetry we have introduced an additional
$\mathbb{Z}_2$ symmetry in the present model.
Under this newly added $\mathbb{Z}_2$ symmetry all the fields
present in the model except $\Phi$ are even. Moreover, since
the doublet $\Phi$ does not acquire any VEV, $\mathbb{Z}_2$
symmetry remains preserved i.e. all the interactions
are $\mathbb{Z}_2$ conserving. This automatically ensures
that all heavier $\mathbb{Z}_2$ odd particles
will decay to the lightest odd particle (LOP).
Hence, the LOP becomes naturally stable over the
cosmological time scale and can be treated as a
viable dark matter candidate of the Universe. In the
present scenario anyone between the two neutral components
of $\Phi$ namely, $\phi^0$, $a^0$ can be an LOP. For definiteness,
in this work we have considered $\phi^0$ as LOP. 
Now to test the viability of $\phi^0$ as a cold dark matter
candidate, the primary task is to calculate its relic abundance
at the present epoch. In order to compute the relic abundance of
a thermal dark matter candidate, we need to solve the Boltzmann
equation involving comoving number density $Y_{i}=
\dfrac{n_{\phi^0}}{\rm s}$, where $n_{i}$ is the actual
number density of a species $i$ while ${\rm s}$ is the entropy
density of the Universe. The relevant Boltzmann equation
for the computation of comoving number density of dark matter
at the present epoch is given by\cite{Griest:1990kh, Edsjo:1997bg},

\begin{eqnarray}
\dfrac{dY}{dx} = -\left(\dfrac{45\,G}{\pi}\right)^{-\frac{1}{2}}
\dfrac{M_{\phi^0}\,\sqrt{g_{\star}}}{x^2}
\sum_{i\,j}\langle{\sigma_{i\,j} {\rm v}_{i\,j}}\rangle\,
(Y_i\,Y_j-Y_i^{\rm eq}\,Y_j^{\rm eq})\,,
\label{eq:BE}
\end{eqnarray} 

where $Y=\sum_i Y_{i}$ is the comoving number density of
dark matter and the summation is over all three $\mathbb{Z}_2$
odd particles, i.e. $\phi^+$, $a^0$ and $\phi^0$. The dimensionless
variable $x$ is defined as $x=\dfrac{M_{\phi^0}}{T}$, where
$M_{\phi^0}$ is the mass of the LOP $\phi^0$ and $T$ is the
temperature of the Universe. The expression for the mass of
$\phi^0$ in terms of parameters appearing in the Lagrangian
is given in Eq.\,\ref{eq:dmmass}. Moreover, $G = M^{-2}_{pl}$
is the Newton's gravitational constant and $M_{pl} = 1.22\times10^{19}$
GeV, is the Planck mass. The expression of $g_{\star}$ is given by
\begin{eqnarray}
\sqrt{g_{\star}} = \dfrac{{g_{\rm s}}}{\sqrt{g_\rho}}
\left(1+\dfrac{1}{3}\,\dfrac{d\ln g_{\rm s}}{d\ln T}\right)\,,
\label{eq:gstar}
\end{eqnarray}      
where $g_{\rm s}$ and $g_{\rho}$ are degrees of freedom related
to the entropy density (${\rm s}$) and energy density ($\rho$) of
the Universe.

Now, the above Eq.\,\ref{eq:BE} will be in a simplified form,
if we use an approximation $\dfrac{Y_{i}}{Y}\simeq
\dfrac{Y^{\rm eq}_{i}}{Y^{\rm eq}}$ \cite{Griest:1990kh,
Edsjo:1997bg} i.e. the fraction of a species $i$ to the total
comoving number density $Y$ of odd sector particles always
maintains its equilibrium value. Using this relation, the Boltzmann
equation can be written in the following form \cite{Griest:1990kh,
Edsjo:1997bg}
\begin{eqnarray}
\dfrac{dY}{dx} = -\left(\dfrac{45\,G}{\pi}\right)^{-\frac{1}{2}}
\dfrac{M_{\phi^0}\,\sqrt{g_{\star}}}{x^2}
\langle{\sigma {\rm v}}\rangle_{\rm eff}\,
(Y^2-(Y^{\rm eq})^2)\,.
\label{eq:BEapprox}
\end{eqnarray}
The quantity $\langle{\sigma {\rm v}}\rangle_{\rm eff}$
is defined as
\begin{eqnarray}
\langle{\sigma {\rm v}}\rangle_{\rm eff} =
\sum_{i\,j}\langle{\sigma_{i\,j} {\rm v}_{i\,j}}\rangle
\times r_i\,r_j\,,
\label{eq:sigmaveff}
\end{eqnarray}
where
\begin{eqnarray}
r_i = \dfrac{Y^{\rm eq}_i}{Y} = \dfrac{n^{\rm eq}_i}{n}=
\dfrac{g_i\left(1+\Delta_i\right)^{3/2}\exp[-\Delta_i\,x]}
{\sum_{i} g_i\left(1+\Delta_i\right)^{3/2}\exp[-\Delta_i\,x]}\,.
\label{eq:ri}
\end{eqnarray}
Here $\Delta_i = \dfrac{M_i-M_{\phi^0}}{M_{\phi^0}}$,
$g_i$ is the internal degrees of freedom of odd sector
particle $i$ ($i=\phi^\pm,\,a^0,\,\phi^0$) and $n = \sum_{i} n_i$
is the total number density of $\mathbb{Z}_2$ odd particles.
For equilibrium number density $n^{\rm eq}_i$ of a species $i$,
one can use the Maxwell Boltzmann distribution. Finally, the
quantity $\langle{\sigma_{i\,j} {\rm v}_{i\,j}}\rangle$
in the above equations represents the thermally averaged
annihilation cross section between the odd sector particles
and ${\rm v}_{i\,j}$ is the relative velocity between the
two annihilating initial state particles. For two identical initial
state particles ($i=j$), $\sigma_{ii}$ denotes the self-annihilation
cross section of a species $i$ to all possible final state
particles allowed by the symmetries of the Lagrangian while
the co-annihilation between these particles occurs when
annihilating particles are not identical i.e. $i\neq j$.
Besides the self annihilation processes of LOP ($i=j=\phi^0$),
the co-annihilation between an odd sector particle $i$
and LOP as well as the self annihilation of the species $i$
will have a significant effect on dark matter relic abundance
at the present epoch if the mass splitting between LOP and other
odd sector particle $i$ is very small i.e. $\Delta_{i}=
\dfrac{M_i-M_{\phi^0}}{M_{\phi^0}}<0.1$ \cite{Griest:1990kh}. The expression
of $\langle{\sigma_{i\,j} {\rm v}_{i\,j}}\rangle$ is given by
\cite{Biswas:2016yjr}
\begin{eqnarray}
\langle{\sigma_{i\,j} {\rm v}_{i\,j}}\rangle =
\frac{1}{8\,M^2_{i}\,M^2_{j}\,T\,{\rm K}_2\left(\dfrac{M_i}{T}\right)\,
{\rm K}_2\left(\dfrac{M_j}{T}\right)} \times \int^{\infty}_{(M_i+M_j)^2}
\dfrac{\sigma_{ij}}{\sqrt{s}}\,f_1\,f_2\,{\rm K}_1
\left(\frac{\sqrt{s}}{T}\right)\,ds\,,
\label{eq:sigmavij}
\end{eqnarray}
with
\begin{eqnarray}
f_1 = \sqrt{s^2 + (M^2_i-M^2_j)^2-2\,s\,(M^2_i+M^2_j)}\,,\nonumber\\
f_2 = \sqrt{s - (M^2_i-M^2_j)^2}\sqrt{s-(M^2_i+M^2_j)^2}\,.
\end{eqnarray}
In the above, $s$ is the Mandelstam variable and ${\rm K}_i$
is the $i$th order Modified Bessel function of second kind.
All the relevant couplings required to calculate $\sigma_{i\,j}$ are given
in Appendix \ref{dm_coupling}. To compute relic density of dark matter we
need the value of comoving number density $Y$ at the present
epoch $T=T_0$, which can be found by solving the Boltzmann equation
given in Eq.\,\ref{eq:BEapprox}. We have solved this equation
using {\tt micrOMEGAs}~\cite{Belanger:2014vza} package where the
information about the present model has been implemented
using {\tt FeynRules}~\cite{Alloul:2013bka} package. After finding
the value of $Y(T_0)$, the dark matter relic density can now
be obtained from the following relation \cite{Edsjo:1997bg}
\begin{eqnarray}
\Omega h^2 = 2.755\times 10^8\,\left(\dfrac{M_{\phi^0}}
{\rm GeV}\right)\,Y(T_0)\,.
\label{eq:omega}
\end{eqnarray}

In addition to all the theoretical as well as experimental constraints
mentioned in previous sections, we have also considered few more
experimental bounds which are indispensable to the dark matter
phenomenology. These are discussed below,
\begin{itemize}
\item {\bf Relic density of dark matter:}
Various satellite borne experiments viz. WMAP \cite{Hinshaw:2012aka},
Planck \cite{Ade:2015xua} have precisely measured the abundance of dark matter
in the Universe at the present epoch, which is
\begin{eqnarray}
0.1172\leq\Omega h^2\leq0.1226\,\,\,\,{\rm at}\,\,68\%\,\,{\rm C.L.}\,. 
\end{eqnarray}
\item {\bf Spin independent elastic scattering cross section:} In the
present model dark matter candidate $\phi^0$ can elastically scatter off
the terrestrial detector nuclei by exchanging CP even scalar
bosons $h^0$ and $H^0$. This is known as the spin independent
elastic scattering cross section of $\phi^0$ which is assumed to be
responsible for its direct signature in the earth based detectors.
The spin independent elastic scattering cross section
for the process $\phi^0 + N \rightarrow \phi^0 + N$ with $N$ being
a nucleon is given by
\begin{eqnarray}
\sigma_{\rm SI} = \dfrac{\mu^2_{\rm red}}{4\pi}\left[\dfrac{M_N\,f_N}
{M_{\phi^0}\,v_d}\left(\dfrac{\,g_{\phi^0\phi^0 h^0}}{M^2_{h^0}}
-\dfrac{\,g_{\phi^0\phi^0 H^0}}{M^2_{H^0}}\right)\right]^2\,,
\label{eq:sigmasi}
\end{eqnarray}
where $\mu_{\rm red} = \frac{M_N M_{\phi^0}}{M_N + M_{\phi^0}}$
is the reduced mass of nucleon $N$ and $\phi^0$ while
$f_N$ is the nuclear form factor for scalar mediated
interactions and its value is $\sim 0.3$ \cite{1306.4710}. In
Eq.\,\ref{eq:sigmasi}, the negative sign arises due to
the opposite sign of couplings of $h^0$ and $H^0$ with
quarks $q$ (i.e. $h^0(H^0)\bar{q}q$ coupling).
The coupling between two dark matter particles and a CP even
scalar $h^0$($H^0$) is represented by $g_{\phi^0\phi^0 h^0(H^0)}$,
which can be decomposed into two parts. One is coming form the $\mathbb{Z}_2$
even doublet $H$ while other part is from the triplet $\Delta$.
These coupling can be written as
\begin{eqnarray}
g_{\phi^0\phi^0 h^0} &=& -\bar{\lambda}_1\,v_d-\bar{\lambda}_2\,v_t\,,
\label{eq:phphhcoup}\\
g_{\phi^0\phi^0 H^0} &=& \tilde{\lambda}_1\,v_d-\tilde{\lambda}_2\,v_t\,,
\label{eq:phphHcoup}
\end{eqnarray}
with
\begin{eqnarray}
\bar{\lambda}_1 &=& \left(\lambda_5+\lambda_6+2\lambda_9\right)\,
\cos \alpha\,,
\\
\bar{\lambda}_2 &=& \left(\lambda_7+\lambda_8-\frac{\sqrt{2}\,\tilde{\mu}}
{v_t}\right)\,\sin \alpha \,,
\\
\tilde{\lambda}_1 &=& \left(\lambda_5+\lambda_6+2\lambda_9\right)
\,\sin \alpha \,,
\\
\tilde{\lambda}_2 &=& \left(\lambda_7+\lambda_8-\frac{\sqrt{2}\,
\tilde{\mu}}{v_t}\right)\,\cos \alpha \,,
\label{eq:lambar}
\end{eqnarray}
and the quantity $\tilde{\mu}$ can be defined in terms free parameters
of the present model as
\begin{eqnarray}
\tilde{\mu} = \dfrac{(\lambda_6+2\lambda_9)\,v^2_d+ \lambda_8 v^2_t+
2\,\Delta M^2_\pm}{2\sqrt{2}\,v_t}\,,
\label{eq:mutilde}
\end{eqnarray}
while $\Delta M^2_\pm$ is the mass squared difference between
the LOP and inert charged scalar $\phi^\pm$ i.e.
\begin{equation}
\Delta M^2_\pm = M^2_{\phi^\pm} - M^2_{\phi^0} \,.
\label{eq:mdeltap} 
\end{equation}
Later, we will see that $g_{\phi^0\phi^0 h^0}$
coupling will play a significant role in the freeze-out
process for low mass range of $\phi^0$ ($\phi^0<80$ GeV).

The non-observation of any dark matter signal in direct
detection experiments has severely constrained the
dark matter-nucleon spin independent scattering cross
section and at present the most stringent bound on
$\sigma_{\rm SI}$ has been reported by XENON 1T collaboration
\cite{Aprile:2017iyp}. Therefore, a viable dark matter model requires
$\sigma_{\rm SI} < \sigma_{exp}$ where $\sigma_{exp}$ being
the upper bound on $\sigma_{\rm SI}$ obtained from
the XENON 1T direct detection experiment.  
\item {\bf Lower bounds on BSM scalar masses:} Precise
measurement of $Z$ boson decay width at LEP \cite{ALEPH:2005ab}
forbids any additional invisible decay modes of $Z$
boson i.e. $Z\rightarrow \phi^0\,a^0$. This puts a
lower bound on the sum of the two masses i.e.
\begin{eqnarray}
M_{\phi^0} + M_{a^0} > M_{Z}\,.
\end{eqnarray} 
Apart from this, the LEP II data also ruled out a significant
portion of odd sector scalar masses which satisfy following
inequalities \cite{0810.3924}
\begin{eqnarray}
M_{\phi^0} < 80\,\,{\rm GeV}, M_{a^0}<100\,\,{\rm GeV}\,\,{\rm and}\,\,
M_{a^0}-M_{\phi^0}> 8\,\,{\rm GeV}\,. 
\end{eqnarray}
Moreover, there exists a lower bound of 80 GeV at 95\% C.L.
\cite{Patrignani:2016xqp}
on charged scalar mass from LEP. Keeping in mind all these
experimental results, in this work we have considered the
masses of $\phi^\pm$ and $a^0$ larger than 100 GeV i.e.
$M_{\phi^\pm}\,,M_{a^0}>100$ GeV.

\item {\bf Invisible decay width of SM Higgs boson:} In
the present model, since we have $h^0\phi^0\phi^0$ coupling,
a pair of $\phi^0$ can be produced from the decay of SM-like
Higgs boson $h^0$ at LHC, if the kinematical condition $M_{h^0}\geq
2\,M_{\phi^0}$ holds. This non-standard decay channel is
known as the invisible decay mode of SM-like Higgs boson. Current
experimental lower limits on the masses of $\mathbb{Z}_2$ particles
allow only one invisible decay mode of $h^0$ in the
present scenario which is $h^0\rightarrow \phi^0\phi^0$.
The decay width of this process is given by
\begin{eqnarray}
\Gamma_{h^0\rightarrow \phi^0\phi^0} =
\dfrac{g^2_{h^0\rightarrow \phi^0\phi^0}}{32 \pi M_{h^0}}
\sqrt{1-\dfrac{4\,M^2_{\phi^0}}{M^2_{h^0}}}\,.
\end{eqnarray}
Throughout this work we have restricted ourselves
to that portion of the parameter space where
the invisible decay width of SM-like Higgs boson is
less than $20\%$ of its total decay width
(invisible branching ratio ${\rm BR}_{\rm inv}< 0.2$)
\cite{Khachatryan:2016whc}.
\end{itemize} 

\begin{figure}[h!]
\centering
\includegraphics[height=5cm,width=7.5cm,angle=0]{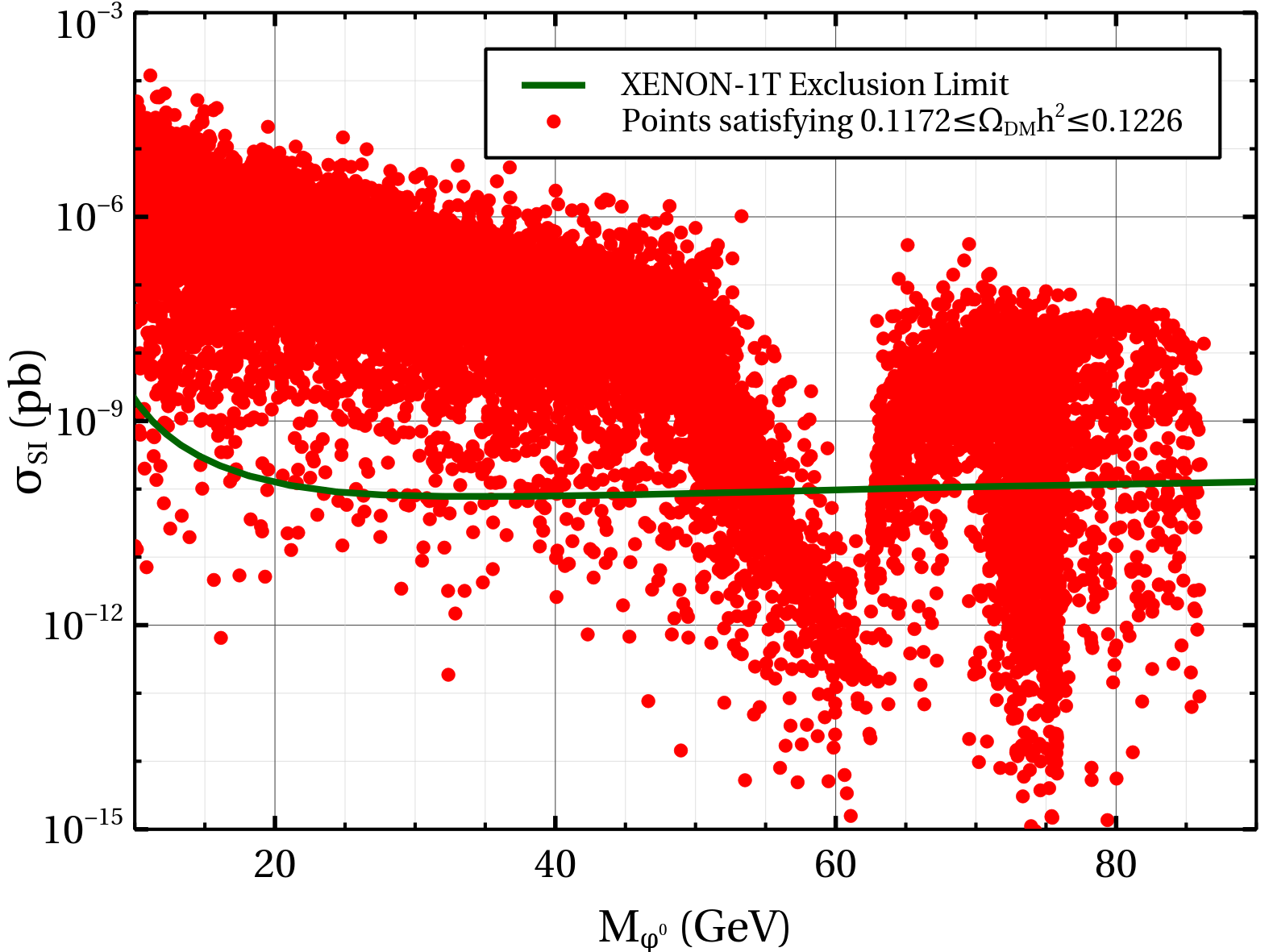}
\includegraphics[height=5cm,width=7.5cm,angle=0]{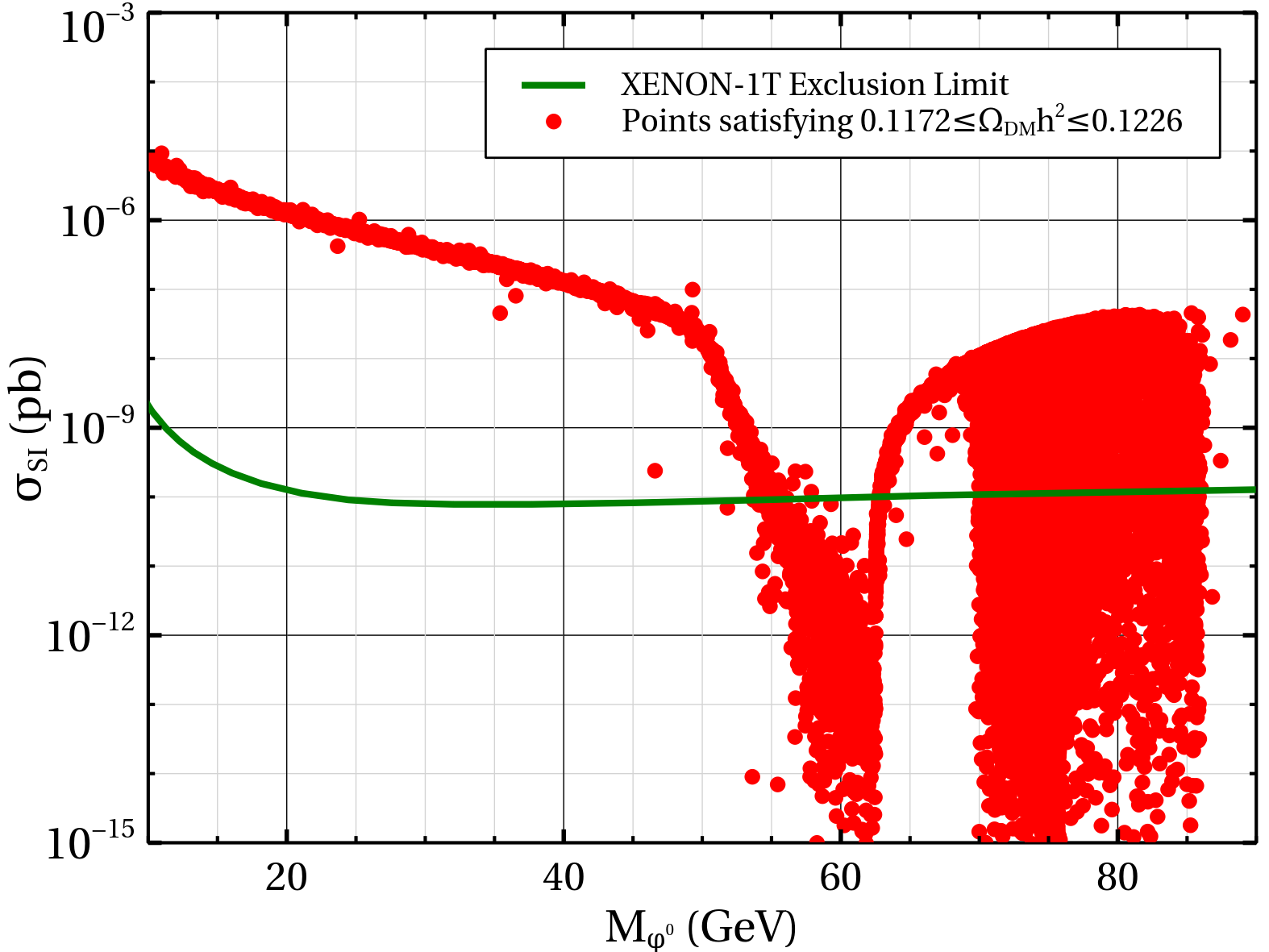}
\caption{Variation of spin independent scattering cross section
$\sigma_{\rm SI}$ with the mass of $\phi^0$ for two different
values of triplet VEV namely $v_t=3$ GeV (left panel) and $v_t=1$ MeV
(right panel).}
\label{plot:mdm-sigmaSI-lowmass}
\end{figure}

Now, we will present the results of dark matter phenomenology of
the present model considering $\phi^0$ as our dark matter candidate.
In this work, we have varied the mass of $\phi^0$ between 10 GeV
to 1 TeV. In order to determine the allowed parameter space of this
model which will satisfy both theoretical as well as experimental constraints
we have scanned the free parameters (mentioned in Eq.\,\ref{eq:para})
in the following ranges
\begin{eqnarray}
\begin{array}{cccccc}
1.0\times 10^{-5}&\le& \left|\lambda_i\right|\,(i=5-9)&\le& 1.0\,\,,\\ 
1.0\times 10^{-4}&\le& \lambda_{\Phi} &\le& 1.0\,\,,\\
1.0\times 10^{-6}&\le& \left|\alpha\right|\,({\rm rad})& <
& 1.0\times 10^{-2}\,\, ,\\
120&\le & M_{H^{\pm\pm}}\,\text{(GeV)} & \le& 350\,\, ,\\
110&\le & M_{H^\pm}\,\text{(GeV)} & \le& 330\,\, ,\\
100&\le & M_{H^0}\,\text{(GeV)} & \le& 300\,\, ,\\
10&\le & M_{\phi^0}\,\text{(GeV)} & \le& 1000\,\, ,\\
100&< & M_{\phi^\pm}
\,\text{(GeV)} & \le& 1050\,,
\label{eq:para-ranges}
\end{array}
\end{eqnarray}
with $M_{\phi^\pm}>M_{\phi^0}$. Furthermore, in the present scenario
mass of the inert CP odd scalar $a^0$ is not a free parameter as one can
easily express $M_{a^0}$ in terms of our chosen free parameters, i.e.
\begin{eqnarray}
M^2_{a^0} = \lambda_8 v^2_t + \lambda_6 v^2_d
+(2\,M^2_{\phi^\pm} -M^2_{\phi^0})\,.
\label{eq:Ma0}
\end{eqnarray}
Throughout this work, we have used the condition $M_{a^0}>M_{\phi^0}$,
in other word, $\phi^0$ is the lightest particle of the odd sector (LOP).
Using the above ranges of model parameters,
we have found that the dark matter relic density
satisfies the Planck limit ($0.1172\leq\Omega h^2\leq
0.1226$ \cite{Ade:2015xua}) only in two distinct mass ranges of $\phi^0$. One
of them is the low mass region where $M_{\phi^0}$ lies below
90 GeV ($M_{\phi^0}<90$ GeV) while in the high mass region
$M_{\phi^0}$ is larger than 535 GeV ($M_{\phi^0}>535\text{ GeV}$).
The allowed region in $M_{\phi^0}-\sigma_{\rm SI}$ plane
for the low mass range of $\phi^0$ is shown
Fig.\,\ref{plot:mdm-sigmaSI-lowmass}. The left panel of
Fig.\,\ref{plot:mdm-sigmaSI-lowmass} has been generated for
the triplet VEV $v_t=3$ GeV while the right panel is for $v_t=1$ MeV.
In both panels, all red colours points in $M_{\phi^0}-\sigma_{\rm SI}$ plane
satisfy the relic density bound as well as the other theoretical
constraints considered in this work like unitarity, vacuum stability etc.  
The green solid line denotes the most severe upper bounds on the
dark matter spin independent scattering cross section till date.
This experimental upper limits on $\sigma_{\rm SI}$ has been reported
recently by the XENON 1T collaboration \cite{Aprile:2017iyp}. From
both panels of Fig.\,\ref{plot:mdm-sigmaSI-lowmass}, one
can see that the current limits on $\sigma_{\rm SI}$ from
XENON 1T have ruled out maximum portion of the region
with $M_{\phi^0}\la50$ GeV.

\begin{figure}[h!]
\centering
\includegraphics[height=5cm,width=8.0cm,angle=0]{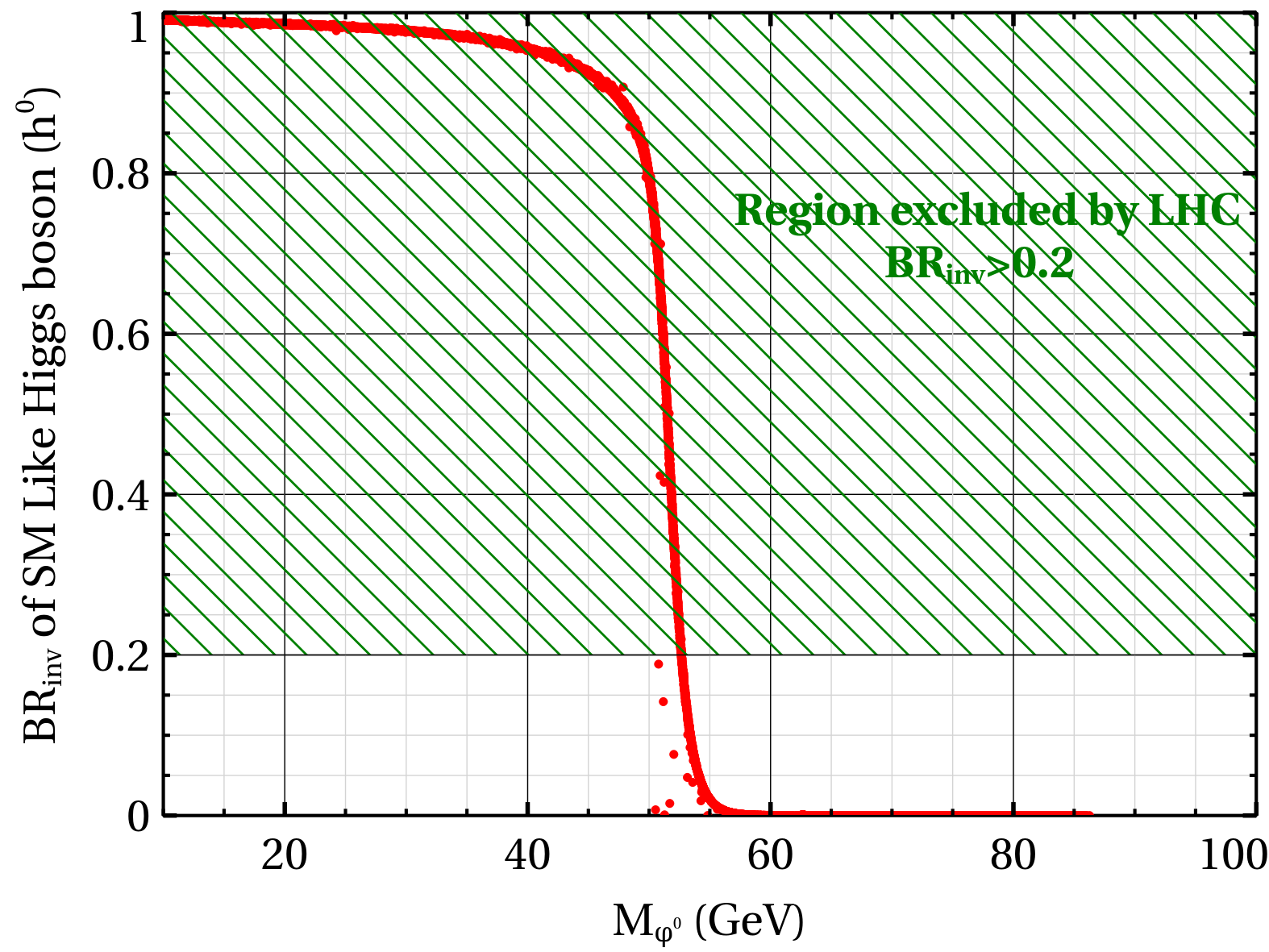}
\caption{Variation of invisible branching ratio of SM-like Higgs boson
$h^0$ with the mass of LOP $\phi^0$.}
\label{plot:invbr}
\end{figure}

Although, for $v_t=3$ GeV
there are still some allowed parameter space with $M_{\phi^0}\la50$ GeV,
however those regions will be forbidden if we impose the constraint on
invisible branching ratio of the SM-like Higgs boson $h^0$ (see
Fig.\,\ref{plot:invbr}). In the low mass region, the dark matter
particle $\phi^0$ annihilates to SM fermion and antifermion, $W^+W^-$
pairs. Most of the contribution to $\langle\sigma {\rm v}\rangle$
arises from $\phi^0\phi^0\rightarrow b\bar{b}$ ($W^+W^-$) channel
for $M_{\phi^0}\la70$ GeV ($\ga 70$ GeV). Since
we have always considered $M_{\phi^\pm}, M_{a^0}>100$
GeV, the co-annihilations among the inert sector particles
have no significant effect on the dark matter relic density in
the low mass region of $\phi^0$. The sudden dip in $\sigma_{\rm SI}$
around $M_{\phi^0}\simeq 60$ GeV is due to the resonance effect of
SM-like Higgs boson $h^0$ of mass 125.5 GeV. In the resonance region
of $h^0$ ($M_{\phi^0}\sim M_{h^0}/2$), the annihilation cross section
of $\phi^0$ mediated by $h^0$ increases sharply, which has an inverse
effect on $\Omega h^2$. Hence, to generate dark matter relic density
within the desired ballpark of Plank limit, the $\phi^0\phi^0 h^0$
coupling (defined in Eq.\,\ref{eq:phphhcoup}) has to be decreased
accordingly. As the same coupling also enters into the
expression of $\sigma_{\rm SI}$ (see Eq. \ref{eq:sigmasi}),
there exits a sharp decrease in $\sigma_{\rm SI}$
around the resonance region of $h^0$.  

It has been already mentioned that in the present scenario, the
only source of invisible decay mode of $h^0$ is $h^0\rightarrow \phi^0\phi^0$.
Therefore, in Fig.\,\ref{plot:invbr} invisible branching ratio
of the SM-like Higgs boson has been plotted with respect
to the mass of $\phi^0$. Here the green dashed region represents
${\rm BR}_{\rm inv}\geq 0.2$, which is excluded by
the current LHC data \cite{Khachatryan:2016whc}. All the red points in
$M_{\phi^0}-{\rm BR}_{\rm inv}$ plane reproduce the
dark matter relic density within the allowed ballpark of the Planck limit.
It is also evident from Fig.\,\ref{plot:invbr} that
initially when $M_{\phi^0}<40$ GeV, the invisible
branching ratio of $h^0$ is very high and thereafter
there is a sharp fall of ${\rm BR}_{\rm inv}$ for $M_{\phi^0}$
lying between 40 GeV to 60 GeV. This nature of ${\rm BR}_{\rm inv}$
is due to the phase space suppression i.e. the available phase
space for this decay mode deceases as the mass of $\phi^0$
increases and eventually when $M_{\phi^0}$ becomes larger
than $M_{h^0}/2 \sim 62.5$ GeV the decay mode $h^0\rightarrow
\phi^0\phi^0$ becomes kinematically forbidden and hence
${\rm BR}_{\rm inv}$ of $h^0$ vanishes.    
Most importantly, from this plot it is clearly evident
that the dark matter mass $M_{\phi^0}\la 50$ GeV is
excluded by the invisible decay width constraint of $h^0$. 

\begin{figure}[h!]
\centering
\includegraphics[height=5cm,width=5.5cm,angle=0]{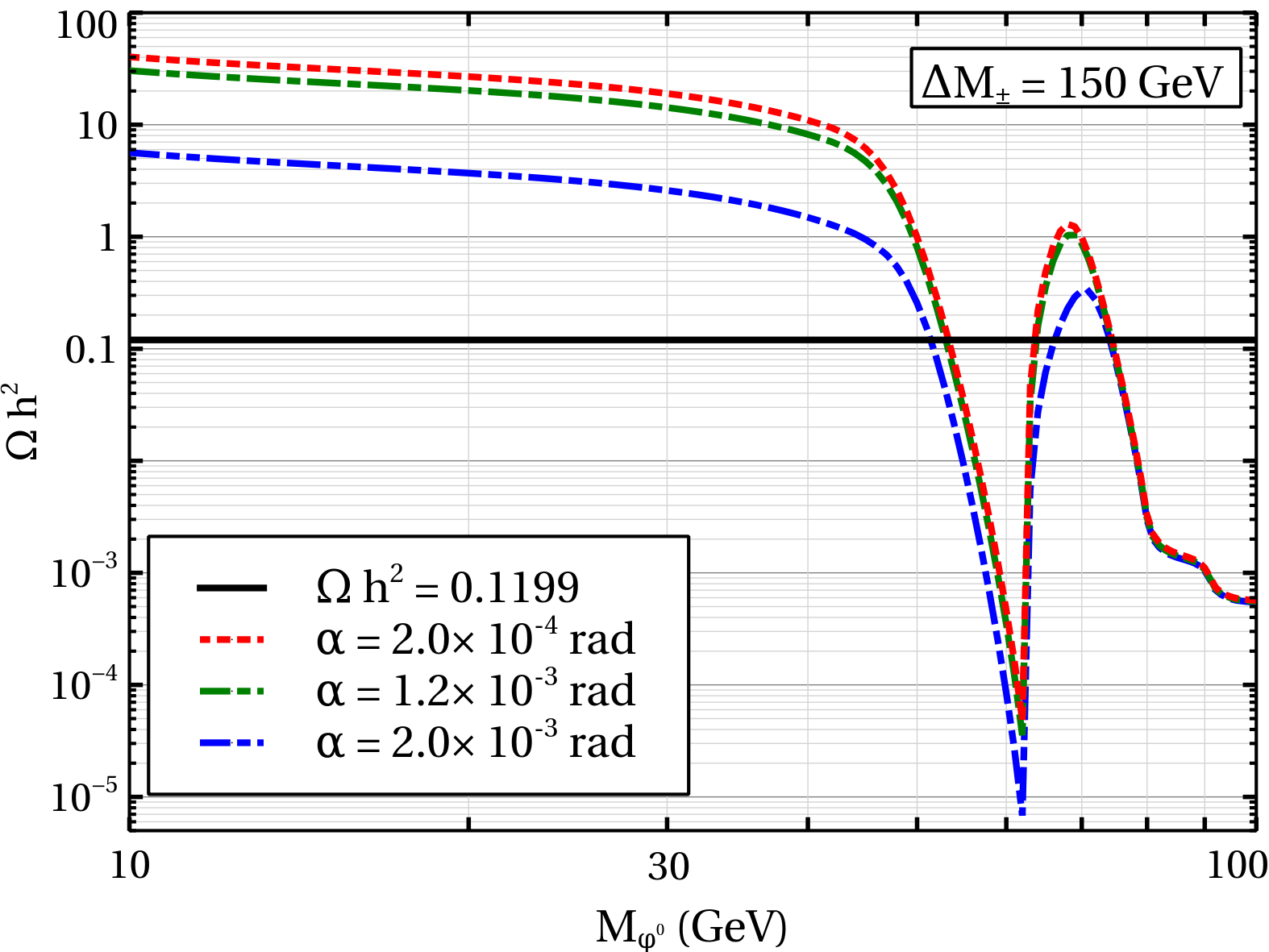}
\includegraphics[height=5cm,width=5.5cm,angle=0]{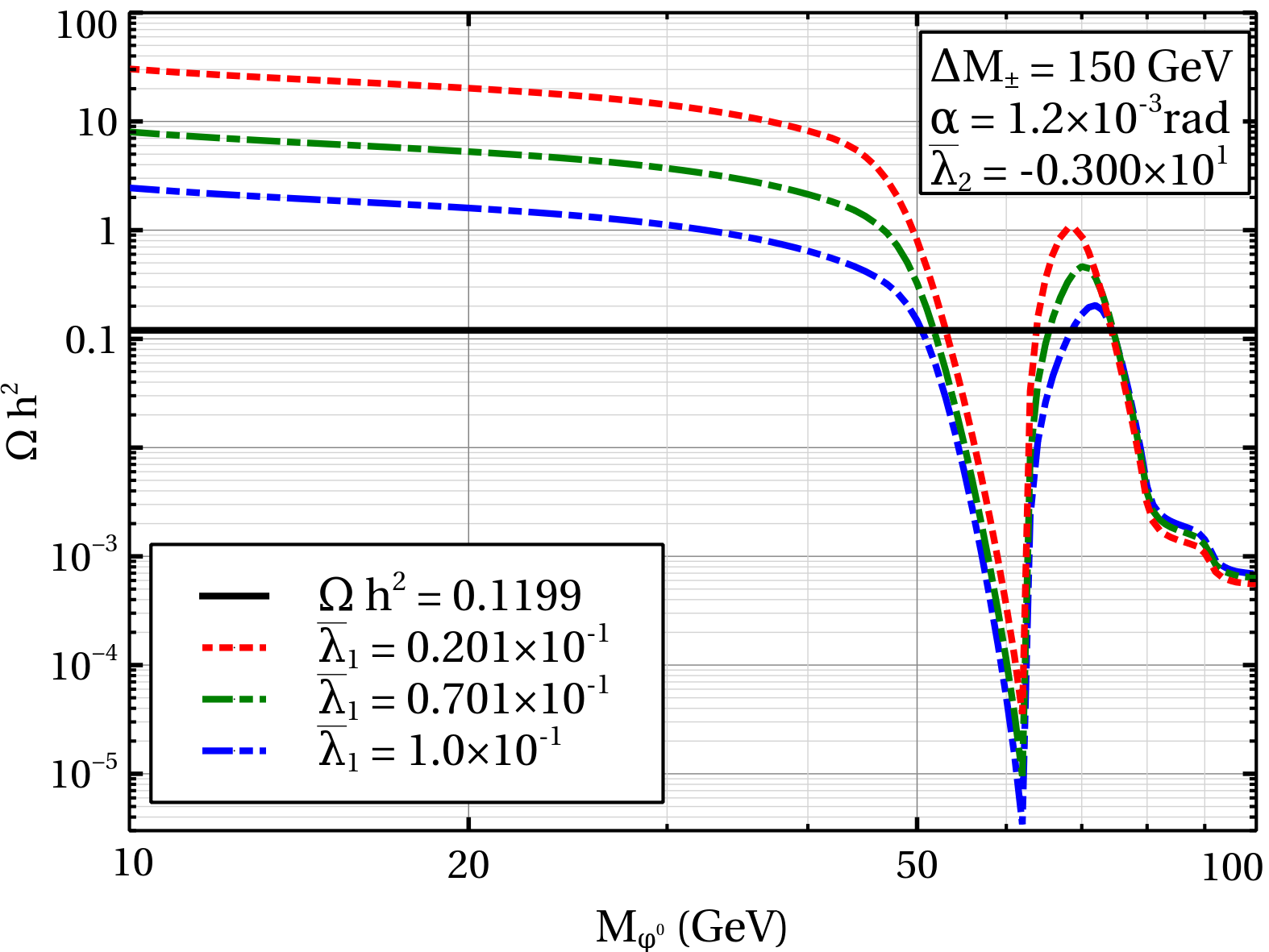}
\includegraphics[height=5cm,width=5.5cm,angle=0]{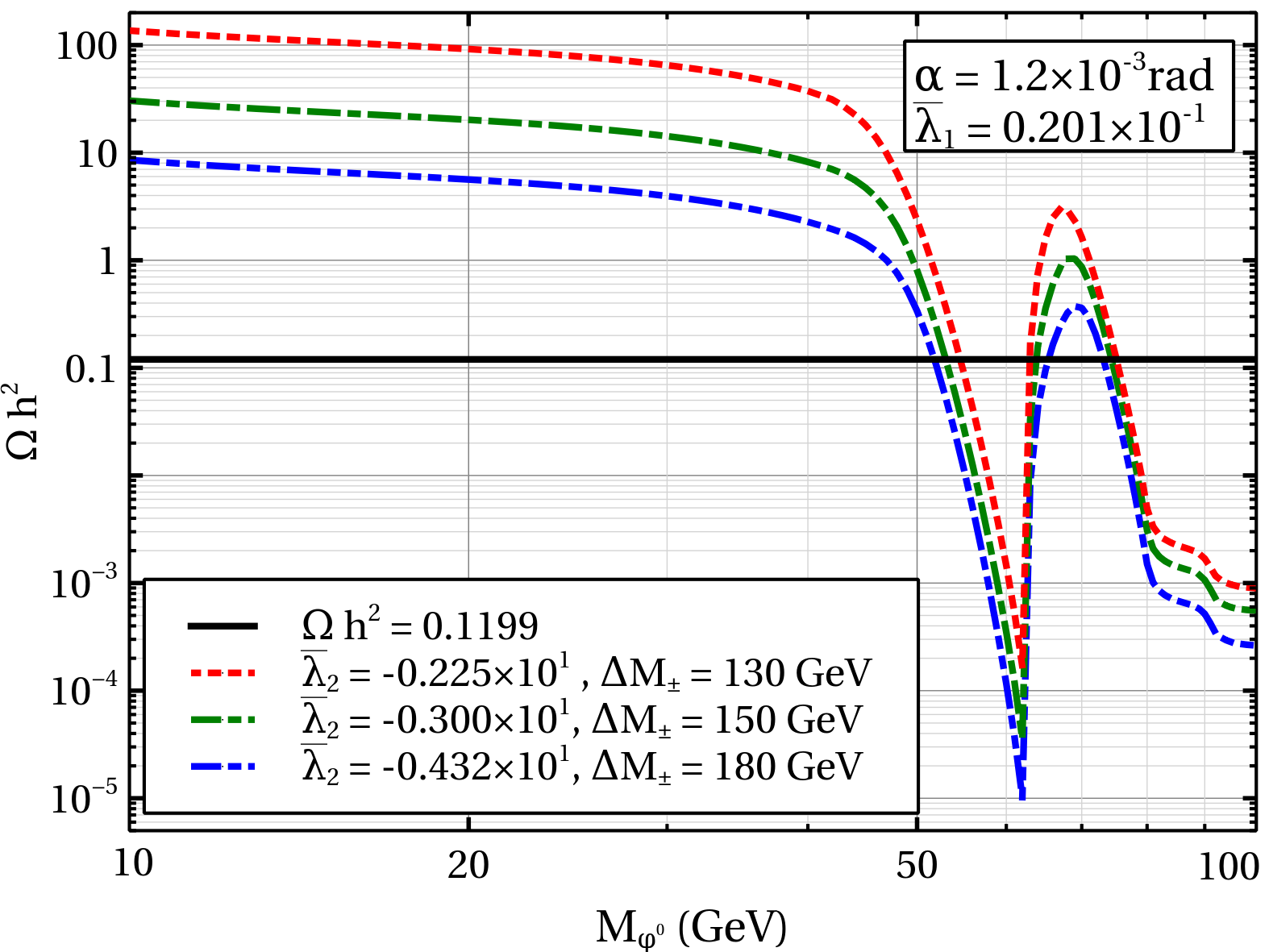}
\caption{Variation of $\Omega h^2$ with $M_{\phi^0}$ for three
different values of model parameters namely, $\alpha$ (left panel),
$\bar{\lambda}_1$ (central panel) and $\bar{\lambda}_2$ (right panel).}
\label{plot:lineplot-lowmass}
\end{figure}

Next, in order to illustrate the effects of some relevant parameters
on dark matter relic density, we show the variation of $\Omega h^2$
with respect to the mass of $\phi^0$ in Fig.\,\ref{plot:lineplot-lowmass}
for three different values of model parameters namely $\alpha$,
$\bar{\lambda}_1$ and $\bar{\lambda}_2$ respectively.   
In each panel of Fig.\,\ref{plot:lineplot-lowmass},
three lines represent the variation of relic density for
three different values of a chosen model parameter while the horizontal
black solid line denotes the central value of dark matter
relic density i.e. $\Omega h^2=0.1199$ as observed by the Planck satellite.
In the leftmost panel, we have considered three different values of
CP even scalar mixing angle $\alpha = 2.0\times 10^{-4}$ (red dashed line),
$1.2\times 10^{-3}$ (green dashed dot line) and $2.0\times 10^{-3}$ (blue
dashed dot dot line) respectively. The values of other free parameters
have been kept fixed at $M_{H^{\pm\pm}}=156.90$ GeV, $M_{H^\pm}=203.39$ GeV,
$M_{H^0}=M_{A^0}=241.03$ GeV, $\Delta M_\pm = 150$ GeV,
$\lambda_{\Phi}=0.0203$, $\lambda_5=0.02$, $\lambda_6 = 0.4\times10^{-4}$,
$\lambda_7=0.41\times10^{-3}$, $\lambda_8=-0.646\times10^{-2}$,
$\lambda_9 = 0.4\times10^{-4}$ and $v_t=3$ GeV. 
From the leftmost panel one can see that relic density decreases
when we increase $\alpha$. This can be understood in the following
way. As mentioned earlier, in the low dark matter mass region
where $M_{\phi^0}\la70$ GeV, the main annihilation channel is
$\phi^0\phi^0\rightarrow b\bar{b}$ mediated by the SM-like Higgs
boson $h^0$. The coupling $g_{\phi^0\phi^0h^0}$ (Eq.\,\ref{eq:phphhcoup})
has two parts one of them is proportional to $\cos \alpha$ while other
part has $\sin \alpha$. In this plot we have varied $\alpha$ between
$2.0\times 10^{-4}$ rad to $2.0\times 10^{-3}$ rad. In this small range
of $\alpha$, the coupling $g_{\phi^0\phi^0 h^0}$ as well as
$\langle \sigma {\rm v}\rangle$ increase with
$\alpha$ and hence the relic density behaves oppositely 
as it is inversely proportional to $\langle \sigma {\rm v}\rangle$.
However, beyond the $h^0$ resonance region i.e
for $M_{\phi^0}\ga 70$ GeV, $\phi^0\phi^0\rightarrow W^+W^-$
becomes the main annihilation channel and in this case dominant
contribution comes from the diagrams mediated by $H^0$
instead of $h^0$. The coupling $g_{\phi^0\phi^0H^0}$ has
exactly opposite angular dependence compared to $g_{\phi^0\phi^0h^0}$
(Eq.\,\ref{eq:phphHcoup}). Now for the considered values of
model parameters $\left|\left(\lambda_7+\lambda_8
-\dfrac{\sqrt{2}\,\tilde{\mu}}{v_t}\right)\right|>>
\left|\left(\lambda_5+\lambda_6+2\,\lambda_9\right)\right|$,
hence the dominant process becomes practically independent
of $\alpha$. As a result we have not observed any significant change in
$\Omega h^2$ for three different values of $\alpha$ when
$M_{\phi^0}\ga 70$ GeV. Similarly, the effects of $\bar{\lambda}_1$
and $\bar{\lambda}_2$ on $\Omega h^2$ can be easily understood
using Eqs.\,\ref{eq:phphhcoup}-\ref{eq:lambar}. One 
should note that in the present model the parameter
$\tilde{\mu}$ (defined in Eq.\,\ref{eq:mutilde})
has a profound effect on relic density, e.g. 
for the particular benchmark point both the
couplings $g_{\phi^0\phi^0h^0}$ and $g_{\phi^0\phi^0H^0}$
enhance with the $\tilde{\mu}$ for
$10\text{ GeV}\leq M_{\phi^0}\leq 100\text{ GeV}$ while $\tilde{\mu}$
itself increases with $\Delta M_{\pm}$. 

\begin{figure}[h!]
\centering
\includegraphics[height=5cm,width=7.5cm,angle=0]{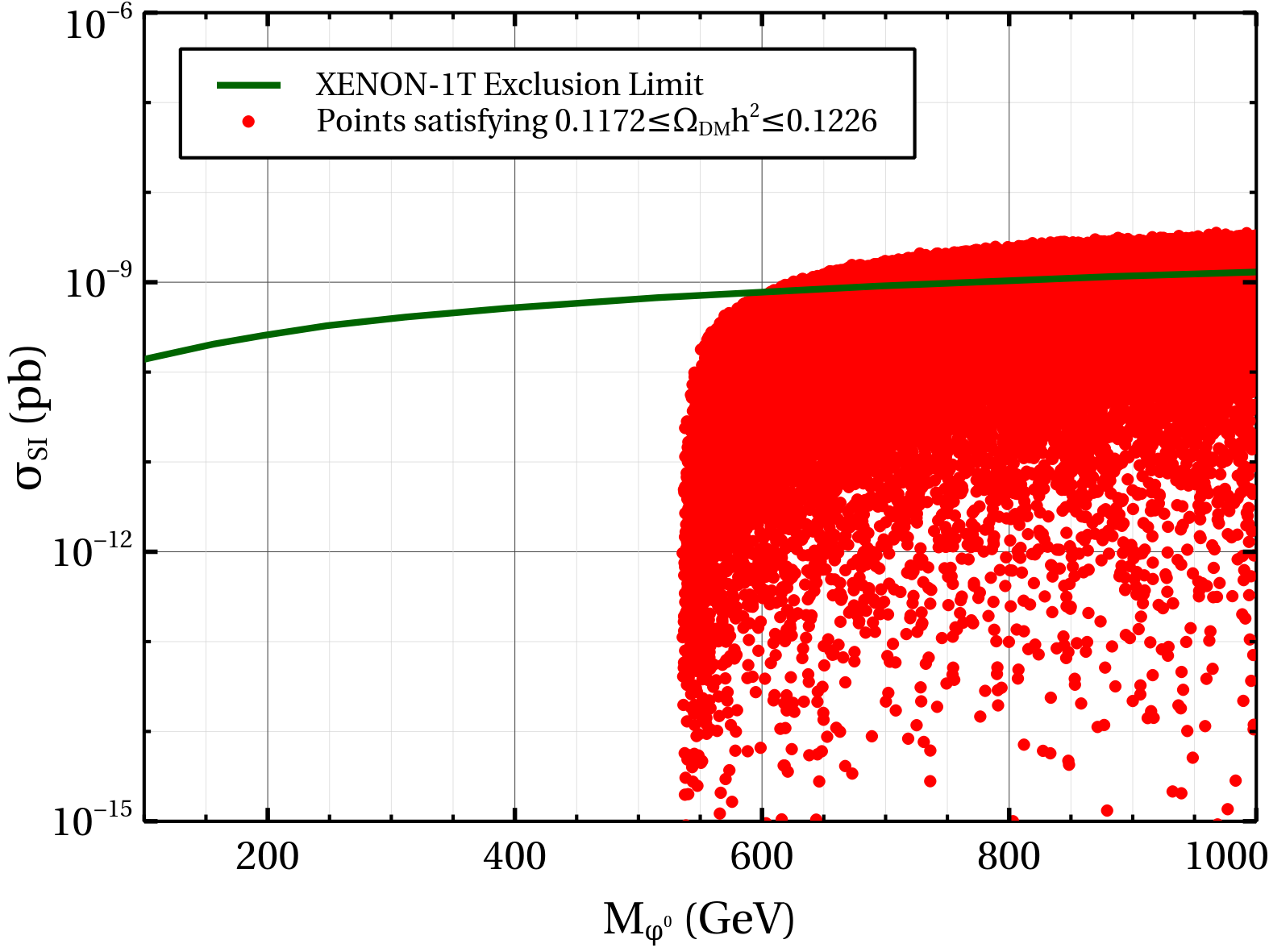}
\includegraphics[height=5cm,width=7.5cm,angle=0]{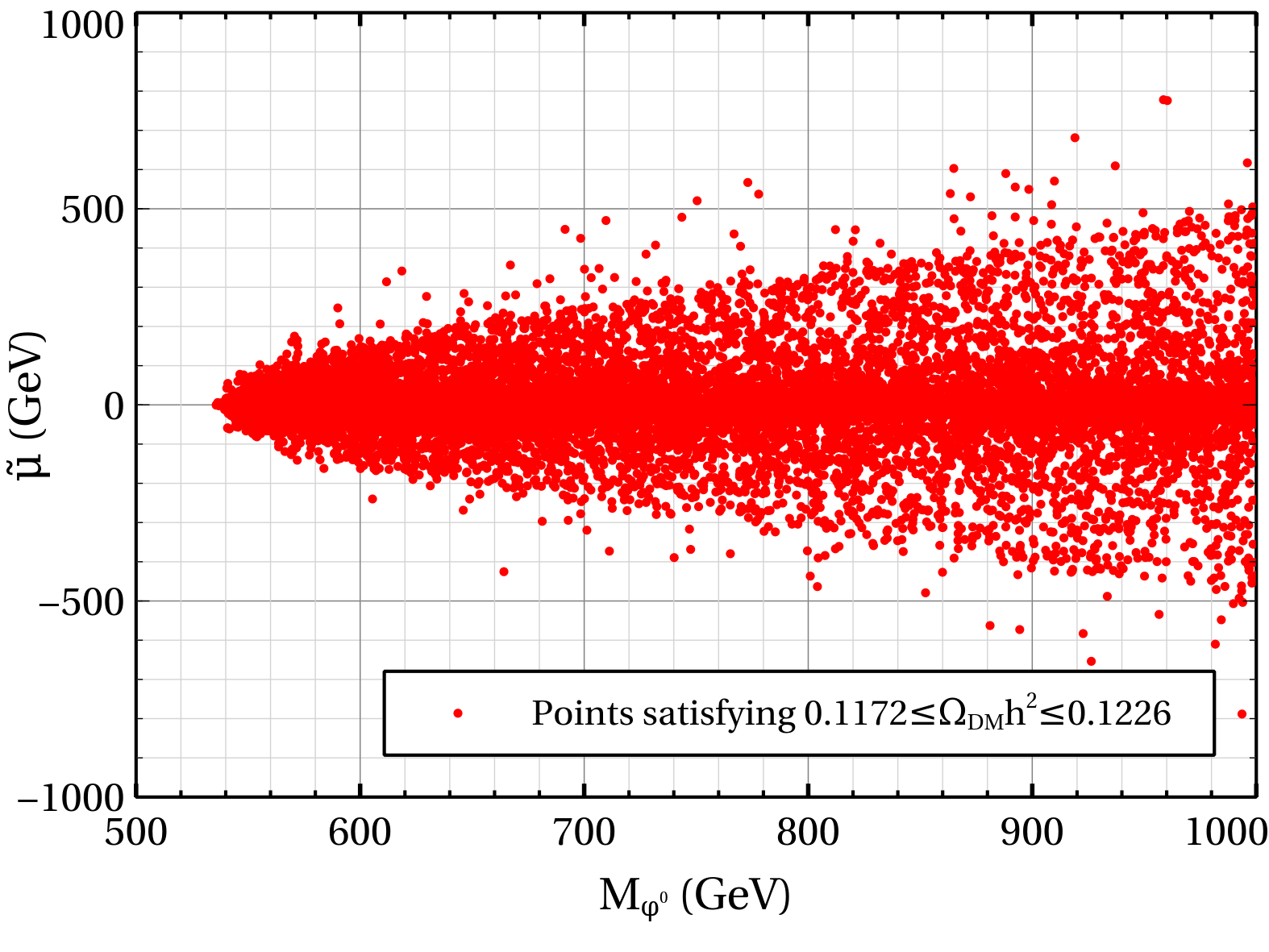}
\caption{Left panel: Allowed region in $M_{\phi^0}-\sigma_{\rm SI}$
plane for the high mass range of $\phi^0$, Right panel: Allowed values
of trilinear coupling $\tilde{\mu}$, which satisfy Planck relic density
bound.}
\label{plot:mdm-sigmaSI-highmass}
\end{figure}


Instead of low mass region of $\phi^0$, let us now concentrate on
the high mass region. Therefore, in the left panel of
Fig.\;\ref{plot:mdm-sigmaSI-highmass} we show the variation 
$\sigma_{\rm SI}$ with $M_{\phi^0}$ for the high mass region.
Analogous to the previous case, here also all the red coloured points in
$M_{\phi^0}-\sigma_{\rm SI}$ plane satisfy the Plank limit of relic density
while green solid line represents the current exclusion limit
on $\sigma_{\rm SI}$ from the XENON 1T collaboration. From this
plot it is evidently seen that the high mass region starts from
$M_{\phi^0}\simeq 535$ GeV. Therefore, in the present model we have not
found any parameter space for the dark matter mass lies between $\sim 90$ GeV
to $\sim 534$ GeV. A possible reason is that in this intermediated
mass region the main dark matter annihilation channels are
$\phi^0\phi^0\rightarrow W^+W^-\,,ZZ$ and the annihilation
cross sections for these processes are too high to maintain
relic abundance in the right ballpark. On the other hand, in the higher mass
region, one can only satisfy the Planck bound for low values
of $\tilde{\mu}$ i.e. $|\tilde{\mu}|\la 500$ GeV
(see plot in the right panel of Fig.\,\ref{plot:mdm-sigmaSI-highmass}).
Otherwise, $\phi^0$ will dominantly self annihilate
to the components of scalar triplet
$\Delta$, which will jeopardise the viability $\phi^0$ as a
DM candidate by reducing its relic density severely. Therefore,
one has to lower the value of $\tilde{\mu}$ by properly adjusting
the parameters $\Delta M_{\pm}$ and $\lambda$'s
(mainly $\lambda_6$ and $\lambda_9$), which critically constrain
$\Delta M_{\pm} = \sqrt{M^2_{\phi^{\pm}}-M^2_{\phi^0}}$
within a very small range. As a result for the low value of
$\tilde{\mu}$ ($\Delta M^2_{\pm}$), similar to the
intermediate mass range of $M_{\phi^0}$ ($90 \text{ GeV} \leq M_{\phi^0}
\leq 534$ GeV), the annihilation channels $W^+W^-$ and $ZZ$
again become the dominant processes in the higher mass region
as well. However, in this case low value of $\Delta M^2_{\pm}$ induces significant
self annihilation as well as co-annihilation among the odd sector
particles which substantially increase the relic density $\Omega h^2$
so that it satisfies the Plank limit.  

\begin{figure}[h!]
\centering
\includegraphics[height=5cm,width=7.5cm,angle=0]{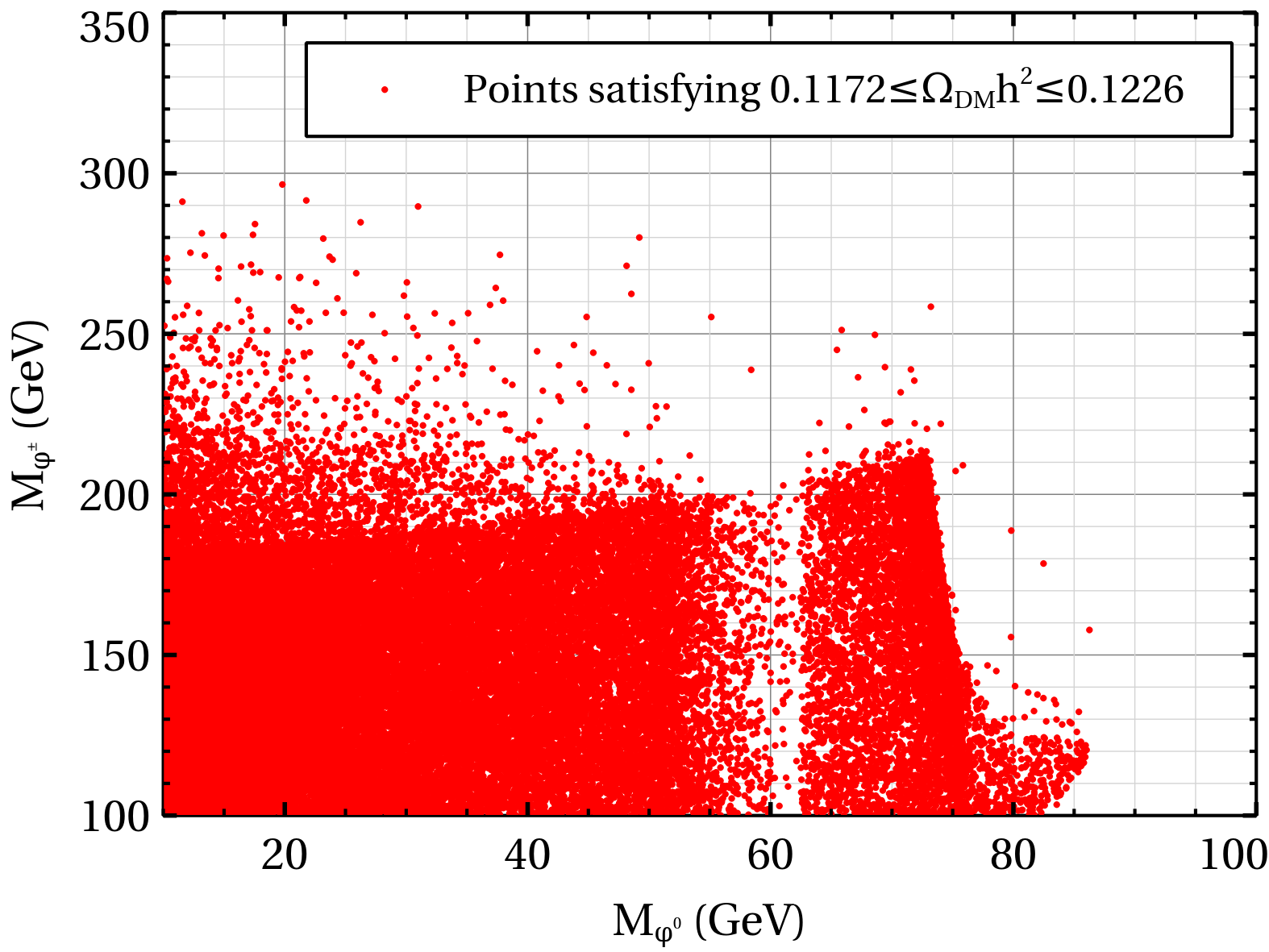}
\includegraphics[height=5cm,width=7.5cm,angle=0]{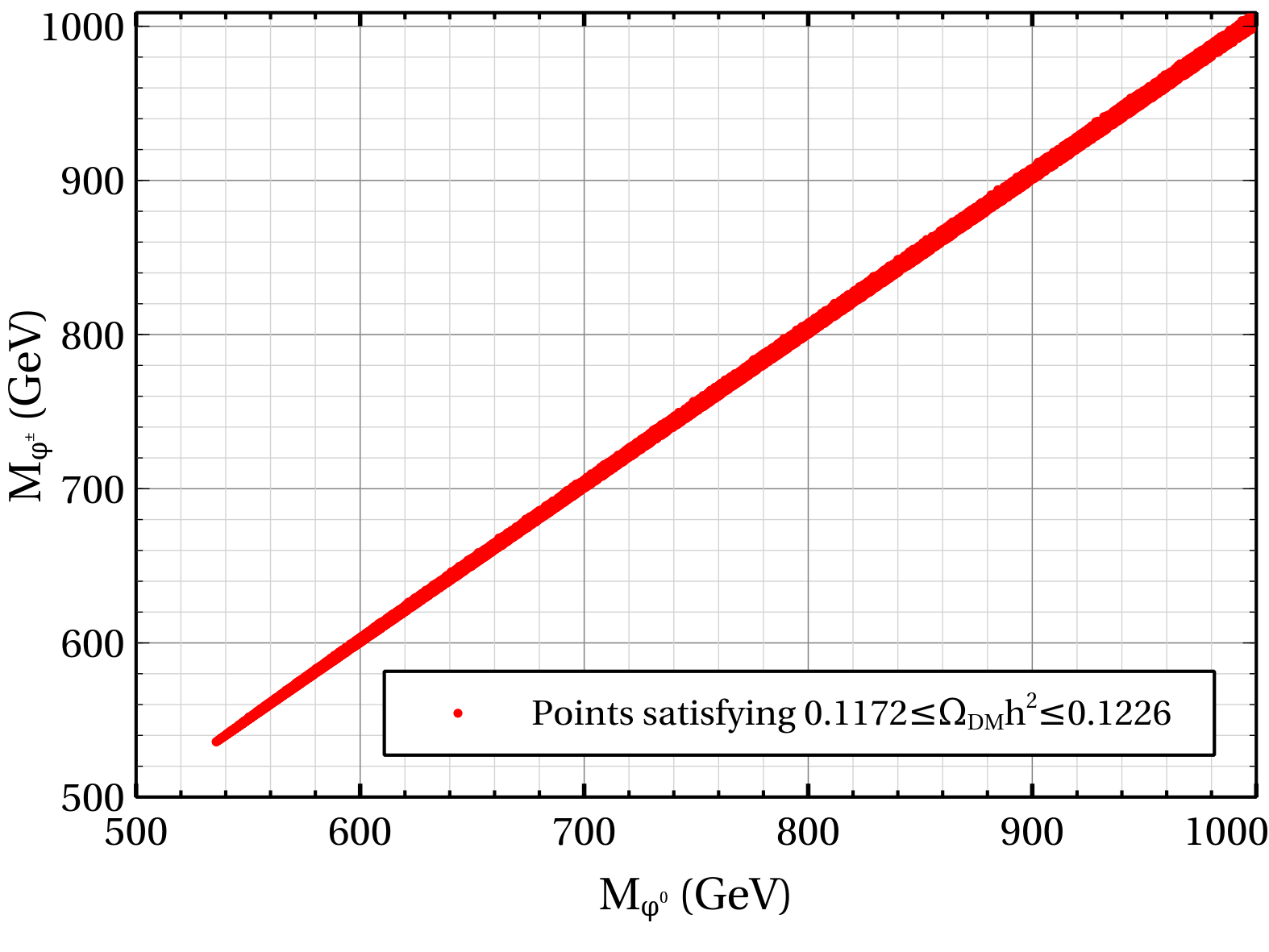}
\caption{Allowed values of inert charged scalar mass with respect to
the mass of $\phi^0$ for low mass region (left panel) and high mass
region (right panel).}
\label{plot:oddsectormassrange}
\end{figure}

The effect of co-annihilation can be clearly understood by
comparing plots in both the panels of Fig.\,\ref{plot:oddsectormassrange}.
The left panel shows the allowed values of $\phi^{\pm}$ which satisfy
the relic density for the low mass region of $\phi^0$. It is seen that
for a particular value of $M_{\phi^0}$, mass of $\phi^{\pm}$ can vary
between its lowest possible value of 100 GeV to $\sim 250$ GeV and
for such a large mass difference between $\phi^0$ and $\phi^\pm$
there is practically no effect of co-annihilation. On the other hand,
the right panel of Fig.\,\ref{plot:oddsectormassrange} clearly
depicts that to satisfy the Planck limit on dark matter relic density,
one needs very small mass difference between LOP and $\phi^{\pm}$, which
evidently illustrate the effect of co-annihilation on $\Omega h^2$ in
the high mass region of $\phi^0$. 
\begin{figure}[h!]
\centering
\includegraphics[height=5cm,width=7.5cm,angle=0]{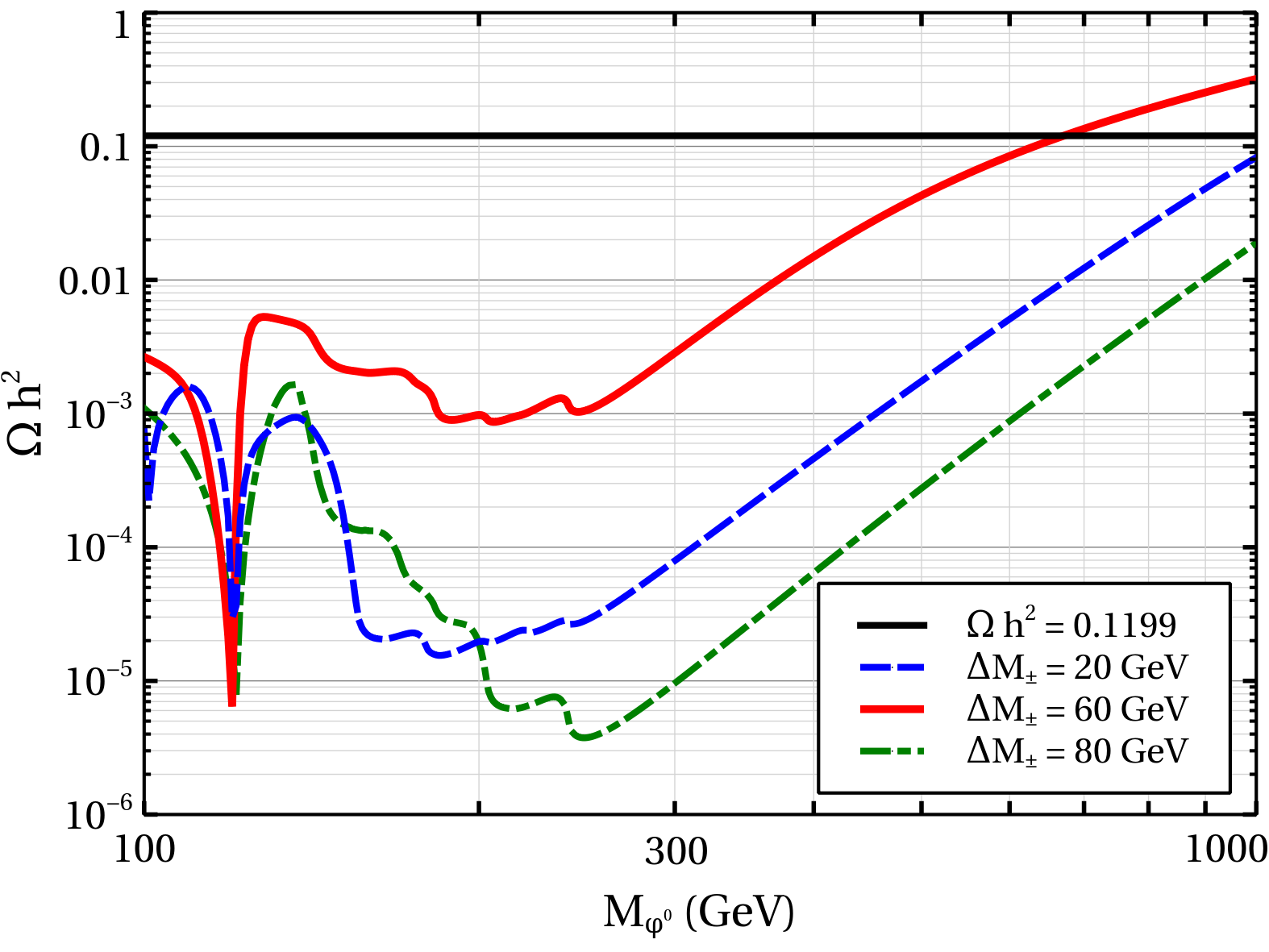}
\includegraphics[height=5cm,width=7.5cm,angle=0]{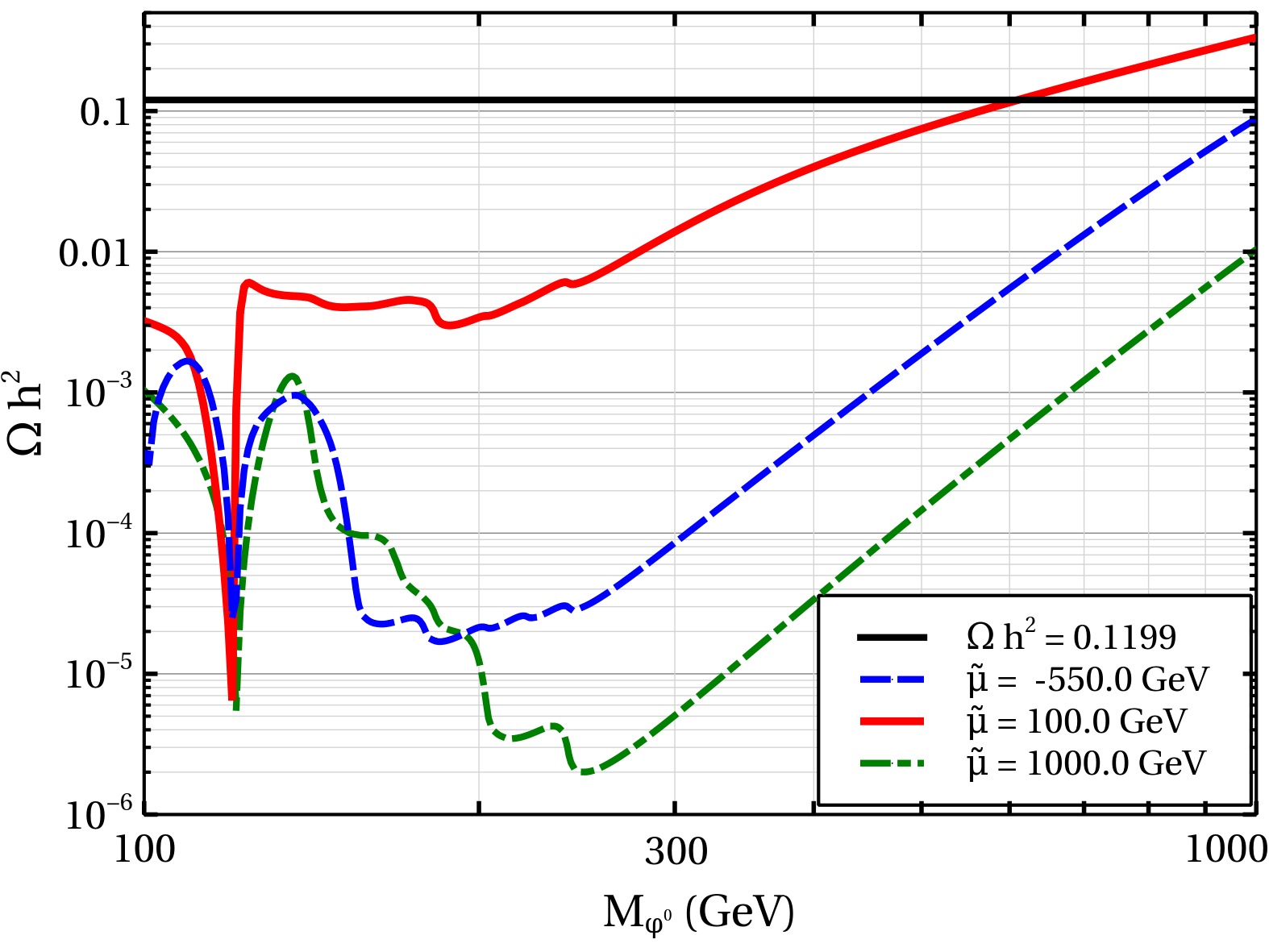}
\caption{Variation of dark matter relic density with the mass of $\phi^0$
for three different values of $\Delta M_{\pm}$ (left panel) and $\tilde{\mu}$
(right panel).}
\label{plot:lineplot-highmass}
\end{figure}

Now, to understand the effects of the trilinear scalar coupling
$\tilde{\mu}$ and mass splitting $\Delta M_{\pm}$ on the dark matter
relic density we have plotted the variation $\Omega h^2$ with $M_{\phi^0}$
for three different values of $\tilde{\mu}$ and $\Delta M_{\pm}$ respectively
in Fig.\,\ref{plot:lineplot-highmass}. In the left panel, three different
coloured lines in $M_{\phi^0}-\Omega h^2$ plane represent three differently
chosen values of $\Delta M_{\pm}$ i.e. $\Delta M_{\pm} = 20$ GeV (blue 
dashed line),  60 GeV (red solid line) and 80 GeV (green dashed dot line).
This figure has been generated for $M_{H^{\pm\pm}}=156.90$ GeV,
$M_{H^\pm}=203.39$ GeV, $M_{H^0}=M_{A^0}=241.03$ GeV,
$\lambda_{\Phi}=0.0203$, $\lambda_5=0.02$, $\lambda_6 = -0.1\times10^{-1}$,
$\lambda_7=0.41\times10^{-3}$, $\lambda_8=0.1\times10^{-1}$,
$\lambda_9 = 0.41\times10^{-1}$ and $v_t=3$ GeV. From this plot
it appears that for the above mentioned set of model parameters
there is a unique value of $\Delta M_{\pm}=60$ GeV for which dark matter
relic density satisfies Planck limit when $M_{\phi^0}=671$ GeV.
In this particular situation, $\Delta M_{\pm}=60$ GeV corresponds
to a mass difference of $2.677$ GeV between $\phi^{\pm}$ and LOP ($\phi^0$)
while the mass of CP odd inert scalar $a^0$ is 675.896 GeV.
In this case, the annihilation and co-annihilation channels which have
dominant contributions to the dark matter relic density are
$\phi^0\phi^0\rightarrow W^+W^-,\,ZZ$, $\phi^0\phi^+\rightarrow W^+\gamma,\,W^+Z$,
$\phi^+\phi^-\rightarrow W^+W^-,\,\gamma\gamma,\,Z\gamma,\,H^{++}H^{--}$,
$a^0a^0\rightarrow W^+W^-,\,ZZ$, $\phi^+a^0\rightarrow W^+ \gamma$.
Now if we fix the mass of LOP to a particular value then one can not reproduce
the observed relic density by lowering the mass gap between
$\phi^0$ and $\phi^\pm$ arbitrarily small. This can well be
understood if we see the expression of trilinear scalar
coupling between the triplet $\Delta$ and inert doublet $\Phi$
given in Eq.\,\ref{eq:mutilde}. From this equation, one can notice that
for a particular chosen values of $\lambda_6$,
$\lambda_8$, $\lambda_9$ and $M_{\phi^0}$ there exists a definite
range of $\Delta M_{\pm}$ for which the absolute value
of $\tilde{\mu}$ lies within the limit specified by the plot in the right
panel of Fig.\,\ref{plot:mdm-sigmaSI-highmass}. The value of
$\Delta M_{\pm}$ beyond this range will enhance the absolute value
of $\tilde{\mu}$, which will eventually reduce the dark matter
relic density by increasing the annihilation cross section.
This feature has been illustrated in the right panel of
Fig.\,\ref{plot:lineplot-highmass}, where three adopted 
values of $\tilde{\mu}$ correspond to $\Delta M_{\pm}=
21.21$ GeV (blue dashed line), 56.64 GeV (red solid line) and
83.82 GeV (green dashed dot line). Moreover, for a particular
set of model parameters one can easily find the allowed
values of $\Delta {M_{\pm}}$ for $535\text{ GeV}\leq M_{\phi^0}\leq
1000\text{ GeV}$ by setting $|\tilde{\mu}|\la 500$ GeV (from the
right panel of Fig.\,\ref{plot:lineplot-highmass}). Using
this upper limit on the absolute value of $\tilde{\mu}$,
we find a range of allowed values of $\Delta M_{\pm}$
lying between 25.8 GeV to 70.03 GeV which satisfy the Planck
limit on relic density for the chosen set of model parameters
mention above. Now both panels of Fig.\,\ref{plot:mdm-sigmaSI-highmass}
reveal that, this range of $\Delta M_{\pm}$ is indeed
true for this set of model parameters as in both
panels the blue and green lines do not satisfy the
Planck limit since the parameter $\Delta M_{\pm}$ corresponding
to these lines lie outside the above specified range.

\begin{figure}[h!]
\centering
\includegraphics[height=5cm,width=7.5cm,angle=0]{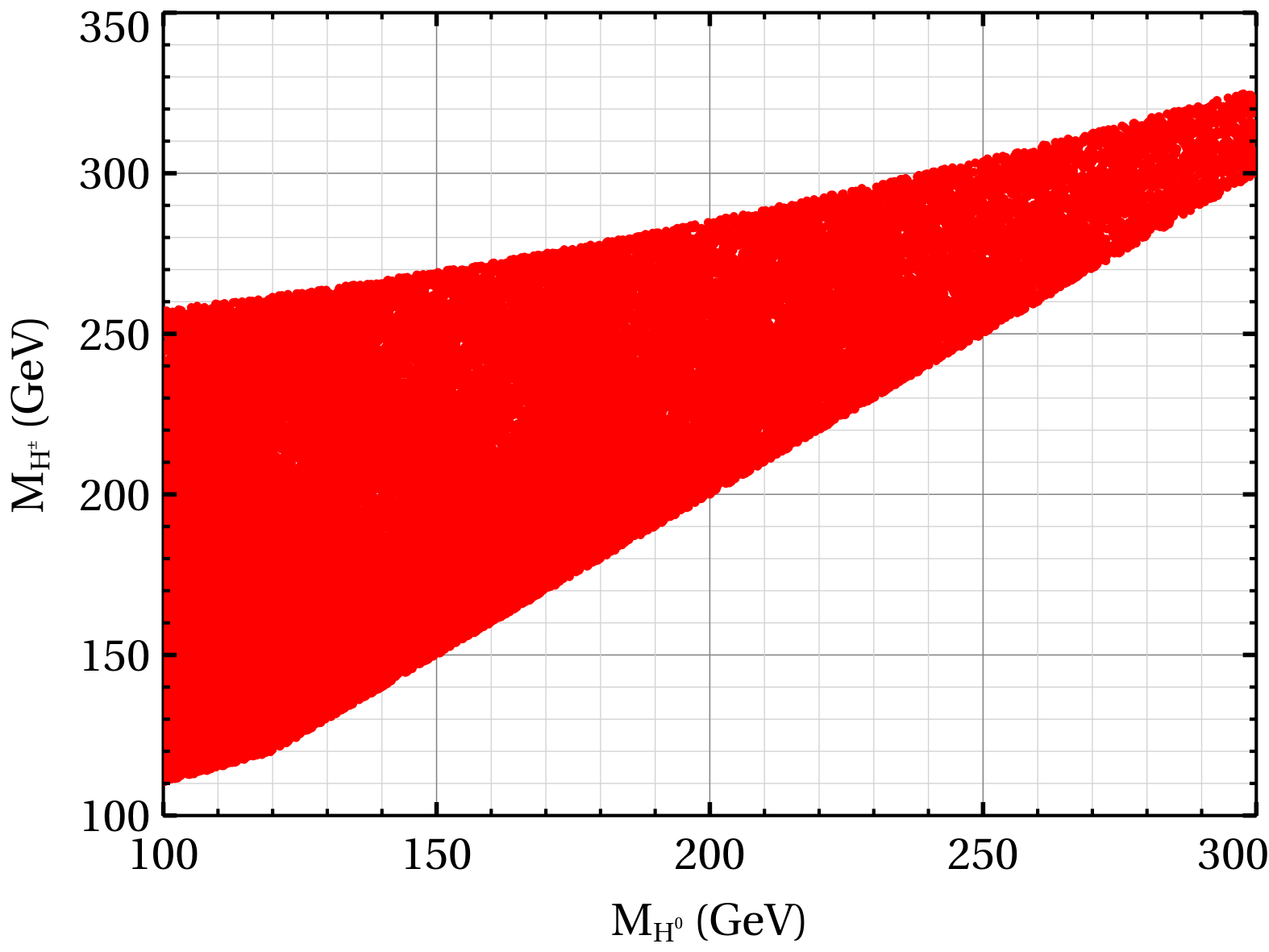}
\includegraphics[height=5cm,width=7.5cm,angle=0]{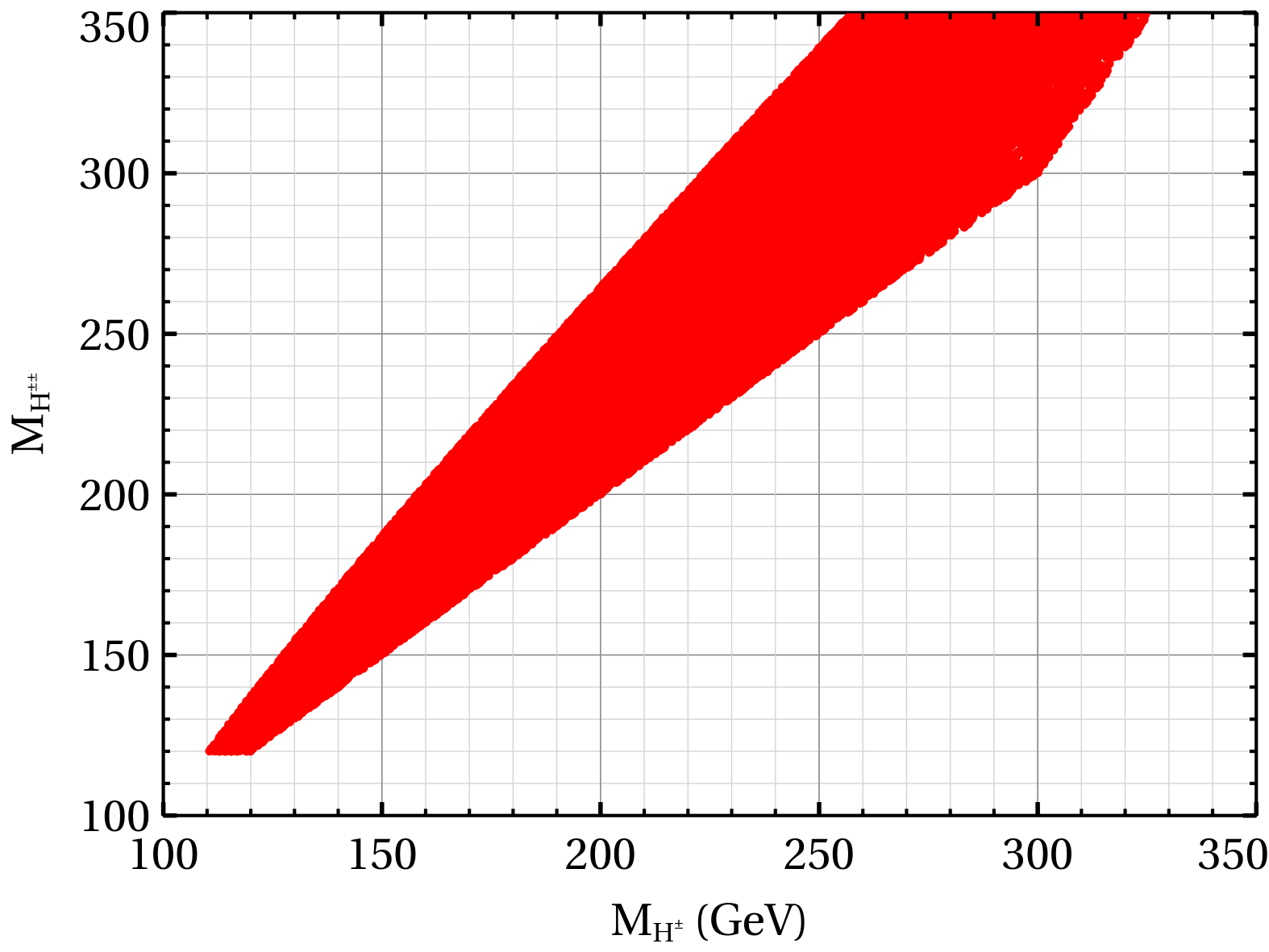}
\caption{Allowed mass ranges for the BSM scalars, which satisfy all
the theoretical and experimental constrains considered in this work.}
\label{plot:tripletmassrange}
\end{figure}

Next, in the both panels of Fig.\,\ref{plot:tripletmassrange}
we demonstrate the allowed mass ranges of the components
of triplet scalar $\Delta$, which satisfy all the theoretical
constrains such as unitarity, vacuum stability etc. and the relevant
experimental bounds mainly from LEP, LHC etc. 
In the left panel, we have presented the allowed ranges
of $M_{H^0}$ and $M_{H^\pm}$ while the region allowed
in $M_{H^0}-M_{H^{\pm\pm}}$ plane has been shown in the
right panel. We have checked that these allowed
mass ranges of BSM scalars also satisfy the
dark matter relic density (in both the allowed regions) and the
experimental upper bound obtain from the non-observation of flavour
violating decay like $\mu \rightarrow e\,\gamma$ (see Eq.\,\ref{eq:mutoegamma}
and related discussions in Section \ref{neu}). Note that the nature
of these two plots (aligned around a line with slope $45^\circ$)
mainly arise due to the unitarity constrains discussed in
Section \ref{sec:unitarity}.

\begin{figure}[h!]
\centering
\includegraphics[height=5cm,width=7.5cm,angle=0]{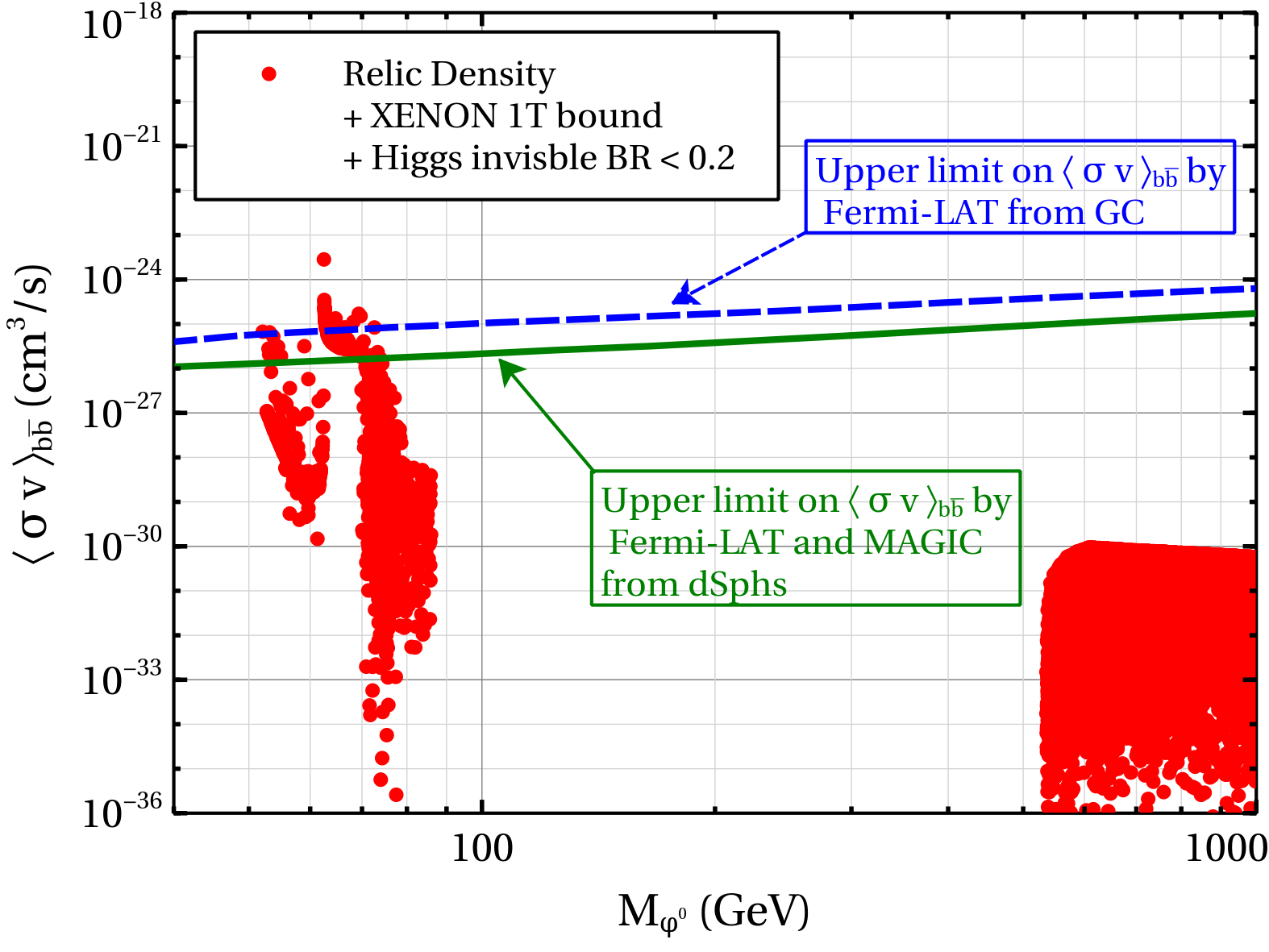}
\includegraphics[height=5cm,width=7.5cm,angle=0]{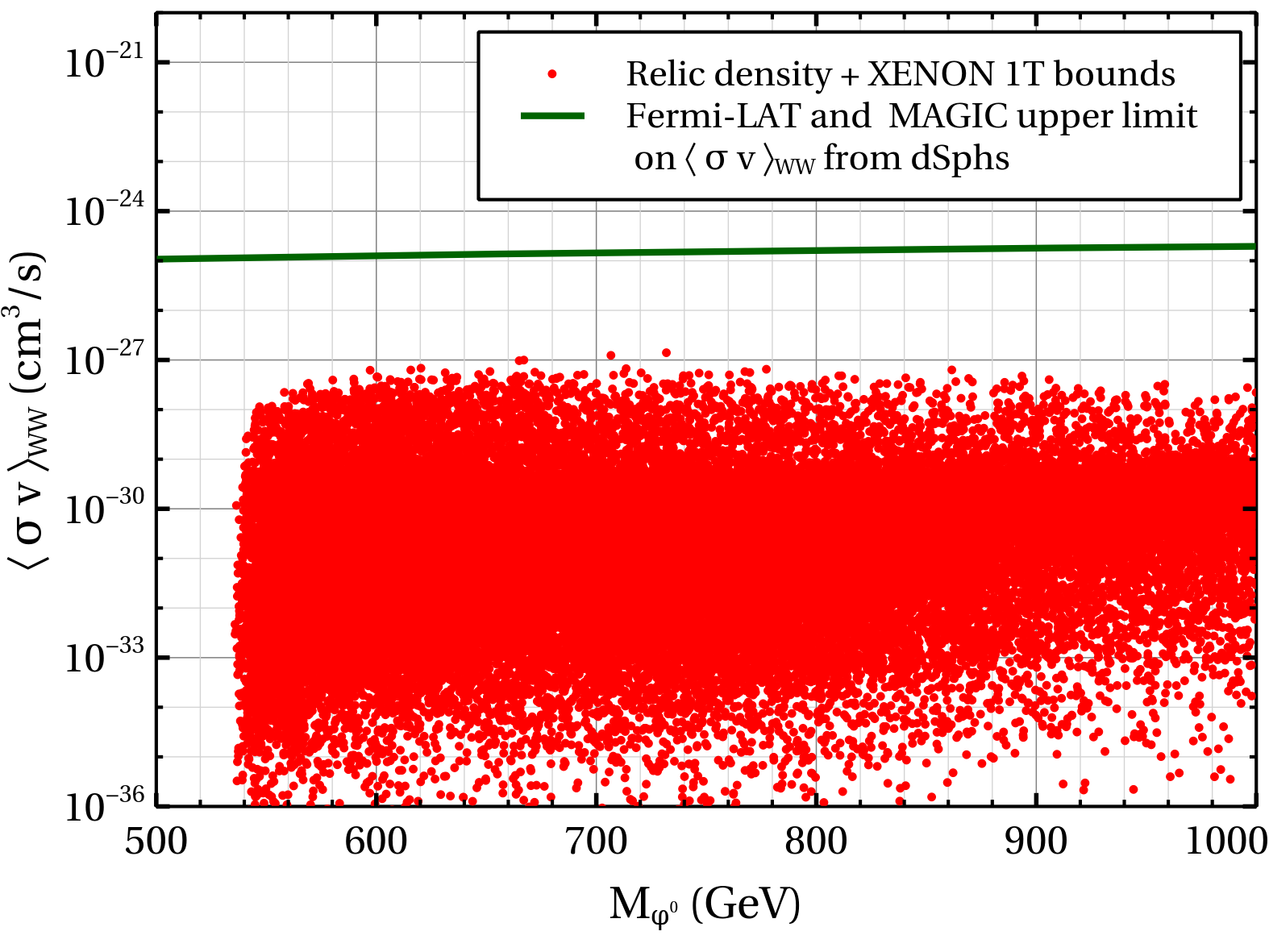}
\caption{Annihilation cross sections of $\phi^0$ at the present
epoch for $b\bar{b}$ (left panel) and $W^+W^-$ (right panel)
final states.}
\label{plot:indirect}
\end{figure}

Finally, we have computed the annihilation cross section
of LOP ($\phi^0$) at the present epoch for $b\bar{b}$ and
$W^+W^-$ final states and these are plotted in the left
and right panels of Fig.\,\ref{plot:indirect} respectively.
In each panel, annihilation cross section for a particular
channel is computed for the model parameter space which
has satisfied all the theoretical and experimental constraints
i.e. unitarity, vacuum stability, Planck limit on relic density,
bounds on $\sigma_{\rm SI}$ from the XENON 1T collaboration,
constrains on the invisible decay width and signal strength
of $h^0$ from LHC etc. The dark matter annihilation cross sections 
computed in the present model for two different final states
are compared with the existing observational bounds on the
same quantities. The non-observation of any significant gamma-ray
excess from the dwarf spheroidal galaxies (dSphs) has put an upper limit
on the dark matter annihilation cross sections for various
final states. Recently a joint analysis \cite{1601.06590} by the Fermi-LAT and
the MAGIC collaborations have provided upper limits on
dark matter annihilation cross section for different final 
states like $b\bar{b}$, $W^+W^-$, $\mu^+\mu^-$, $\tau^+\tau^-$
from the observations of 15 dSphs by the Fermi-LAT for 6 years
and also 158 hours of observation of Segue 1 satellite galaxy
by the MAGIC collaboration. These upper limits on a specific
final state are indicated in each panel by a green solid line.
On the other hand, a more recent analysis by the Fermi-LAT collaboration
\cite{1704.03910} on the Galactic centre gamma-ray excess (GCE)
disfavours the dark matter interpretation of this long standing
puzzle as they have also found gamma-ray excess from a region
where the dark matter signal is not expected i.e. along the Galactic
plane. Thus considering a different astrophysical origin
of this gamma-ray excess (other than dark matter), they have
also reported upper limits on the dark matter annihilation
cross sections for $b\bar{b}$ and $\tau^+\tau^-$ channels.
In the left panel, for comparison, we have also plotted the upper
limits on $b\bar{b}$ final state (blue dashed line) from the GCE.  
From the left panel of Fig.\,\ref{plot:indirect}, it appears
that some portions of the allowed parameter space are
already ruled by these indirect observations. However, in the
high mass region for both $b\bar{b}$ and $W^+W^-$ final
states the limits from indirect detection are not as strong
as in the low mass region.
\section{Collider signature of dark matter at the 13 TeV LHC}\label{cldr}
In this section we perform dark matter searches in a qualitative way at the LHC of centre of momentum energy ($\sqrt{s}$) 13 TeV. Since the dark matter particles are invisible, therefore they reveal their presence only as missing transverse energy ($\mET\,\,$). Furthermore, the inert sector is odd under $\mathbb{Z}_2$ symmetry, consequently the inert scalars are produced in pairs e.g. $\phi^{+} \phi^{-}$, $\phi^0 a^0$, $\phi^\pm a^0$, $\phi^\pm \phi^0$ and $a^0 a^0$, where $\phi^0$ is the stable cold dark matter candidate. For the purpose of collider study we set the triplet VEV $v_t$ at 3 GeV. In the following, we consider some benchmark points (given in Table\;\ref{table:bp}) which we have chosen from the parameter space permitted by the all constraints including dark matter relic abundance and direct detection data. Further, it is evident from the earlier discussions given in the introduction that the masses of the non-standard scalars of the chosen benchmark points also satisfy the present collider bounds obtained from the LHC data. Moreover, with these given benchmark points we are trying to show the variation of significance of dark matter search at the future LHC experiments. For the selected benchmark points $\phi^\pm$ and $a^0$ decay into $W^\pm \phi^0$ and $Z \phi^0$ respectively with 100\% branching ratio. Based on the decay channels of $W^\pm$, $Z$, several final states (e.g. ${\rm jets} + \mET$, ${\rm leptons} + \mET$, ${\rm jets} + {\rm leptons} + \mET$) can be observed at future LHC experiments.



\begin{table}[htbp!]
	\centering
\resizebox{17cm}{!}{
	\begin{tabular}{|p{2cm}|c|c|c|c|c|c|c|c|c|c|c|c|c|c|c|}
		\hline
		Benchmark Points & $\tan{\alpha}$ & $M_{H^{\pm\pm}}$ & $M_{H^{\pm}}$ & $M_{H^0}(=M_{A^0})$ & $M_{\phi^0}$ & $M_{\phi^\pm}$& $M_{a^0}$ & $\lambda_\Phi$ & $\lambda_5$& $\lambda_6$& $\lambda_7$& $\lambda_8$& $\lambda_9$ & $\Omega h^2$ &  $\sigma_{\rm SI}$  \\ 
		&  & $(\rm GeV)$ & $(\rm GeV)$ & $(\rm GeV)$ & $(\rm GeV)$ & $(\rm GeV)$ & $(\rm GeV)$ & & & & & & & &   ($pb$) \\ \hline
		BP1 & $-1.6\times10^{-5}$ & 130.96 & 196.05 & 244.42 & 56.08 & 151.73 & 207.54 & $0.15\times10^{-2}$& $-0.23\times10^{-3}$& $0.28\times10^{-2}$& 0.43 & $0.55\times10^{-1}$ &$0.28\times10^{-3}$ & 0.1212 &  $3.46\times10^{-11}$ \\ \hline
		BP2 & $ 2.1\times10^{-3}$ & 161.82 & 191.16 & 216.65 & 52.24 & 157.90 & 222.83 & $0.12\times10^{-3}$& $-0.16\times10^{-3}$& $0.42\times10^{-1}$& -0.20 & $-0.54\times10^{-1}$ &$-0.60\times10^{-4}$ & 0.1217 &  $2.65\times10^{-11}$ \\ \hline
		BP3 & $2.0\times10^{-6}$ & 187.07 & 204.95 & 221.34 & 74.00 & 178.00 & 240.62 & $0.14$& $0.86\times10^{-3}$& $0.70\times10^{-4}$& $0.40\times10^{-4}$ & $0.17$ &$-0.14\times10^{-2}$ & 0.1177 &  $6.01\times10^{-12}$ \\ \hline
	\end{tabular}}
	\caption{Benchmark points for dark matter searches at the LHC with corresponding
		dark matter relic density and direct detection cross section.}
	\label{table:bp}
\end{table}

On the basis of best possible decay modes and production cross
sections we have selected the following processes
(Eq.\,\ref{s1}-\ref{s4}). After that
depending on the significance we have studied some final states
at the LHC. At this point we would like to mention that due to
the mixing of the triplet scalar fields with the SM doublet fields, in each
of the following processes there are extra contributions coming
from $\mathbb{Z}_2$ even non-standard scalars which are absent
in typical Inert Doublet Model (IDM). However, their contributions
have no practical significance in the processes.  
\begin{subequations}
\begin{eqnarray} 
a)~p~p&\rightarrow &\phi^{+} \phi^{-}\rightarrow W^{+}\phi^0+W^{-}\phi^0 \equiv W^{+}W^{-}\mET\,, \label{s1} \\
b)~p~p&\rightarrow & \phi^0 a^0\rightarrow Z \phi^0 \phi^0 \equiv Z \mET\,,  \label{s2}\\
c)~p~p&\rightarrow & \phi^\pm a^0 \rightarrow W^{\pm} Z \phi^0 \phi^0\equiv W^{\pm} Z \mET\,, \label{s3} \\
d)~p~p&\rightarrow & \phi^\pm \phi^0 \rightarrow W^{\pm} \phi^0\phi^0 \equiv W^{\pm} \mET\,. \label{s4}
\end{eqnarray} 
\end{subequations}

In Table.\;\ref{table:cross} we have alluded the gross production
cross sections at 13 TeV LHC for different processes (given in above)
for the chosen benchmark points. Note that the values of cross sections
for the signal events have been calculated at leading order (LO)
therefore in this sense our choice is conservative as the $K$-factors
for the next to leading order (NLO) corrections are larger than unity.

\begin{table}[H]
	\centering
	\begin{tabular}{|p{2cm}|c|c|c|c|}
		\hline
Benchmark Points & {process (a) ($pb$) } & {process (b) ($pb$) } & {process (c) ($pb$) } & {process (d) ($pb$)} \\ \hline
BP1 & {0.04333}  & {0.05861} & {0.04009} &{0.2756} \\ \hline
BP2 & {0.03738}  & {0.04787} & {0.03204} &{0.2570} \\ \hline
BP3 & {0.02390}  & {0.03105} & {0.02236} &{0.1394} \\ \hline
\end{tabular}
	\caption{LO cross sections (in $pb$) for the processes given in Eq.\;\ref{s1}-\ref{s4} at $\sqrt{s}=13$ TeV.}
	\label{table:cross}
\end{table}

In our analysis, we use {\tt FeynRules}~\cite{Alloul:2013bka}
from which we have produced UFO model files required in {\tt Madgraph5}~\cite{Alwall:2014hca} to generate the signal events at the LO
parton level. For the purpose of SM background processes,
we have generated events using {\tt Madgraph5}. 
To simulate showering and hadronisation effects, we have passed
the unweighted parton level through the {\tt Pythia}(v6.4)~\cite{Sjostrand:2006za}, including fragmentation. We have done the detector simulation using the {\tt
Delphes}(v3)~\cite{deFavereau:2013fsa}. Jets are constructed using {\tt Fastjet}~\cite{Cacciari:2011ma} with anti-$k_T$ \cite{Cacciari:2008gp} jet clustering algorithm with proper MLM matching scheme chosen for background processes.
The production cross sections are calculated using the {\tt NNPDF3.0} parton distributions. 
Finally, we perform the cut analyses using {\tt MadAnalysis5}~\cite{Conte:2012fm}.

Before we proceed to simulate the events we 
should impose some basic cuts as our final states under consideration may also result from hard subprocesses associated with initial and final state
radiation, or  soft decays. Hence, we demand that
\begin{subequations}
\begin{eqnarray}
\Delta R_{j j} > 0.4,~~~~\Delta R_{\ell \ell} > 0.7,~~~~\Delta R_{j \ell} > 0.7, \label{c1} \\
\Delta R_{bj} > 0.7,~~~~\Delta R_{b \ell} > 0.2, \label{c2}\\ 
p_T^j > 20 ~{\rm GeV},~~~~|\eta_j| < 2.5, \label{c3}\\
p_T^\ell > 10 ~{\rm GeV},~~~~|\eta_\ell| < 2.5, \label{c4}\\
p_T^\gamma > 10 ~{\rm GeV},~~~~|\eta_\gamma| < 2.5.
\end{eqnarray}\label{basic_cuts}
\end{subequations}

Moreover, we consider the following $p_T$-dependent $b$-tagging
efficiency given by the ATLAS collaboration~\cite{ATLAS:2014pla}, 
\begin{eqnarray}
\epsilon_b = \begin{cases}
0 & p_T^b \leq 30 ~\rm GeV \;,\\
0.6 & 30 ~{\rm GeV} < p_T^b < 50 ~{\rm GeV} \;,\\
0.75 & 50 ~{\rm GeV} < p_T^b < 400 ~{\rm GeV}\;, \\
0.5 & p_T^b > 400 ~\rm GeV\;.
\end{cases} 
\label{b-tag}
\end{eqnarray}
Apart from this, we also introduce a mistagging probability
of 10\% (1\%) for charm-jets (light-quark and gluon jets).
Further, the absolute rapidity of $b$-jets are demanded
to be less than 2.5 $(|\eta_b| < 2.5)$.

\subsection{Cut Analysis:}
All processes given in Eq.\;\ref{s1}-\ref{s4} may contribute
to several final states which we are going to study in this work.
These final sates can be tagged as $Signal$ while the SM processes
which mimic the signal are considered as $Background$. In order to
improve the signal to background ratio we will impose some selection
cuts in addition to the basic cuts given in Eq.\;\ref{basic_cuts}.
After imposing the selection cuts, if there exists $N_S$ and $N_B$
number of events for signal and background respectively, then we
can calculate the significance ($S$) of any particular final
states using the following relation
\begin{eqnarray}
S &=& \frac{N_S}{\sqrt{N_S+N_B}} \, \ .
\label{significance}
\end{eqnarray}
Now we are in a position where we can study some final states
which arise from four sub-processes given in Eq.\;\ref{s1}-\ref{s4},
\begin{subequations}
\begin{eqnarray}	
&(i)&	2\ell + \mET  \,, \label{f1} \\
&(ii)&	2j + \mET \,, \label{f2} \\
&(iii)&	3\ell + \mET \,, \label{f3}
\end{eqnarray}\label{final_states}
\end{subequations}

where $\ell\equiv e, \mu$ and $j$ represents ordinary light jets. For
the practical purpose we need to consider the following SM subprocesses
as backgrounds to the aforementioned final states. 
\begin{itemize}

\item $W^\pm + \rm jets$: We consider this process with up to two
hard jets as this process serves as the dominant background for
the signal which contains hard $\mET$ in the final state
and ordinary light jets.
\item $Z + \rm jets$: This can be considered as significant
background for the signal with two charged leptons in the
final state. In this case also we consider two hard jets
for semi-inclusive cross section.  
\item $V V+ \rm jets$ (where $V = W^\pm, Z$): We have estimated the
processes with two hard jets because they exactly mimic some of
the gross production channels. 
\item $t\bar t (+ \rm jets)$: $t\bar t$ production with two additional
hard jets, can play as one of the major dominant background for
some of the three final states. 
\item Single top + jets: This will contribute mainly to final state $(ii)$.
\item $t\bar t + (W^\pm/Z/\gamma) $: Analogous to $t\bar t (+ \rm jets)$,
these processes could also contribute to the total SM background,
but with much smaller production cross sections.
\item $V V V$ (where $V = W^\pm, Z$): In the case of leptonic decays
of $W^\pm, Z$ we could consider this process in the SM background.
\item QCD($\leq3$ jets): Pure QCD processes play as the most dominant
SM-background for hadronic final states such as multi-jet production
where $\mET$~comes either from the jets fragmenting into neutrinos
or simply from the mismeasurement of the jet energy.
\end{itemize}

(i)	${\underline {2\ell + \mET }}$\, : This final state can be
produced from the processes (\ref{s1}), (\ref{s2}) and ({\ref{s3}}).
In this final state we have two charged leptons with $\mET$.
Therefore, this signal is relatively clean however production
cross section is small. Here in the background events, the
$\mET$~~comes only from the neutrinos, while for the signal
events, it arises from $\phi^0$. The mass difference between
$\phi^0$ and the decaying $a^0$ significantly enhances the
$\mET$. Therefore, by choosing $\mET> 100$ GeV we may inhibit
the background and consequently improve the signal
significance. Below we mention the selection criteria for this signal. 

\begin{table}[H]
	\centering
	\begin{tabular}{|c|c|} \hline \hline
		Cuts Name & Selection Criteria \\ \hline
		C1-1 & Reject number of $b$-tagged jets with $p_T(b) > 20 $ GeV \\ \hline
		C1-2 & Select lepton with $p_T(\ell_2) > 10 $ GeV \\ \hline
		C1-3 & Reject lepton with $p_T(\ell_3) > 10 $ GeV \\ \hline
        C1-4 & Select $\Delta R (\ell_i \ell_j) < $ 2.8\\ \hline
		C1-5 & Reject  additional jet with $p_T(j_1) > 20 $ GeV\\ \hline
		C1-6 &  Select $\mET > 100$ GeV \\ \hline
	\end{tabular}
	\caption{Required selection cuts for $2\ell + \mET$ final state.}
	\label{2lep_mis}
\end{table}

With the above mentioned criteria, we have obtained the following
significance for this signal for the chosen benchmark points.
\begin{table}[H]
\centering
\resizebox{15cm}{!}{
\begin{tabular}{|p{2cm}|c|c|c|c|c|}
\hline 
Benchmark Points & {process (a) ($fb$) } & {process (b) ($fb$) } & {process (c) ($fb$)}& {Total ($fb$)} & {Significance ($S$)}\\ \hline
BP1 & {0.0618}  & {0.3516} & {0.0307} & {0.4441} & {3.72}\\ \hline
BP2 & {0.0537}  & {0.3128} & {0.0264} & {0.3929} & {3.29}\\ \hline
BP3 & {0.0388}  & {0.2192} & {0.0201} & {0.2781} & {2.33}\\ \hline
\end{tabular}
}
\caption{After imposing all cuts the cross sections (in $fb$) for
the processes given in Eq.\;\ref{s1}-\ref{s4} which are
contributed in this signal at $\sqrt{s}=13$ TeV. Also the
corresponding statistical significances for an integrated
luminosity of 3000 $fb^{-1}$ are given for all benchmark
points. The total background cross section for this signal
after all cuts is 42.33 $fb$.}
	\label{table:2lep_mis_sig}
\end{table}
After passing the signal and background events through the selection
criteria (given in Table \;\ref{2lep_mis}) we estimate the corresponding
significance reach at the highest plausible integrated luminosity
that can be achieved at the LHC. It can be seen from this
Table\;\ref{table:2lep_mis_sig}, the maximum significance
greater than $ 3.5\sigma$ is attained for the BP1, due to
the largest production cross section. The signal significance
is small for the benchmark BP3. This analysis shows that there
will be a finite chances to search dark matter via this signal
at the future LHC running at 13 TeV of integrated luminosity 3000$fb^{-1}$.

(ii) ${\underline {2j + \mET}}$  \,: This final state comes from
the processes (\ref{s2}), (\ref{s3}) and (\ref{s4}). The same signal
has been studied in the context of IDM in \cite{Poulose:2016lvz}
at the 13 TeV LHC. As far as the cross section is concerned, this
final state possesses larger value with respect to other final
states. However pure QCD process with large cross section which
we have considered in the SM background suppress the significance.
Nevertheless, in the following we are trying to probe the signal by
imposing some judicious criteria which may improve the signal
efficiencies with respect to the SM background. 
 
\begin{table}[H]
	\centering
	\begin{tabular}{|c|c|} \hline \hline
		Cuts Name & Selection Criteria \\ \hline
		C2-1 & Reject number of $b$-tagged jets with $p_T(b) > 20 $ GeV \\ \hline
		C2-2 & Select $2^{nd}$ jet with $p_T(j_2) > 30 $ GeV \\ \hline
		C2-3 & Reject $3^{rd}$ jet with $p_T(j_3) > 20 $ GeV \\ \hline
		C2-4 & Reject $1^{st}$ lepton with $p_T(\ell_1) > 10 $ GeV\\ \hline
		C2-5 &  Select $\mET > 110$ GeV \\ \hline
	\end{tabular}
	\caption{Required selection cuts for $2j + \mET$ final state.}
	\label{2jet_mis}
\end{table}
These criteria (given in Table\;\ref{2jet_mis}) ensure that in the
signal we have two leading jets with  $p_T(j)$ greater than 30 GeV.
Fourth cut ensures that we have no leptons in our signal. Also the
selection of $\mET$~~greater than 110 GeV helps us to reduce the
dominant backgrounds along with pure QCD background significantly.
Below we have given the statistics for the signal for the selected
benchmark points with respect to SM backgrounds. Specifically we have
given (in Table\;\ref{table:2jet_mis_sig}) the cross sections for
the individual channel after the selection cuts for the selected
benchmark points and also the corresponding significances. For the
first two benchmark points we can have the significance up to
$\gtrsim 1.4 \sigma$ but require high luminosity 3000 $fb^{-1}$.
So it is hard to probe this model via this signal at the LHC
even with very high luminosity. 
\begin{table}[H]
	\centering
\resizebox{15cm}{!}{
	\begin{tabular}{|p{2cm}|c|c|c|c|c|}
		\hline
Benchmark Points & {process (b) ($fb$) } & {process (c) ($fb$) } & {process (d) ($fb$) } & {Total ($fb$)} & {Significance ($S$)}\\ \hline
BP1 & {4.204}  & {1.392} & {13.464} &{19.06} & {1.49}\\ \hline
BP2 & {3.786}  & {1.271} & {12.824} &{17.88} & {1.40}\\ \hline
BP3 & {2.637}  & {0.885} & {8.206} &{11.73} & {0.92}\\ \hline
\end{tabular}
}
	\caption{After using all cuts the cross sections (in $fb$)
for the processes given in Eq.\;\ref{s1}-\ref{s4} which are
contributed in this signal at $\sqrt{s}=13$ TeV. Also the corresponding
statistical significances for an integrated luminosity of 3000 $fb^{-1}$
are given for all benchmark points. The total background cross section
for this signal after all cut is 492728.60 $fb$.}
	\label{table:2jet_mis_sig}
\end{table}

(iii)	${\underline {3\ell + \mET}}$ \,: Finally we consider this signal
which arises from the process (\ref{s3}) only. In Ref.\;\cite{Datta:2016nfz}
one can find the multilepton signature of IDM including trilepton
+$\mET$~~signal at the 13 TeV LHC. In our present model we are also
trying to search dark matter at the LHC via this final state.
By considering the all relevant SM-backgrounds we have calculated
the significance the for this final state.

In the following (Table\;\ref{3lep_mis}) we have shown the selection criteria by which we can extract signal with respect to background.  
\begin{table}[H]
	\centering
	\begin{tabular}{|c|c|} \hline \hline
		Cuts Name & Selection Criteria \\ \hline
		C3-1 & Reject number of $b$-tagged jets with $p_T(b) > 20 $ GeV \\ \hline
		C3-2 & Select $3^{rd}$ lepton with $p_T(\ell_3) > 10 $ GeV \\ \hline
		C3-3 & Reject $1^{st}$ lepton with $p_T(j_1) > 20 $ GeV\\ \hline
		C3-4 &  Select $\mET > 100$ GeV \\ \hline
	\end{tabular}
	\caption{Required selection cuts for ${3\ell + \mET}$ final state.}
	\label{3lep_mis}
\end{table}

As we demand that in our signal we require only three
lepton so we have select third leading lepton with $p_T({l})>$
10 GeV. We have also rejected any jet with $p_T({j})>$ 20 GeV as
the signal does not contain any jet. Finally selection of $\mET > 100$
GeV reduces the background substantially. 
\begin{table}[H]
\centering
\resizebox{10cm}{!}{
\begin{tabular}{|p{2cm}|c|c|}
\hline
Benchmark Points & {process (c) ($fb$) } & {Significance ($S$)}\\ \hline
BP1  & {0.0320} & {1.09}\\ \hline
BP2  & {0.0313} & {1.07}\\ \hline
BP3  & {0.0235} & {0.80}\\ \hline
\end{tabular}
}
\caption{After implementing all cuts the cross sections (in $fb$)
for the processes given in Eq.\;\ref{s1}-\ref{s4} which are
contributed in this signal at $\sqrt{s}=13$ TeV. Also the corresponding
statistical significances for an integrated luminosity of 3000
$fb^{-1}$ are given for all benchmark points. The total background
cross section for this signal is 2.54 $fb$.}
\label{table:3lep_mis_sig}
\end{table}
Finally with the above selection criteria we have significance
$\gtrsim 1\sigma$ for the first two benchmark points. Hence in
this case also, to find the dark matter at LHC through this
signal is less attractive at LHC even for high integrated
luminosity 3000 $fb^{-1}$.

Before we conclude, it would be relevant to discuss some issues
on a region of parameter space which we have not considered
in our collider analysis. In general, one can have the region
of parameter space (which satisfies all the theoretical constraints
as well as dark matter relic abundance and direct detection data)
where  $\phi^\pm$ and $a^0$ decay into $H^\pm \phi^0$ and $A^0\phi^0$
respectively, in fact with 100\% branching ratio. However, we have
not considered this region of parameter space in our collider study.
First of all in this region of parameter space, the production cross
sections of the $\mathbb{Z}_2$ odd scalars at the 13 TeV LHC are
lower than that of given in the Table \ref{table:cross} due
to phase space suppression. Because, in this case for the above
decay modes to become kinematically feasible one requires
$M_{\phi^\pm}> (M_{H^\pm} + M_{\phi^0})$ and
$M_{a^0}> (M_{A^0} + M_{\phi^0})$. Hence, the masses
of $\phi^\pm$ and $a^0$ are larger with respect to
the values of masses given in the Table \ref{table:bp}.
This will drop the signal efficiency. 

Additionally, in this region of parameter space, the non-standard
scalars $H^\pm$ or $A^0$ which are produced from the decay of
inert scalars are dominantly decaying to various three body
decay modes each of which possesses small branching fraction.
Moreover, the remaining two body decay modes of these triplet scalars
also acquire very tiny branching fractions. Because in the case
of leptonic decay modes, the coupling between singly charged Higgs
($H^\pm$) and leptons are suppressed due to the tiny neutrino
Yukawa coupling while for the hadronic decay modes the coupling
of $H^\pm$ with quarks are suppressed by the small mixing angle $\beta'$. Therefore, if we consider the two body decay modes of $H^\pm$
for any particular final state then we will end up with a very
small effective cross section due to small branching fractions.
Consequently, the signal significance will be very low even at
the very high luminosity future collider experiment. 

Further, in the case of three body decay modes one may
consider the following process (for our chosen value of $v_t$) 
\begin{eqnarray}
~p~p\rightarrow \phi^{+} \phi^{-}&\rightarrow& H^{+}
\phi^0+H^{-}\phi^0  \nonumber \\ 
   H^{\pm} &\hookrightarrow & H^{\pm\pm} q \bar{q'}
   \to W^\pm  W^\pm (W^\pm  W^{*\pm}) + q \bar{q'} \nonumber\;.
\end{eqnarray} 

Now, the $W^\pm$ produced from $H^{\pm \pm}$ will further decays
to either leptons or jets. Hence, in this situation one can reach
a stable final state after several decay steps and for each step there
will be a suppression due to small branching fraction. We have already
mentioned earlier that in this region of parameter space the production
cross sections are small and also the small branching fractions
(at each decay step) will further suppress the effective cross
section for any particular final state. Consequently, in the case of
three body decay modes one has the very small signal significances
in comparison to the values what we have obtained from our analysis.

Due to the above mentioned facts we have considered the region of
parameter space where the $\mathbb{Z}_2$ odd scalars $\phi^\pm$ and
$a^0$ decay into $W^\pm \phi^0$ and $Z\phi^0$ with 100\% branching ratio.
\section{Conclusions}
\label{con}
In order to convey the existence of the non-luminous dark matter
of the Universe and the origin of tiny neutrino masses, we have
considered an extension of the Type-II seesaw model by adding a
$\mathbb{Z}_2$ odd doublet $\Phi$. We ensure that the CP even
component of $\Phi$ is the WIMP dark matter candidate which
is stable due to the $\mathbb{Z}_2$ symmetry. On the other hand,
Higgs triplet scalar field generates the masses of neutrinos via
the Type-II seesaw mechanism. Furthermore, in this framework we
have derived the full set of unitarity and vacuum stability
conditions which have always been very important if one deals
with the scalar sector. 

In the Type-II seesaw scenario the triplet VEV is very small
($10^{-9}$ GeV to  ${\cal O} (1)$ GeV) by the electroweak precision constraint. Hence, for the purpose of neutrino mass generation we set the value
of triplet VEV at $10^{-3}$ GeV. We have alluded the absolute values
of neutrino masses allowed by the neutrino oscillation data at 3$\sigma$ range. Then we have shown that the sum of masses of three neutrinos
for the normal(inverted) hierarchical scenario is around $\sim$ 0.06 eV
to 0.1 eV(0.1 eV to 0.2 eV) which respects the bound from cosmological observations
(i.e. $\sum m_{\nu}<0.23$ eV). Furthermore, we have calculated the effective Majorana mass parameter $m_{\beta \beta}$ associated with the neutrinoless
double $\beta$ decay process. We have derived the upper limit
on the mass of the lightest neutrino $m_1(m_3)$ of normal(inverted) hierarchy
by satisfying the combined results of cosmological upper bound
and neutrino oscillation data.  We have also computed the 
Dirac CP phase $\delta$ that resides in the first and fourth
quadrant for the normal hierarchy while it lies between
$90^\circ-140^\circ$ and $220^\circ-270^\circ$ for the
inverted hierarchical scenario. However, the recent results of T2K
experiment are favourable for the values of $\delta$ which lie
in the third and fourth quadrant instead of the first two quadrants.
Finally, we also evaluated the Jarlskog invariant $J_{CP}$
using the model parameters which satisfy the neutrino oscillation data
in $3\sigma$ range. We find that it lies below 0.039 irrespective
of the neutrino mass hierarchy.

We have also explored the dark matter phenomenology in a great detail by considering $\phi^0$ as a WIMP type dark matter candidate of the present scenario. We have considered all possible annihilation channels while calculating the dark matter relic abundance. One should note that, in our case the dark matter particle satisfies the Planck limit ($0.1172\leq\Omega h^2\leq 0.1226$ \cite{Ade:2015xua}) of relic density only for two distinct mass ranges of $\phi^0$. For example, in the low mass range where $M_{\phi^0}$ lies below 90 GeV while in the high mass range $M_{\phi^0}$ is larger than 535 GeV. For the low mass region we have done our analysis for two different values of triplet scalar VEV $v_t$, e.g., $10^{-3}$ GeV and 3 GeV. When $v_t=3$ GeV, we have observed that the dark matter with mass
$\la50$ GeV also satisfies the relic density. However, those regions are forbidden
if one imposes the constraint of invisible branching ratio of the SM-like Higgs boson $h^0$. In the low mass range, the dominant contribution to
$\langle\sigma {\rm v}\rangle$ comes from $\phi^0\phi^0\rightarrow b\bar{b}$
($W^+W^-$) channel for $M_{\phi^0}\la70$ GeV ($\ga 70$ GeV). Also in the low mass range the co-annihilations among the inert sector particles have no considerable effect on the dark matter relic density, as we have taken $M_{\phi^\pm}, M_{a^0}>100$ GeV. Furthermore, we would like to mention that, there is a distinct
feature of the present scenario, with respect to the conventional Inert Doublet Model. The trilinear coupling ($\tilde{\mu}$) between triplet scalar field and inert doublet plays a crucial role in our dark matter analysis. The parameter
$\tilde{\mu}$ effectively proportional to the mass difference between the dark matter and the inert charged scalar. Therefore, depending on the mass gap,
$\tilde{\mu}$ controls the dark matter annihilation processes.
For example, in the low mass region the absolute values of
$\tilde{\mu}$ can vary from 0 to $\sim 10^4$ GeV as in this case the mass
difference between $\phi^+$ and $\phi^0$ varies between
$\sim (20-270)$ GeV. On the other hand,
in the high mass region the mass gaps between the inert scalars are
required to be very small as a consequence of the significant contributions
of different co-annihilation channels to dark matter relic density. 
Hence, in this case to satisfy the Plank limit one should vary the
relevant model parameters in a fine tune way which in turn controls
the trilinear coupling $\tilde{\mu}$ (i.e. $|\tilde{\mu}|\la 500$ GeV).
Apart from that, we have also
evaluated the spin-independent
scattering cross section of dark matter off the detector nuclei.
Our estimations indicate that a major portion of dark matter
parameter space in the low mass region has already
been ruled out by the XENON 1T experiment. However, there are still some
region left which can be tested by the ongoing and future direct
detection experiments. Further, we have also noticed that the high mass
region is still comparatively less constrained by the exclusion
limits from XENON 1T experiments and this region can be a
{\it potential dark matter search region} for the future ``ton scale''
direct detection experiments.   
On the other hand, we have found the similar results from the perspective
of indirect search of dark matter. Here also some portions of the
allowed parameter space in the low dark mass region has been excluded
by the recent analyses of gamma-ray flux from dwarf spheroidal galaxies
and also from the Galactic Centre by the Fermi-LAT and MAGIC
collaborations. However, similar to the case of direct search,
the limits from the indirect detection are also much more relaxed in
the high mass region.

Finally, we have investigated the collider signature of the dark matter in terms of missing transverse energy (~$\mET$~) at the 13 TeV LHC. Due to the $\mathbb{Z}_2$ symmetry the odd particles are produced in pairs. Furthermore, for our chosen benchmark points, satisfying all theoretical and experimental constraints including dark matter relic density, the heavier odd particles decay into the SM gauge bosons and dark matter. Depending on the production channels and the branching fractions of the odd particles one has several final states which can be probed at the current and the future colliders. In our case, we have analysed three final states namely, $2\ell + \mET  \,,~ 2j + \mET  \,$ and $3\ell + \mET  \,,$ at the 13 TeV LHC. For each of the final states we have calculated relevant SM backgrounds. With judicious cut selection, we have evaluated the signal significance for an integrated luminosity 3000$fb^{-1}$. From our simulation study it is clearly evident that for the two benchmark point we can have the significance greater than $3\sigma$ for the final state $2\ell + \mET\,$. Hence, with this signal there will be a finite chance to search dark matter at the future LHC running at 13 TeV of integrated luminosity 3000$fb^{-1}$.

\vskip 3cm
\noindent{\bf Acknowledgments}
Both the authors A.B. and A.S. thank Abhisek Dey for computational help.
They are also grateful to Subhadeep Mondal for the useful
discussions regarding collider study. One of the authors A.B.
would like to acknowledge the cluster computing facility
(http://cluster.hri.res.in) of Harish-Chandra Research Institute.
A.B. would also like to thank the Department of Atomic Energy (DAE),
Govt.\,of INDIA for financial assistance.
\begin{appendices}
\renewcommand{\thesection}{\Alph{section}}
\renewcommand{\theequation}{\thesection-\arabic{equation}} 
\setcounter{equation}{0}  

\section{The eigenvalue equations which are being solved using numerical technique}\label{vac_neu}
\begin{eqnarray}
&& 4\mathbb{X}^3-\bigg(6\lambda+32\lambda_2+24\lambda_3+24\lambda_\Phi\bigg)\mathbb{X}^2+
\bigg(-24 \lambda^2_1+48 \lambda\lambda_2+36 \lambda\lambda_3-24\lambda_1\lambda_4-6\lambda^2_4-16\lambda^2_5\nonumber \\
&& -16\lambda_5\lambda_6-4\lambda^2_6-24\lambda^2_7-24\lambda_7\lambda_8-6\lambda^2_8+36 \lambda\lambda_\Phi +192\lambda_2\lambda_\Phi +144 \lambda_3\lambda_\Phi \bigg)\mathbb{X} \nonumber \\
&& +\bigg(128\lambda_2\lambda^2_5+96\lambda_3\lambda^2_5+128\lambda_2\lambda_5\lambda_6+96\lambda_3\lambda_5\lambda_6+32\lambda_2\lambda^2_6+24 \lambda_3\lambda^2_6-96\lambda_1\lambda_5\lambda_7-48\lambda_4\lambda_5\lambda_7\nonumber \\
&&-48\lambda_1\lambda_6\lambda_7-24\lambda_4\lambda_6\lambda_7+36 \lambda\lambda^2_7-48\lambda_1\lambda_5\lambda_8-24\lambda_4\lambda_5\lambda_8-24\lambda_1\lambda_6\lambda_8-12\lambda_4\lambda_6\lambda_8+36 \lambda\lambda_7\lambda_8\nonumber \\
&&+9 \lambda\lambda^2_8+144\lambda^2_1\lambda_\Phi -288\lambda\lambda_2\lambda_\Phi -216\lambda\lambda_3\lambda_\Phi +144 \lambda_1\lambda_4\lambda_\Phi +36\lambda^2_4\lambda_\Phi\bigg)=0.
\label{eq:qubic1}
\end{eqnarray}

\begin{eqnarray}
&2\mathbb{X}^3-\bigg(\lambda+4\lambda_2+8\lambda_3+4 \lambda_\Phi \bigg)\mathbb{X}^2+
\bigg(2 \lambda\lambda_2+4 \lambda\lambda_3-2\lambda^2_4-2\lambda^2_6-2\lambda^2_8+2 \lambda\lambda_\Phi+8\lambda_2\lambda_\Phi \nonumber \\
& +16\lambda_3 \lambda_\Phi \bigg)\mathbb{X}+\bigg(4\lambda_2\lambda^2_6+8\lambda_3\lambda^2_6-4\lambda_4\lambda_6\lambda_8+\lambda\lambda^2_8-4 \lambda\lambda_2\lambda_\Phi -8 \lambda\lambda_3\lambda_\Phi +4\lambda^2_4\lambda_\Phi\bigg)=0.
\label{eq:qubic2}
\end{eqnarray}

By solving the above cubic Eqs.\;\ref{eq:qubic1} and \ref{eq:qubic2} using numerical technique we have obtained the eigenvalues of $\mathcal{M}_2$. Whereas, from the Eq.\;\ref{eq:qubic2} we have evaluated the eigenvalues of $\mathcal{M}_4$ using the same numerical technique.

\section{Explicit form of BFB conditions}\label{vac_cond}

{\tiny
\begin{eqnarray}\label{eq:lstab1}
&&\lambda  \geq 0,~~\lambda _2+\lambda _3\geq 0, ~~\lambda _2+\frac{\lambda _3}{2} \geq 0,\lambda_\Phi \geq 0, \\
&&\lambda _1+ \sqrt{\lambda (\lambda _2+\lambda _3)}\geq 0,~~\lambda _1+ \sqrt{\lambda \left(\lambda _2+\frac{\lambda _3}{2}\right)} \geq 0, \\
&& \lambda _1+\lambda _4+\sqrt{\lambda (\lambda _2+\lambda _3)}\geq 0,~~\lambda _1+\lambda _4+ \sqrt{\lambda \left(\lambda _2+\frac{\lambda _3}{2}\right)} \geq 0, \\
&&\lambda _7+ 2\sqrt{\lambda_\Phi (\lambda _2+\lambda _3)}\geq 0,\lambda _7+ 2\sqrt{\lambda_\Phi\left(\lambda _2+\frac{\lambda _3}{2}\right)} \geq 0, \\
&& \lambda _7+\lambda _8+2\sqrt{\lambda_\Phi(\lambda _2+\lambda _3)}\geq 0,~~\lambda _7+\lambda _8+ 2\sqrt{\lambda_\Phi\left(\lambda _2+\frac{\lambda _3}{2}\right)} \geq 0, \\
&&\lambda _5+\sqrt{\lambda\lambda_\Phi}\geq 0,~~(\lambda _5+\lambda _6-2|\lambda _9|+\sqrt{\lambda\lambda_\Phi}) \geq 0,  
\end{eqnarray}
}

{\tiny
\begin{subequations}
\begin{eqnarray}
&&\sqrt{\frac{\lambda}{4}(\lambda_2+\frac{\lambda_3}{2})\lambda_\Phi}+\frac{\lambda_1}{2}\sqrt{\lambda_\Phi}+\frac{\lambda_5}{2}\sqrt{\lambda_2+\frac{\lambda_3}{2}}+\frac{\lambda_7}{2}\sqrt{\frac{\lambda}{4}} + \nonumber \\ && {\tiny \sqrt{2\left\{\frac{\lambda_1}{2}+\sqrt{\frac{\lambda}{4}(\lambda_2+\frac{\lambda_3}{2})}\right\}\left\{\frac{\lambda_7}{2}+\sqrt{\lambda_\Phi(\lambda_2+\frac{\lambda_3}{2})}\right\}\left\{\frac{\lambda_5}{2}+\sqrt{\frac{\lambda}{4}\lambda_\Phi}\right\}} \geq 0},  \\ 
&&\sqrt{\frac{\lambda}{4}(\lambda_2+\lambda_3)\lambda_\Phi}+\frac{\lambda_1}{2}\sqrt{\lambda_\Phi}+\frac{\lambda_5}{2}\sqrt{\lambda_2+\lambda_3}+\frac{\lambda_7}{2}\sqrt{\frac{\lambda}{4}}+ \nonumber \\ && {\tiny \sqrt{2\left\{\frac{\lambda_1}{2}+\sqrt{\frac{\lambda}{4}(\lambda_2+\lambda_3)}\right\}\left\{\frac{\lambda_7}{2}+\sqrt{\lambda_\Phi(\lambda_2+\lambda_3)}\right\}\left\{\frac{\lambda_5}{2}+\sqrt{\frac{\lambda}{4}\lambda_\Phi}\right\}} \geq 0}, \\
&&\sqrt{\frac{\lambda}{4}(\lambda_2+\frac{\lambda_3}{2})\lambda_\Phi}+\frac{\lambda_1}{2}\sqrt{\lambda_\Phi}+\frac{\lambda_5}{2}\sqrt{\lambda_2+\frac{\lambda_3}{2}}+\frac{(\lambda_7+\lambda_8)}{2}\sqrt{\frac{\lambda}{4}} + \nonumber \\ && {\tiny \sqrt{2\left\{\frac{\lambda_1}{2}+\sqrt{\frac{\lambda}{4}(\lambda_2+\frac{\lambda_3}{2})}\right\}\left\{\frac{(\lambda_7+\lambda_8)}{2}+\sqrt{\lambda_\Phi(\lambda_2+\frac{\lambda_3}{2})}\right\}\left\{\frac{\lambda_5}{2}+\sqrt{\frac{\lambda}{4}\lambda_\Phi}\right\}} \geq 0}, \\
&&\sqrt{\frac{\lambda}{4}(\lambda_2+\lambda_3)\lambda_\Phi}+\frac{\lambda_1}{2}\sqrt{\lambda_\Phi}+\frac{\lambda_5}{2}\sqrt{\lambda_2+\lambda_3}+\frac{(\lambda_7+\lambda_8)}{2}\sqrt{\frac{\lambda}{4}} + \nonumber \\ && {\tiny \sqrt{2\left\{\frac{\lambda_1}{2}+\sqrt{\frac{\lambda}{4}(\lambda_2+\lambda_3)}\right\}\left\{\frac{(\lambda_7+\lambda_8)}{2}+\sqrt{\lambda_\Phi(\lambda_2+\lambda_3)}\right\}\left\{\frac{\lambda_5}{2}+\sqrt{\frac{\lambda}{4}\lambda_\Phi}\right\}} \geq 0}, \\
&&\sqrt{\frac{\lambda}{4}(\lambda_2+\frac{\lambda_3}{2})\lambda_\Phi}+\frac{(\lambda_1+\lambda_4)}{2}\sqrt{\lambda_\Phi}+\frac{\lambda_5}{2}\sqrt{\lambda_2+\frac{\lambda_3}{2}}+\frac{\lambda_7}{2}\sqrt{\frac{\lambda}{4}} + \nonumber \\ && {\tiny \sqrt{2\left\{\frac{(\lambda_1+\lambda_4)}{2}+\sqrt{\frac{\lambda}{4}(\lambda_2+\frac{\lambda_3}{2})}\right\}\left\{\frac{\lambda_7}{2}+\sqrt{\lambda_\Phi(\lambda_2+\frac{\lambda_3}{2})}\right\}\left\{\frac{\lambda_5}{2}+\sqrt{\frac{\lambda}{4}\lambda_\Phi}\right\}} \geq 0}, \\
&&\sqrt{\frac{\lambda}{4}(\lambda_2+\lambda_3)\lambda_\Phi}+\frac{(\lambda_1+\lambda_4)}{2}\sqrt{\lambda_\Phi}+\frac{\lambda_5}{2}\sqrt{\lambda_2+\lambda_3}+\frac{\lambda_7}{2}\sqrt{\frac{\lambda}{4}} + \nonumber \\ && {\tiny \sqrt{2\left\{\frac{(\lambda_1+\lambda_4)}{2}+\sqrt{\frac{\lambda}{4}(\lambda_2+\lambda_3)}\right\}\left\{\frac{\lambda_7}{2}+\sqrt{\lambda_\Phi(\lambda_2+\lambda_3)}\right\}\left\{\frac{\lambda_5}{2}+\sqrt{\frac{\lambda}{4}\lambda_\Phi}\right\}} \geq 0}, \\
&&\sqrt{\frac{\lambda}{4}(\lambda_2+\frac{\lambda_3}{2})\lambda_\Phi}+\frac{(\lambda_1+\lambda_4)}{2}\sqrt{\lambda_\Phi}+\frac{\lambda_5}{2}\sqrt{\lambda_2+\frac{\lambda_3}{2}}+\frac{(\lambda_7+\lambda_8)}{2}\sqrt{\frac{\lambda}{4}} + \nonumber \\ && {\tiny \sqrt{2\left\{\frac{(\lambda_1+\lambda_4)}{2}+\sqrt{\frac{\lambda}{4}(\lambda_2+\frac{\lambda_3}{2})}\right\}\left\{\frac{(\lambda_7+\lambda_8)}{2}+\sqrt{\lambda_\Phi(\lambda_2+\frac{\lambda_3}{2})}\right\}\left\{\frac{\lambda_5}{2}+\sqrt{\frac{\lambda}{4}\lambda_\Phi}\right\}} \geq 0}, \\
&&\sqrt{\frac{\lambda}{4}(\lambda_2+\lambda_3)\lambda_\Phi}+\frac{(\lambda_1+\lambda_4)}{2}\sqrt{\lambda_\Phi}+\frac{\lambda_5}{2}\sqrt{\lambda_2+\lambda_3}+\frac{(\lambda_7+\lambda_8)}{2}\sqrt{\frac{\lambda}{4}} + \nonumber \\ && {\tiny \sqrt{2\left\{\frac{(\lambda_1+\lambda_4)}{2}+\sqrt{\frac{\lambda}{4}(\lambda_2+\lambda_3)}\right\}\left\{\frac{(\lambda_7+\lambda_8)}{2}+\sqrt{\lambda_\Phi(\lambda_2+\lambda_3)}\right\}\left\{\frac{\lambda_5}{2}+\sqrt{\frac{\lambda}{4}\lambda_\Phi}\right\}} \geq 0}, 
\label{eq:lstab2}
\end{eqnarray} 
\end{subequations}
}

{\tiny
\begin{subequations}
\begin{eqnarray}
&&\sqrt{\frac{\lambda}{4}(\lambda_2+\frac{\lambda_3}{2})\lambda_\Phi}+\frac{\lambda_1}{2}\sqrt{\lambda_\Phi}+\frac{(\lambda_5+\lambda_6-2|\lambda_9|)}{2}\sqrt{\lambda_2+\frac{\lambda_3}{2}}+\frac{\lambda_7}{2}\sqrt{\frac{\lambda}{4}} + \nonumber \\ && {\tiny \sqrt{2\left\{\frac{\lambda_1}{2}+\sqrt{\frac{\lambda}{4}(\lambda_2+\frac{\lambda_3}{2})}\right\}\left\{\frac{\lambda_7}{2}+\sqrt{\lambda_\Phi(\lambda_2+\frac{\lambda_3}{2})}\right\}\left\{\frac{(\lambda_5+\lambda_6-2|\lambda_9|)}{2}+\sqrt{\frac{\lambda}{4}\lambda_\Phi}\right\}} \geq 0}, \\
&&\sqrt{\frac{\lambda}{4}(\lambda_2+\lambda_3)\lambda_\Phi}+\frac{\lambda_1}{2}\sqrt{\lambda_\Phi}+\frac{(\lambda_5+\lambda_6-2|\lambda_9|)}{2}\sqrt{\lambda_2+\lambda_3}+\frac{\lambda_7}{2}\sqrt{\frac{\lambda}{4}} + \nonumber \\ && {\tiny \sqrt{2\left\{\frac{\lambda_1}{2}+\sqrt{\frac{\lambda}{4}(\lambda_2+\lambda_3)}\right\}\left\{\frac{\lambda_7}{2}+\sqrt{\lambda_\Phi(\lambda_2+\lambda_3)}\right\}\left\{\frac{(\lambda_5+\lambda_6-2|\lambda_9|)}{2}+\sqrt{\frac{\lambda}{4}\lambda_\Phi}\right\}} \geq 0}, \\
&&\sqrt{\frac{\lambda}{4}(\lambda_2+\frac{\lambda_3}{2})\lambda_\Phi}+\frac{\lambda_1}{2}\sqrt{\lambda_\Phi}+\frac{(\lambda_5+\lambda_6-2|\lambda_9|)}{2}\sqrt{\lambda_2+\frac{\lambda_3}{2}}+\frac{(\lambda_7+\lambda_8)}{2}\sqrt{\frac{\lambda}{4}}+ \nonumber \\ && {\tiny \sqrt{2\left\{\frac{\lambda_1}{2}+\sqrt{\frac{\lambda}{4}(\lambda_2+\frac{\lambda_3}{2})}\right\}\left\{\frac{(\lambda_7+\lambda_8)}{2}+\sqrt{\lambda_\Phi(\lambda_2+\frac{\lambda_3}{2})}\right\}\left\{\frac{(\lambda_5+\lambda_6-2|\lambda_9|)}{2}+\sqrt{\frac{\lambda}{4}\lambda_\Phi}\right\}} \geq 0}, \\
&&\sqrt{\frac{\lambda}{4}(\lambda_2+\lambda_3)\lambda_\Phi}+\frac{\lambda_1}{2}\sqrt{\lambda_\Phi}+\frac{(\lambda_5+\lambda_6-2|\lambda_9|)}{2}\sqrt{\lambda_2+\lambda_3}+\frac{(\lambda_7+\lambda_8)}{2}\sqrt{\frac{\lambda}{4}}+ \nonumber \\ && {\tiny \sqrt{2\left\{\frac{\lambda_1}{2}+\sqrt{\frac{\lambda}{4}(\lambda_2+\lambda_3)}\right\}\left\{\frac{(\lambda_7+\lambda_8)}{2}+\sqrt{\lambda_\Phi(\lambda_2+\lambda_3)}\right\}\left\{\frac{(\lambda_5+\lambda_6-2|\lambda_9|)}{2}+\sqrt{\frac{\lambda}{4}\lambda_\Phi}\right\}} \geq 0}, \\
&&\sqrt{\frac{\lambda}{4}(\lambda_2+\frac{\lambda_3}{2})\lambda_\Phi}+\frac{(\lambda_1+\lambda_4)}{2}\sqrt{\lambda_\Phi}+\frac{(\lambda_5+\lambda_6-2|\lambda_9|)}{2}\sqrt{\lambda_2+\frac{\lambda_3}{2}}+\frac{\lambda_7}{2}\sqrt{\frac{\lambda}{4}} + \nonumber \\ && {\tiny \sqrt{2\left\{\frac{(\lambda_1+\lambda_4)}{2}+\sqrt{\frac{\lambda}{4}(\lambda_2+\frac{\lambda_3}{2})}\right\}\left\{\frac{\lambda_7}{2}+\sqrt{\lambda_\Phi(\lambda_2+\frac{\lambda_3}{2})}\right\}\left\{\frac{(\lambda_5+\lambda_6-2|\lambda_9|)}{2}+\sqrt{\frac{\lambda}{4}\lambda_\Phi}\right\}} \geq 0}, \\
&&\sqrt{\frac{\lambda}{4}(\lambda_2+\lambda_3)\lambda_\Phi}+\frac{(\lambda_1+\lambda_4)}{2}\sqrt{\lambda_\Phi}+\frac{(\lambda_5+\lambda_6-2|\lambda_9|)}{2}\sqrt{\lambda_2+\lambda_3}+\frac{\lambda_7}{2}\sqrt{\frac{\lambda}{4}}+ \nonumber \\ && {\tiny \sqrt{2\left\{\frac{(\lambda_1+\lambda_4)}{2}+\sqrt{\frac{\lambda}{4}(\lambda_2+\lambda_3)}\right\}\left\{\frac{\lambda_7}{2}+\sqrt{\lambda_\Phi(\lambda_2+\lambda_3)}\right\}\left\{\frac{(\lambda_5+\lambda_6-2|\lambda_9|)}{2}+\sqrt{\frac{\lambda}{4}\lambda_\Phi}\right\}} \geq 0},\\
&&\sqrt{\frac{\lambda}{4}(\lambda_2+\frac{\lambda_3}{2})\lambda_\Phi}+\frac{(\lambda_1+\lambda_4)}{2}\sqrt{\lambda_\Phi}+\frac{(\lambda_5+\lambda_6-2|\lambda_9|)}{2}\sqrt{\lambda_2+\frac{\lambda_3}{2}}+\frac{(\lambda_7+\lambda_8)}{2}\sqrt{\frac{\lambda}{4}}\nonumber \\
&& + {\tiny \sqrt{2\left\{\frac{(\lambda_1+\lambda_4)}{2}+\sqrt{\frac{\lambda}{4}(\lambda_2+\frac{\lambda_3}{2})}\right\}\left\{\frac{(\lambda_7+\lambda_8)}{2}+\sqrt{\lambda_\Phi(\lambda_2+\frac{\lambda_3}{2})}\right\}\left\{\frac{(\lambda_5+\lambda_6-2|\lambda_9|)}{2}+\sqrt{\frac{\lambda}{4}\lambda_\Phi}\right\}} \geq 0}, \\
&&\sqrt{\frac{\lambda}{4}(\lambda_2+\lambda_3)\lambda_\Phi}+\frac{(\lambda_1+\lambda_4)}{2}\sqrt{\lambda_\Phi}+\frac{(\lambda_5+\lambda_6-2|\lambda_9|)}{2}\sqrt{\lambda_2+\lambda_3}+\frac{(\lambda_7+\lambda_8)}{2}\sqrt{\frac{\lambda}{4}}\nonumber \\
&& + {\tiny \sqrt{2\left\{\frac{(\lambda_1+\lambda_4)}{2}+\sqrt{\frac{\lambda}{4}(\lambda_2+\lambda_3)}\right\}\left\{\frac{(\lambda_7+\lambda_8)}{2}+\sqrt{\lambda_\Phi(\lambda_2+\lambda_3)}\right\}\left\{\frac{(\lambda_5+\lambda_6-2|\lambda_9|)}{2}+\sqrt{\frac{\lambda}{4}\lambda_\Phi}\right\}} \geq 0}, 
\label{eq:lstab3}
\end{eqnarray} 
\end{subequations}
}

\section{Relevant couplings of dark matter field with other other fields}\label{dm_coupling}
$\bullet$ \underline {Trilinear coupling of dark matter with other scalar fields:}
\begin{eqnarray}
\phi^0 \phi^0 h^0 &:& -\left(\lambda_5+\lambda_6+2\lambda_9\right)\cos\alpha\,v_d - \bigg(\lambda_7+\lambda_8-\frac{\sqrt{2}\tilde{\mu}}
{v_t}\bigg)\sin\alpha\,v_t\\
\phi^0 \phi^0 H^0 &:& -\left(\lambda_5+\lambda_6+2\lambda_9\right)\sin\alpha\,v_d - \bigg(\lambda_7+\lambda_8-\frac{\sqrt{2}\tilde{\mu}}
{v_t}\bigg)\cos\alpha\,v_t\\
\phi^0 a^0 A^0 &:& (4\lambda_9 v_t + \sqrt{2}\tilde{\mu})\cos\beta\\
\phi^0 \phi^\pm H^\mp &:& \frac{\cos\beta'\left(4\tilde{\mu}+\sqrt{2}(2\lambda_6-\lambda_8+4\lambda_9)v_t\right)}{4}
\end{eqnarray}
$\bullet$ \underline {Quartic coupling of dark matter with other scalar fields:}
\begin{eqnarray}
\phi^0 \phi^0 h^0 h^0 &:& -(\lambda_5+\lambda_6+2\lambda_9)\cos^2\alpha-(\lambda_7+\lambda_8)\sin^2\alpha \\
\phi^0 \phi^0 H^0 H^0 &:& -(\lambda_5+\lambda_6+2\lambda_9)\sin^2\alpha-(\lambda_7+\lambda_8)\cos^2\alpha \\
\phi^0 \phi^0 A^0 A^0 &:& -(\lambda_5+\lambda_6+2\lambda_9)\sin^2\beta-(\lambda_7+\lambda_8)\cos^2\beta \\
\phi^0 \phi^0 H^+ H^- &:& -\lambda_5\sin^2\beta'-(\lambda_7+\frac{\lambda_8}{2})\cos^2\beta'\\
\phi^0 \phi^0 H^{++} H^{--} &:& -\lambda_7\\
\phi^0 \phi^0 h^0 H^0 &:& -(\lambda_5+\lambda_6+2\lambda_9)\sin\alpha\cos\alpha-(\lambda_7+\lambda_8)\sin\alpha\cos\alpha\\
\phi^0 \phi^\pm h^0 H^\mp &:& -\frac{\bigg(\lambda_8\sin\alpha\cos\beta'-\sqrt{2}(\lambda_6+2\lambda_9)\sin\beta'\cos\alpha\bigg)}{2\sqrt{2}}\\
\phi^0 \phi^\pm H^0 H^\mp &:& -\frac{\bigg(\lambda_8\cos\alpha\cos\beta'+\sqrt{2}(\lambda_6+2\lambda_9)\sin\beta'\sin\alpha\bigg)}{2\sqrt{2}}\\
\phi^0 \phi^\pm A^0 H^\mp &:& \mp\frac{i\bigg(\lambda_8\cos\beta\cos\beta'+\sqrt{2}(\lambda_6-2\lambda_9)\sin\beta'\sin\beta\bigg)}{2\sqrt{2}}\\
\phi^0 a^0 h^0 A^0 &:& 2\lambda_9\sin\beta\cos\alpha\\
\phi^0 a^0 H^0 A^0 &:& -2\lambda_9\sin\beta\sin\alpha\\
\phi^0 \phi^\pm H^\pm H^{\mp\mp} &:& \mp\frac{\lambda_8}{2}\cos\beta'\\
\phi^0 \phi^0 \phi^0 \phi^0 &:& -6\lambda_\Phi\\
\phi^0 \phi^0 a^0 a^0 &:& -2\lambda_\Phi\\
\phi^0 \phi^0 \phi^+ \phi^- &:& -2\lambda_\Phi
\end{eqnarray}
$\bullet$ \underline {Trilinear coupling of dark matter with  gauge fields:}
\begin{eqnarray}
\phi^0\phi^\pm W^\mp_\mu &:& \mp\frac{e}{2\sin\theta_W}(p_1-p_2)_\mu\\
\phi^0a^0 Z_\mu &:& i\frac{e}{2\sin\theta_W \cos\theta_W}(p_1-p_2)_\mu
\end{eqnarray}
$\bullet$ \underline {Quartic coupling of dark matter with  gauge fields:}
\begin{eqnarray}
\phi^0\phi^0 W^+_\mu W^-_\nu &:& \frac{e^2}{2\sin^2\theta_W}g_{\mu\nu}\\
\phi^0\phi^0 W^\pm_\mu Z_\nu &:& -\frac{e^2}{2\cos\theta_W}g_{\mu\nu}\\
\phi^0\phi^0 Z_\mu Z_\nu &:& \frac{e^2}{2\sin^2\theta_W\cos^2\theta_W}g_{\mu\nu}
\end{eqnarray}

\end{appendices}

\newpage

\bibliographystyle{jhep}
\bibliography{bs_v1}

\end{document}